\documentclass[aps, prx, reprint, superscriptaddress]{revtex4-2}
\usepackage{amsbsy}
\usepackage{amsfonts}
\usepackage{amsmath}
\usepackage{amstext}
\usepackage{amsthm}
\usepackage{amssymb}
\usepackage{graphicx}
\usepackage{epsfig}
\usepackage{enumerate}
\usepackage{epstopdf}
\usepackage{bm}
\usepackage{bbm}
\usepackage{braket}
\usepackage{color}
\definecolor{prlblue}{rgb}{0.18,0.19,0.57}
\usepackage{multirow}
\usepackage[colorlinks]{hyperref}
\usepackage{cleveref}
\usepackage{ifthen}
\usepackage{tikz}
\usetikzlibrary{through,hobby}
\usetikzlibrary{patterns.meta} 
\usetikzlibrary{calc}

\newcommand{\1}{\text{\uppercase\expandafter{\romannumeral1}}}
\newcommand{\2}{\text{\uppercase\expandafter{\romannumeral2}}}
\newcommand{\3}{\text{\uppercase\expandafter{\romannumeral3}}}
\newcommand{\4}{\text{\uppercase\expandafter{\romannumeral4}}}
\newcommand{\5}{\text{\uppercase\expandafter{\romannumeral5}}}
\newcommand{\6}{\text{\uppercase\expandafter{\romannumeral6}}}

\def\H{\mathcal{H}}

\def\Z{\mathbb{Z}}

\newcommand{\rrangle}{\rangle\!\rangle}
\newcommand{\llangle}{\langle\!\langle}

\newcommand{\llpipe}{|}
\newcommand{\sket}[1]{\ensuremath{\llpipe#1\rrangle}}

\newcommand{\sbraket}[1]{\ensuremath{\llangle#1\rrangle}}

\begin{document}

\title{Locally Purified Density Operators for Symmetry-Protected Topological Phases in Mixed States}

\author{Yuchen Guo}
\thanks{These authors contributed equally to this work, alphabetically sorted.}
\affiliation{State Key Laboratory of Low-Dimensional Quantum Physics, Department of Physics, Tsinghua University, Beijing 100084, China}
\author{Jian-Hao Zhang}
\thanks{These authors contributed equally to this work, alphabetically sorted.}
\affiliation{Department of Physics, The Pennsylvania State University, University Park, Pennsylvania 16802, USA}
\author{Hao-Ran Zhang}
\affiliation{State Key Laboratory of Low-Dimensional Quantum Physics, Department of Physics, Tsinghua University, Beijing 100084, China}
\author{Shuo Yang}
\email{shuoyang@tsinghua.edu.cn}
\affiliation{State Key Laboratory of Low-Dimensional Quantum Physics, Department of Physics, Tsinghua University, Beijing 100084, China}
\affiliation{Frontier Science Center for Quantum Information, Beijing, China}
\author{Zhen Bi}
\email{zjb5184@psu.edu}
\affiliation{Department of Physics, The Pennsylvania State University, University Park, Pennsylvania 16802, USA}

\date{\today}

\begin{abstract}
We propose a tensor network approach known as the locally purified density operator (LPDO) to investigate the classification and characterization of symmetry-protected topological (SPT) phases in open quantum systems.
We extend the concept of injectivity, originally associated with matrix product states and projected entangled pair states, to LPDOs in $(1+1)D$ and $(2+1)D$ systems, unveiling two distinct types of injectivity conditions that are inherent for short-range entangled density matrices.
Within the LPDO framework, we outline a classification scheme for decohered average symmetry-protected topological (ASPT) phases, consistent with earlier results obtained through spectrum sequence techniques.
However, our approach offers an intuitive and explicit construction of ASPT states with the decorated domain wall picture emerging naturally.
We illustrate our framework with ASPT phases protected by a weak global symmetry and strong fermion parity symmetry, then extend it to a general group structure.
Moreover, we derive both the classification data and the explicit forms of the obstruction functions using the LPDO formalism, particularly in the case of nontrivial group extension between strong and weak symmetries, where intrinsic ASPT phases may emerge.
We demonstrate constructions of fixed-point LPDOs for ASPT phases in both $(1+1)D$ and $(2+1)D$, and discuss their physical realization in decohered or disordered systems.
In particular, we construct examples of intrinsic ASPT states in $(1+1)D$ and $(2+1)D$ using the LPDO formalism.
\end{abstract}

\maketitle

\section{Introduction}

Topological phases of matter have emerged as a central focus in condensed matter physics over recent decades.
In particular, the revealing of long-range entanglement patterns in many-body states has become a fundamental physical mechanism for comprehending topologically ordered phases~\cite{Levin2005, Kitaev2006, Chen2010, Fidkowski2011, Zhang2012}.
Furthermore, quantum states without intrinsic topological order can also bear nontrivial topological properties when systems host certain global symmetries, known as \textit{symmetry-protected topological (SPT)} phases~\cite{Chen2011A, Chen2011B, Levin2012, Vishwanath2013, Chen2013, Chen2014, Wang2014, Senthil2015}.
Among the prominent examples of SPT phases lies the topological insulators~\cite{Hasan2010, Qi2011}, protected by both time-reversal symmetry and $U(1)$ charge conservation symmetry.

The study of these novel phases reveals the elegant structure of many-body topology in pure states. However, the inherent exchange of energy and particles between a quantum system and its surrounding environment generally leaves the system as a mixed state. In recent years, open quantum systems have garnered considerable attention across diverse disciplines, encompassing condensed matter theory, quantum computing, and quantum information~\cite{Breuer2007, Rivas2012, Cattaneo2021, Schlimgen2021, Kamakari2022, Guo2022, Cai2023, Guo2023}. Exploring topological phases in mixed states provides insight into the impact of quantum decoherence and the durability of topological phases. These investigations could lead to the development of resilient quantum information processing and error correction strategies in the upcoming technological advances. 
Furthermore, open systems have the potential to uncover new quantum phenomena and phases that are not observable in closed systems~\cite{Weiss2012, Kessler2012, Walter2014, Kessler2021}, thus significantly broadening our fundamental understanding of many-body topology~\cite{Lu2023, Sang2024, Sang2025}.

An intriguing direction of research is the generalization of SPT phases within the realm of open quantum systems. A significant difference between the symmetry structure of pure states and mixed states is that mixed states can host two types of global symmetries that are referred to as \textit{strong (exact) symmetry} (we label it $K$ for the rest of the paper) and \textit{weak (average) symmetry} (labeled as $G$)~\cite{Buca2012, Li2023B, DeGroot2022, Ma2023A, Lee2025}.
A strong symmetry dictates that $\rho$ remains invariant under the action of a symmetry operator $U(k)$ solely on one side, as expressed by
\begin{align}
U(k)\rho=e^{i\theta}\rho.
\end{align} 
In fact, all symmetries within pure states can be thought of as strong symmetries. However, certain local decoherence can disrupt a part of the strong symmetry, leading to a weak symmetry in mixed states. A weak symmetry requires that $\rho$ be not invariant under the one-side action of the symmetry but only remains invariant under the simultaneous actions of the symmetry operators $U(g)$ on both sides, namely
\begin{align}
U(g)\rho U(g)^\dag=\rho.
\end{align}
The weak symmetry can be interpreted as a statistical symmetry within the ensemble of quantum channels the system navigates through.

Recently, the concept of \textit{average symmetry-protected topological} (ASPT) phases has been a pioneering development in the field of symmetry-protected topology in mixed quantum states~\cite{McGinley2020, DeGroot2022, Ma2023A, Lee2025, Zhang2022, Ma2023B, Zhang2023, Ma2024, Guo2024C, Guo2024D}.
These phases in mixed states are jointly protected by strong and weak symmetries. An intriguing phenomenon revealed by a previous study \cite{Ma2023B} is the existence of possible ASPT phases that lack a pure state correspondence, which means that there is no nontrivial SPT phase in pure states protected by the same symmetry group when all symmetries are strong. These phases highlight the intricate landscape of quantum phases in mixed states, revealing complexity and richness that surpass those found in pure states.

For the representation of topological phases in pure states, the tensor network (TN) approaches~\cite{Orus2014, Bridgeman2017, Cirac2021} such as the matrix product state (MPS)~\cite{Verstraete2006, Perez2007, Verstraete2008, Schollwock2011} and the projected entangled pair state (PEPS)~\cite{Schuch2007, Perez2008, Schuch2010}, offer an intuitive understanding and concise representation of the entanglement structure within many-body wave functions.
This framework has greatly facilitated the classification and construction of various topological quantum matter~\cite{Pollmann2010, Chen2011A, Chen2011B, Schuch2011, Cirac2011, Schuch2013, Yang2014, Yang2015, Williamson2016, Sahinoglu2021, Molnar2018}.

In the realm of presenting and simulating open quantum systems, the concept of locally purified density operator (LPDO) has gained prominence~\cite{Verstraete2004, Zwolak2004, Cuevas2013}.
These LPDOs, also known as the locally purified form of matrix product density operators (MPDOs) or simply MPDOs, have found various applications in the study of thermal states~\cite{Verstraete2004}, master equations, or noisy quantum circuits~\cite{Werner2016, Cheng2021, Guo2024B}, as well as quantum state or process tomography~\cite{Guo2022, Li2023A, Torlai2023, Guo2024A}.
In addition to the physical indices affiliated with a density operator, an LPDO encompasses two distinct categories of internal indices for contraction: virtual indices which capture quantum entanglement, and Kraus indices signifying a classical mixture~\cite{Cheng2021}.
The LPDO structure has the advantage of accurately representing the physical scenario where a system interacts locally with the environment, and the environment is subsequently traced out.
Based on previous studies~\cite{Ma2023A, Ma2023B}, we make a physical assumption that the ancillary space inherits the same locality structure as the system under consideration.
As a result of this natural assumption, for an on-site weak symmetry $U(g)$ composed of direct product transformations on each site, the corresponding symmetry action on the ancilla is also factorized at each site.

In this article, we use the LPDO formalism to construct and classify the decohered ASPT phases based on this assumption.
We begin by extending the injectivity condition to LPDO, proposing weak and strong injectivity to prevent the potential emergence of topological or long-range order in mixed states.
Our attention is then directed towards LPDOs that satisfy both injectivity conditions, where the global symmetry condition is encoded into the symmetry action on the local tensor to manifest potential decohered ASPT phases.
We elucidate our methodology by considering $K=\mathbb{Z}_2^f$, i.e., fermion parity symmetry, in $(1+1)D$ systems, as it represents the most pertinent scenario where all other system symmetries are broken to weak, leaving fermion parity as the sole strong symmetry.
Subsequently, we extend this classification to an arbitrary group $K$ and an Abelian group $G$, producing results consistent with previous investigations utilizing spectral sequence techniques~\cite{Ma2023A, Ma2023B}. 
Within our formalism, we systematically treat a novel class of ASPT states, referred to as ``intrinsic ASPT" states.
These intrinsic ASPT phases are short-range entangled mixed states, defined exclusively for open quantum systems, representing a distinct mixed-state quantum phase without a pure-state equivalent.
The nontrivial topological properties are described by the commutation relations between different symmetry actions on local tensors.
Using the LPDO formalism, we construct an $(1+1)D$ intrinsic ASPT phase protected by $\mathbb{Z}_4$ symmetry, a nontrivial group extension of two $\mathbb{Z}_2$ symmetries, which does not have a pure-state SPT counterpart.
Furthermore, for a given LPDO representation, we explore the physical realization of ASPT phases in decohered and disordered systems.
Finally, we broaden our exploration by generalizing the LPDO structure and ASPT classification to higher dimensions, with an illustrative example of a $(2+1)D$ intrinsic ASPT phase on the honeycomb lattice protected by $\Z_2\times \Z_4$ symmetry.
Our proposed framework for representing and classifying ASPT phases utilizing the LPDO structure provides an intuitive graphical representation of the interplay between symmetry and topology in mixed states with a clear physical interpretation.

\section{Matrix product states for SPT phases in $(1+1)D$}\label{sec.pure_SPT}

In this section, we provide a concise review of the classification of the bosonic SPT order in $(1+1)D$ systems using MPS representations.
Specifically, distinct SPT phases are categorized by different equivalent classes of the projective representation of the symmetry group on virtual indices~\cite{Chen2011A}.

\subsection{Matrix product states and injectivity}
In a $(1+1)D$ system, a quantum state demonstrating area-law entanglement can be effectively represented by an MPS~\cite{Verstraete2006, Perez2007, Verstraete2008, Schollwock2011}, expressed as
\begin{align}
|\psi\rangle=\sum\limits_{\{i_j\}}\mathrm{Tr}\left(A^{i_1}\cdots A^{i_N}\right)|i_1\cdots i_N\rangle,
\label{MPS1}
\end{align}
where each $A^{i_j}$ denotes a rank-3 tensor $\mathsf{A}$ featuring a physical index $i$ (with dimension $d_p$) and two virtual indices $\alpha$ and $\beta$ (with dimension $D$) at each site.
The tensor $\mathsf{A}$ can be conceptualized as a mapping from the virtual space to the physical space, i.e.,
\begin{align}
\begin{tikzpicture}[scale=0.8]
\tikzstyle{sergio}=[rectangle,draw=none]
\filldraw[fill=white, draw=black, rounded corners] (0,-0.5)--(1.5,-0.5)--(1.5,0.5)--(0,0.5)--cycle;
\draw[line width=1pt] (1.5,0) -- (2,0);
\draw[line width=1pt] (0,0) -- (-0.5,0);
\draw[line width=2pt, color=red] (0.75,0.5) -- (0.75,1);
\path (0.75,0) node [style=sergio]{\large $\mathsf{A}$};
\path (1.1,1) node [style=sergio]{$|i\rangle$};
\path (-0.5,0.3) node [style=sergio]{$(\alpha|$};
\path (2,0.3) node [style=sergio]{$(\beta|$};
\path (-3.25,-0.1) node [style=sergio]{$\mathsf{A}=\sum\limits_{\alpha\beta i}A_{\alpha\beta}^{i}|i\rangle(\alpha, \beta|=$};
\end{tikzpicture}.
\label{MPS2}
\end{align}

In particular, a $(1+1)D$ short-range correlated quantum state can be efficiently represented by the so-called injective MPS~\cite{Perez2007}.
An injective MPS is defined by the property of its local tensor ($L$ sites grouped together if necessary), namely
\begin{align}
A_{\alpha\beta}^{i}: (\mathbb{C}^D)^{\otimes 2}\rightarrow (\mathbb{C}^{d_p})^{\otimes L},
\end{align}
forms an injective map.
This implies that the physical space of the local tensor can span the entire virtual space.

An injective MPS characterizes a short-range correlated state, where any two-point correlation function of local operators decays exponentially with distance~\cite{Perez2007, Orus2014},
\begin{align}
C(i, j)\equiv \braket{\psi|O_iO_j|\psi}-\braket{\psi|O_i|\psi}\braket{\psi|O_j|\psi} \sim  \text{e}^{-|i-j|/\xi},
\end{align}
where $\xi$ represents the correlation length.
The correlation length $\xi$ can be determined from the spectrum of the transfer matrix of an MPS, defined as
\begin{align}
    \mathbb{E} = \sum_{i}A^i_{\alpha^u\beta^u}{A}^{*i}_{\alpha^l\beta^l},  
\end{align}
with the following graphical representation
\begin{align}
\begin{tikzpicture}[scale=0.75]
\tikzstyle{sergio}=[rectangle,draw=none]
\filldraw[fill=white, draw=black, rounded corners] (0,-0.5)--(1.5,-0.5)--(1.5,0.5)--(0,0.5)--cycle;
\draw[line width=1pt] (1.5,0) -- (2.5,0);
\draw[line width=1pt] (0,0) -- (-1,0);
\path (0.75,0) node [style=sergio]{\large $\mathsf{A}$};
\draw[line width=2pt, color=red] (0.75,-1.5) -- (0.75,-0.5);
\path (-2.5,-1) node [style=sergio]{$\mathbb{E}=$};
\filldraw[fill=white, draw=black, rounded corners] (0,-2.5)--(1.5,-2.5)--(1.5,-1.5)--(0,-1.5)--cycle;
\path (0.75,-2) node [style=sergio]{\large $\mathsf{A}^*$};
\draw[line width=1pt] (1.5,-2) -- (2.5,-2);
\draw[line width=1pt] (0,-2) -- (-1,-2);
\end{tikzpicture}.
\end{align}
The correlation length is determined by $\xi = -1/\log{\left({|\lambda_2|}/{|\lambda_1|}\right)}$, where $\lambda_{1}$ and $\lambda_2$ denote the largest and second largest eigenvalues in magnitude, respectively, of the transfer matrix $\mathbb{E}$. Consequently, a finite correlation length $\xi$ corresponds to a nondegenerate transfer matrix $\mathbb{E}$. Indeed, the nondegeneracy of a transfer matrix serves as an equivalent definition of the injectivity of an MPS~\cite{Perez2007, Orus2014}.

From a general perspective, in $(1+1)D$ systems, intrinsic topological order is absent~\cite{Chen2010}, leading to the categorization of gapped quantum phases into two classes: short-range correlated states and long-range correlated states.
The former class is characterized by injective MPS, including trivial product states, which can be transformed into each other via finite-depth local unitary (FDLU) circuits~\cite{Chen2010}.
By imposing symmetry constraints and studying the symmetry actions on MPS representation, a comprehensive characterization of the SPT phases in $(1+1)D$ can be achieved~\cite{Chen2011A, Chen2011B}. 
Conversely, long-range correlated states are associated with the concept of spontaneous symmetry breaking (SSB). 
An illustrious example within this category is the Greenberger-Horne-Zeilinger (GHZ) state, which can be represented by a non-injective MPS with $D=2$.

\subsection{$(1+1)D$ Bosonic SPT: Projective representation on virtual indices}
When constrained to symmetric FDLU circuits, distinct short-range correlated states manifesting nontrivial SPT order are further distinguished as different quantum phases. 

In a $(1+1)D$ system with symmetry group $K$, the onsite symmetry action on the physical index of the local tensor forms a linear representation of the symmetry group, expressed as
\begin{align}
    U(k_1)U(k_2) = U(k_1k_2), \quad k_1, k_2\in K.\label{linear rep}
\end{align}
For a symmetric MPS, the symmetry action on the physical index induces a gauge transformation $V(k)$ on the virtual indices, depicted as
\begin{align}
\begin{tikzpicture}[scale=0.8]
\tikzstyle{sergio}=[rectangle,draw=none]
\filldraw[fill=white, draw=black, rounded corners] (-0.25,-0.5)--(1.25,-0.5)--(1.25,0.5)--(-0.25,0.5)--cycle;
\draw[line width=1pt] (1.25,0) -- (1.75,0);
\draw[line width=1pt] (-0.25,0) -- (-0.75,0);
\draw[line width=2pt, color=red] (0.5,0.5) -- (0.5,2);
\path (0.5,0) node [style=sergio]{\large $\mathsf{A}$};
\filldraw[fill=white, draw=black] (0.5,1.25)circle (10pt);
\path (0.5,1.25) node [style=sergio]{\small $U_k$};
\path (2.5,0) node [style=sergio]{$=e^{i\theta_k}$};
\filldraw[fill=white, draw=black, rounded corners] (4.75,-0.5)--(6.25,-0.5)--(6.25,0.5)--(4.75,0.5)--cycle;
\draw[line width=1pt] (4.75,0) -- (3.25,0);
\filldraw[fill=white, draw=black] (4,0)circle (10pt);
\path (4,0) node [style=sergio]{\scriptsize $V^{-1}_k$};
\path (5.5,0) node [style=sergio]{\large $\mathsf{A}$};
\draw[line width=1pt] (6.25,0) -- (7.75,0);
\filldraw[fill=white, draw=black] (7,0)circle (10pt);
\path (7,0) node [style=sergio]{\small $V_k$};
\draw[line width=2pt, color=red] (5.5,0.5) -- (5.5,1);
\label{symmetry_mps}
\end{tikzpicture}.
\end{align}
With the injectivity condition, the transformation $V(k)$ on the virtual index is uniquely determined up to a phase for a given MPS.

Applying Eq.~\eqref{linear rep} to the physical index of a local tensor induces transformations $V(k_1)V(k_2)$ and $V(k_1k_2)$ on the right virtual index, respectively.
As these two tensors are identical, transformations on the virtual index should be equivalent up to a $U(1)$ phase
\begin{align}
V(k_1)V(k_2)=\mu_2(k_1, k_2)V(k_1k_2), \quad \mu_2(k_1, k_2)\in U(1).\label{Vk}
\end{align}
Thus, $V(k)$ adheres to group multiplication up to a phase factor, constituting a projective representation of the symmetry group $K$.
Moreover, sequential application of $U(k_3)$, $U(k_2)$, and $U(k_1)$ results in the transformation on the virtual index as $V(k_1)V(k_2)V(k_3)$.
Ensuring consistency via the associativity condition yields
\begin{align}
\frac{\mu_2(k_1,k_2)\mu_2(k_1k_2,k_3)}{\mu_2(k_1,k_2k_3)\mu_2(k_2,k_3)}=1,\label{2-cocycle}
\end{align}
indicating that $\mu_2$ satisfies the 2-cocycle condition.
We assert that $\mu_2(k_1, k_2)$ serves as the topological invariant of $(1+1)D$ SPT phases, which may differ by a 2-coboundary $\eta_2(k_1, k_2)$.
Here $\eta_2(k_1, k_2)$ arises as the image of a 1-cochain $\mu_1(k):~K\rightarrow U(1)$ under a coboundary map
\begin{align}
    \eta_2(k_1, k_2)=\frac{\mu_1(k_1)\mu_1(k_2)}{\mu_1(k_1k_2)}.
\end{align}
Different classes of phase structure $\mu_2(k_1, k_2)$ correspond to different SPT phases, identified by the second group cohomology $\mu_2\in \H^2{[K, U(1)]}$~\cite{Chen2013}.

In the following, we prove that $\mu_2$ is a topological invariant under symmetric local unitary transformations and, as such, can be used to distinguish different SPT phases.
Consider a local unitary operator $E$.
For simplicity, we will examine the case of a $3$-local unitary, although the argument applies equally to an $n$-local unitary.
A general unitary $E$ can be represented by the local tensor form with finite bound dimensions as
\begin{align}
\label{unitarydecomp}
\begin{tikzpicture}[scale=0.8]
\tikzstyle{sergio}=[rectangle,draw=none]
\filldraw[fill=white, draw=black, rounded corners] (-7.5,-1.5)--(-3.5,-1.5)--(-3.5,-0.5)--(-7.5,-0.5)--cycle;
\draw[line width=2pt, color=red] (-7,-0.5) -- (-7,0);
\draw[line width=2pt, color=red] (-7,-2) -- (-7,-1.5);
\draw[line width=2pt, color=red] (-4,-1.5) -- (-4,-2);
\draw[line width=2pt, color=red] (-5.5,-1.5) -- (-5.5,-2);
\draw[line width=2pt, color=red] (-4,0) -- (-4,-0.5);
\draw[line width=2pt, color=red] (-5.5,-0.5) -- (-5.5,0);
\path (-5.5,-1) node [style=sergio]{\large $E$};
\path (-3,-1) node [style=sergio]{$=$};
\draw[line width=1pt] (0.5,-1) -- (-1.5,-1);
\filldraw[fill=white, draw=black, rounded corners] (-2.5,-1.5)--(-1.5,-1.5)--(-1.5,-0.5)--(-2.5,-0.5)--cycle;
\filldraw[fill=white, draw=black, rounded corners] (-1,-1.5)--(0,-1.5)--(0,-0.5)--(-1,-0.5)--cycle;
\filldraw[fill=white, draw=black, rounded corners] (0.5,-1.5)--(1.5,-1.5)--(1.5,-0.5)--(0.5,-0.5)--cycle;
\draw[line width=2pt, color=red] (-2,-0.5) -- (-2,0);
\draw[line width=2pt, color=red] (-2,-2) -- (-2,-1.5);
\draw[line width=2pt, color=red] (1,-1.5) -- (1,-2);
\draw[line width=2pt, color=red] (-0.5,-1.5) -- (-0.5,-2);
\draw[line width=2pt, color=red] (1,0) -- (1,-0.5);
\draw[line width=2pt, color=red] (-0.5,-0.5) -- (-0.5,0);
\path (-2,-1) node [style=sergio]{\large $E_1$};
\path (-0.5,-1) node [style=sergio]{\large $E_2$};
\path (1,-1) node [style=sergio]{\large $E_3$};
\end{tikzpicture}.
\end{align}
The effect of the local unitary is to modify the local tensor of the MPS as follows
\begin{align}
\begin{tikzpicture}[scale=0.8]
\tikzstyle{sergio}=[rectangle,draw=none]
\draw[line width=1pt] (8,-2.5) -- (7.5,-2.5);
\draw[line width=1pt] (3.5,-2.5) -- (3,-2.5);
\draw[line width=2pt] (6.5,-2.5) -- (4.5,-2.5);
\filldraw[fill=white, draw=black, rounded corners] (3.5,-3)--(4.5,-3)--(4.5,-2)--(3.5,-2)--cycle;
\filldraw[fill=white, draw=black, rounded corners] (5,-3)--(6,-3)--(6,-2)--(5,-2)--cycle;
\filldraw[fill=white, draw=black, rounded corners] (6.5,-3)--(7.5,-3)--(7.5,-2)--(6.5,-2)--cycle;
\draw[line width=2pt, color=red] (4,-2) -- (4,-1.5);
\draw[line width=2pt, color=red] (7,-1.5) -- (7,-2);
\draw[line width=2pt, color=red] (5.5,-2) -- (5.5,-1.5);
\path (4,-2.5) node [style=sergio]{\large $A_1$};
\path (5.5,-2.5) node [style=sergio]{\large $A_2$};
\path (7,-2.5) node [style=sergio]{\large $A_3$};
\path (2.5,-2.5) node [style=sergio]{$=$};
\draw[line width=1pt] (2,-3) -- (-3,-3);
\filldraw[fill=white, draw=black, rounded corners] (-2.5,-3.5)--(-1.5,-3.5)--(-1.5,-2.5)--(-2.5,-2.5)--cycle;
\filldraw[fill=white, draw=black, rounded corners] (-1,-3.5)--(0,-3.5)--(0,-2.5)--(-1,-2.5)--cycle;
\filldraw[fill=white, draw=black, rounded corners] (0.5,-3.5)--(1.5,-3.5)--(1.5,-2.5)--(0.5,-2.5)--cycle;
\path (-2,-3) node [style=sergio]{\large $A$};
\path (-0.5,-3) node [style=sergio]{\large $A$};
\path (1,-3) node [style=sergio]{\large $A$};
\draw[line width=1pt] (0.5,-1.5) -- (-1.5,-1.5);
\filldraw[fill=white, draw=black, rounded corners] (-2.5,-2)--(-1.5,-2)--(-1.5,-1)--(-2.5,-1)--cycle;
\filldraw[fill=white, draw=black, rounded corners] (-1,-2)--(0,-2)--(0,-1)--(-1,-1)--cycle;
\filldraw[fill=white, draw=black, rounded corners] (0.5,-2)--(1.5,-2)--(1.5,-1)--(0.5,-1)--cycle;
\draw[line width=2pt, color=red] (-2,-1) -- (-2,-0.5);
\draw[line width=2pt, color=red] (-2,-2.5) -- (-2,-2);
\draw[line width=2pt, color=red] (1,-2) -- (1,-2.5);
\draw[line width=2pt, color=red] (-0.5,-2) -- (-0.5,-2.5);
\draw[line width=2pt, color=red] (1,-0.5) -- (1,-1);
\draw[line width=2pt, color=red] (-0.5,-1) -- (-0.5,-0.5);
\path (-2,-1.5) node [style=sergio]{\large $E_1$};
\path (-0.5,-1.5) node [style=sergio]{\large $E_2$};
\path (1,-1.5) node [style=sergio]{\large $E_3$};
\end{tikzpicture}.
\end{align}
We will in the following argue that the virtual legs of the modified tensors have the same projective representation as the original tensors.
Therefore, the projective representation on the virtual legs is a topological invariant under local unitaries. 

Demanding $E$ to be symmetric, namely $UEU^{\dagger}=E$, we have the following tensor network equation
\begin{align}
\begin{tikzpicture}[scale=0.8]
\tikzstyle{sergio}=[rectangle,draw=none]
\draw[line width=1pt] (-4.5,-1) -- (-6.5,-1);
\filldraw[fill=white, draw=black, rounded corners] (-7.5,-1.5)--(-6.5,-1.5)--(-6.5,-0.5)--(-7.5,-0.5)--cycle;
\filldraw[fill=white, draw=black, rounded corners] (-6,-1.5)--(-5,-1.5)--(-5,-0.5)--(-6,-0.5)--cycle;
\filldraw[fill=white, draw=black, rounded corners] (-4.5,-1.5)--(-3.5,-1.5)--(-3.5,-0.5)--(-4.5,-0.5)--cycle;
\draw[line width=2pt, color=red] (-7,-0.5) -- (-7,1);
\draw[line width=2pt, color=red] (-7,-3) -- (-7,-1.5);
\draw[line width=2pt, color=red] (-4,-1.5) -- (-4,-3);
\draw[line width=2pt, color=red] (-5.5,-1.5) -- (-5.5,-3);
\draw[line width=2pt, color=red] (-4,1) -- (-4,-0.5);
\draw[line width=2pt, color=red] (-5.5,-0.5) -- (-5.5,1);
\path (-7,-1) node [style=sergio]{\large $E_1$};
\path (-5.5,-1) node [style=sergio]{\large $E_2$};
\path (-4,-1) node [style=sergio]{\large $E_3$};
\path (-3,-1) node [style=sergio]{$=$};
\filldraw[fill=white, draw=black] (-7,0.25)circle (10pt);
\path (-7,0.25) node [style=sergio]{\small $U_k$};
\filldraw[fill=white, draw=black] (-5.5,0.25)circle (10pt);
\path (-5.5,0.25) node [style=sergio]{\small $U_k$};
\filldraw[fill=white, draw=black] (-4,0.25)circle (10pt);
\path (-4,0.25) node [style=sergio]{\small $U_k$};
\filldraw[fill=white, draw=black] (-7,-2.25)circle (10pt);
\path (-7,-2.25) node [style=sergio]{\small $U_k^*$};
\filldraw[fill=white, draw=black] (-5.5,-2.25)circle (10pt);
\path (-5.5,-2.25) node [style=sergio]{\small $U_k^*$};
\filldraw[fill=white, draw=black] (-4,-2.25)circle (10pt);
\path (-4,-2.25) node [style=sergio]{\small $U_k^*$};
\draw[line width=1pt] (0.5,-1) -- (-1.5,-1);
\filldraw[fill=white, draw=black, rounded corners] (-2.5,-1.5)--(-1.5,-1.5)--(-1.5,-0.5)--(-2.5,-0.5)--cycle;
\filldraw[fill=white, draw=black, rounded corners] (-1,-1.5)--(0,-1.5)--(0,-0.5)--(-1,-0.5)--cycle;
\filldraw[fill=white, draw=black, rounded corners] (0.5,-1.5)--(1.5,-1.5)--(1.5,-0.5)--(0.5,-0.5)--cycle;
\draw[line width=2pt, color=red] (-2,-0.5) -- (-2,0);
\draw[line width=2pt, color=red] (-2,-2) -- (-2,-1.5);
\draw[line width=2pt, color=red] (1,-1.5) -- (1,-2);
\draw[line width=2pt, color=red] (-0.5,-1.5) -- (-0.5,-2);
\draw[line width=2pt, color=red] (1,0) -- (1,-0.5);
\draw[line width=2pt, color=red] (-0.5,-0.5) -- (-0.5,0);
\path (-2,-1) node [style=sergio]{\large $E_1$};
\path (-0.5,-1) node [style=sergio]{\large $E_2$};
\path (1,-1) node [style=sergio]{\large $E_3$};
\end{tikzpicture}.
\label{3-local symmetry}
\end{align}
The local tensors of $E$ transform in the following way
\begin{align}
\begin{tikzpicture}[scale=0.8]
\tikzstyle{sergio}=[rectangle,draw=none]
\filldraw[fill=white, draw=black, rounded corners] (-7.5,3)--(-6.5,3)--(-6.5,4)--(-7.5,4)--cycle;
\draw[line width=2pt, color=red] (-7,4) -- (-7,5.5);
\draw[line width=2pt, color=red] (-7,1.5) -- (-7,3);
\path (-7,3.5) node [style=sergio]{\large $E_1$};
\filldraw[fill=white, draw=black] (-7,4.75)circle (10pt);
\path (-7,4.75) node [style=sergio]{\small $U_k$};
\filldraw[fill=white, draw=black] (-7,2.25)circle (10pt);
\path (-7,2.25) node [style=sergio]{\small $U_k^*$};
\draw[line width=1pt] (-6,3.5) -- (-6.5,3.5);
\path (-5.5,3.5) node [style=sergio]{$=$};
\draw[line width=1pt] (-1,3.5) -- (-2.5,3.5);
\filldraw[fill=white, draw=black, rounded corners] (-3.5,3)--(-2.5,3)--(-2.5,4)--(-3.5,4)--cycle;
\draw[line width=2pt, color=red] (-3,4) -- (-3,4.5);
\draw[line width=2pt, color=red] (-3,2.5) -- (-3,3);
\path (-3,3.5) node [style=sergio]{\large $E_1$};
\filldraw[fill=white, draw=black] (-1.75,3.5)circle (10pt);
\path (-1.75,3.5) node [style=sergio]{\small $V_{1k}$};
\filldraw[fill=white, draw=black, rounded corners] (-7.5,-1.5)--(-6.5,-1.5)--(-6.5,-0.5)--(-7.5,-0.5)--cycle;
\draw[line width=2pt, color=red] (-7,-0.5) -- (-7,1);
\draw[line width=2pt, color=red] (-7,-3) -- (-7,-1.5);
\path (-7,-1) node [style=sergio]{\large $E_2$};
\filldraw[fill=white, draw=black] (-7,0.25)circle (10pt);
\path (-7,0.25) node [style=sergio]{\small $U_k$};
\filldraw[fill=white, draw=black] (-7,-2.25)circle (10pt);
\path (-7,-2.25) node [style=sergio]{\small $U_k^*$};
\draw[line width=1pt] (-6,-1) -- (-6.5,-1);
\draw[line width=1pt] (-7.5,-1) -- (-8,-1);
\path (-5.5,-1) node [style=sergio]{$=$};
\draw[line width=1pt] (-3.5,-1) -- (-5,-1);
\filldraw[fill=white, draw=black, rounded corners] (-3.5,-1.5)--(-2.5,-1.5)--(-2.5,-0.5)--(-3.5,-0.5)--cycle;
\draw[line width=2pt, color=red] (-3,-0.5) -- (-3,0);
\draw[line width=2pt, color=red] (-3,-2) -- (-3,-1.5);
\path (-3,-1) node [style=sergio]{\large $E_2$};
\filldraw[fill=white, draw=black] (-4.25,-1)circle (10pt);
\path (-4.25,-1) node [style=sergio]{\small $V_{1k}^{-1}$};
\draw[line width=1pt] (-1,-1) -- (-2.5,-1);
\filldraw[fill=white, draw=black] (-1.75,-1)circle (10pt);
\path (-1.75,-1) node [style=sergio]{\small $V_{2k}$};
\filldraw[fill=white, draw=black, rounded corners] (-7.5,-6)--(-6.5,-6)--(-6.5,-5)--(-7.5,-5)--cycle;
\draw[line width=2pt, color=red] (-7,-5) -- (-7,-3.5);
\draw[line width=2pt, color=red] (-7,-7.5) -- (-7,-6);
\path (-7,-5.5) node [style=sergio]{\large $E_3$};
\filldraw[fill=white, draw=black] (-7,-4.25)circle (10pt);
\path (-7,-4.25) node [style=sergio]{\small $U_k$};
\filldraw[fill=white, draw=black] (-7,-6.75)circle (10pt);
\path (-7,-6.75) node [style=sergio]{\small $U_k^*$};
\draw[line width=1pt] (-7.5,-5.5) -- (-8,-5.5);
\path (-5.5,-5.5) node [style=sergio]{$=$};
\draw[line width=1pt] (-3.5,-5.5) -- (-5,-5.5);
\filldraw[fill=white, draw=black, rounded corners] (-3.5,-6)--(-2.5,-6)--(-2.5,-5)--(-3.5,-5)--cycle;
\draw[line width=2pt, color=red] (-3,-5) -- (-3,-4.5);
\draw[line width=2pt, color=red] (-3,-6.5) -- (-3,-6);
\path (-3,-5.5) node [style=sergio]{\large $E_3$};
\filldraw[fill=white, draw=black] (-4.25,-5.5)circle (10pt);
\path (-4.25,-5.5) node [style=sergio]{\small $V_{2k}^{-1}$};
\end{tikzpicture}.
\end{align}
First, the symmetry transformations on the virtual legs from two neighboring tensors must cancel each other to preserve the overall symmetry, i.e., $V_i$ and $V_i^{-1}$.
Next, the representations $V_i$ and $V_{i+1}$ associated with the same bulk tensor $E_i$ must belong to the same projective representation.
Otherwise, the projective phase for combining or commuting two symmetry actions from the left and right virtual legs would not cancel, rendering the tensor $E_i$ inconsistent with the symmetry.
Finally, the boundary tensors introduce an additional constraint.
For the boundary tensor to align with the symmetry, the representations on the first and last virtual legs must be trivial projective representations, i.e., linear representations of the group $K$.
As a result, all representations on the virtual legs of the local unitary become linear representations.
Therefore, the class of projective representations on the virtual indices of an MPS remains unchanged after applying a local symmetric unitary.

In summary, the symmetry action on the virtual indices of an injective MPS offers an intuitive and comprehensive approach to the classification and construction of SPT phases. 

\section{Locally purified density operators in $(1+1)D$}

In this section, we discuss the definition of LPDO in $(1+1)D$ systems, representing a universal tensor network structure that is adept at describing short-range correlated density matrices.

\subsection{Definition of LPDO}
A mixed-state density matrix $\rho$ can be effectively represented by an MPDO as~\cite{Verstraete2004}
\begin{align}
\rho=\sum\limits_{\{i_j,i_j'\}}\mathrm{Tr}\left(A^{i_1i_1'}\cdots A^{i_Ni_N'}\right)|i_1\cdots i_N\rangle\langle i_1'\cdots i_N'|,
\end{align}
where $\mathsf{A}$ is a rank-4 tensor featuring two physical indices $i$ and $i'$, along with two virtual indices $\alpha$ and $\beta$ at each site
\begin{align}
\begin{tikzpicture}[scale=0.8]
\tikzstyle{sergio}=[rectangle,draw=none]
\filldraw[fill=white, draw=black, rounded corners] (0,-0.5)--(1.5,-0.5)--(1.5,0.5)--(0,0.5)--cycle;
\draw[line width=1pt] (1.5,0) -- (2,0);
\draw[line width=1pt] (0,0) -- (-0.5,0);
\draw[line width=2pt, color=red] (0.75,0.5) -- (0.75,1);
\draw[line width=2pt, color=red] (0.75,-0.5) -- (0.75,-1);
\path (0.75,0) node [style=sergio]{\large $\mathsf{A}$};
\path (1.1,1) node [style=sergio]{$|i\rangle$};
\path (1.1,-1) node [style=sergio]{$\langle i'|$};
\path (-0.5,0.3) node [style=sergio]{$(\alpha|$};
\path (2,0.3) node [style=sergio]{$(\beta|$};
\path (-3.5,-0.1) node [style=sergio]{$\mathsf{A}=\sum\limits_{\alpha\beta ii'}A_{\alpha\beta}^{ii'}|i\rangle\langle i'|(\alpha, \beta|=$};
\end{tikzpicture}.
\end{align}
However, a general MPDO lacks the capability to accurately capture essential properties of a density matrix, such as Hermiticity and semidefinite positivity, solely through conditions imposed on local tensors.
Furthermore, the entanglement structure of a mixed state is inadequately represented by the virtual indices of an MPDO, unlike the clarity achieved in the pure state case with MPS, as both quantum entanglement and classical correlation contribute to the virtual spaces.
Hence, while the MPDO structure offers a compact representation and efficient algorithms for studying and simulating many-body open systems~\cite{Cui2015, Noh2020}, its ambiguous physical interpretation, particularly concerning the role of virtual indices, hampers its utility for the classification of quantum phases and topological matter in open systems.

Fortunately, a significant subclass of MPDOs exists, known as LPDOs, which exhibit a much clearer entanglement structure and can be interpreted as locally purifying a mixed state~\cite{Verstraete2004, Zwolak2004, Cuevas2013}.
To elucidate further, let us commence with a pure state $|\psi_{p\otimes a}\rangle$ belonging to the composite Hilbert space $\mathcal{H}_p \otimes \mathcal{H}_a$, taking the form of the following MPS
\begin{align}
\begin{tikzpicture}[scale=0.7]
\tikzstyle{sergio}=[rectangle,draw=none]
\filldraw[fill=white, draw=black, rounded corners] (0,-0.5)--(1.5,-0.5)--(1.5,0.5)--(0,0.5)--cycle;
\filldraw[fill=white, draw=black, rounded corners] (2.5,-0.5)--(4,-0.5)--(4,0.5)--(2.5,0.5)--cycle;
\draw[line width=1pt] (1.5,0) -- (2.5,0);
\draw[line width=1pt] (0,0) -- (-1,0);
\draw[line width=1pt] (4,0) -- (5,0);
\draw[line width=2pt, color=red] (0.5,0.5) -- (0.5,1);
\draw[line width=3pt, color=blue] (1,0.5) -- (1,1);
\path (0.25,1.25) node [style=sergio]{$p$};
\path (1.25,1.25) node [style=sergio]{$a$};
\draw[line width=2pt, color=red] (3,0.5) -- (3,1);
\draw[line width=3pt, color=blue] (3.5,0.5) -- (3.5,1);
\path (0.75,0) node [style=sergio]{\large $\mathsf{A}$};
\path (3.25,0) node [style=sergio]{\large $\mathsf{A}$};
\path (2.75,1.25) node [style=sergio]{$p$};
\path (3.75,1.25) node [style=sergio]{$a$};
\path (-1.5,0) node [style=sergio]{$\cdots$};
\path (5.5,0) node [style=sergio]{$\cdots$};
\path (-3.25,0) node [style=sergio]{$|\psi_{p\otimes a}\rangle=$};
\end{tikzpicture},
\label{1d MPS}
\end{align}
where $p$ denotes the physical degrees of freedom and $a$ denotes the ancillary degrees of freedom.
Subsequently, an LPDO is constructed by tracing out the ancillae $a$, resulting in
\begin{align}
    \rho=\mathrm{Tr}_a\left(|\psi_{p\otimes a}\rangle\langle\psi_{p\otimes a}|\right).
\end{align}
In other words, an LPDO has an alternative graphical representation
\begin{align}
\begin{tikzpicture}[scale=0.75]
\tikzstyle{sergio}=[rectangle,draw=none]
\filldraw[fill=white, draw=black, rounded corners] (0,-0.5)--(1.5,-0.5)--(1.5,0.5)--(0,0.5)--cycle;
\filldraw[fill=white, draw=black, rounded corners] (2.5,-0.5)--(4,-0.5)--(4,0.5)--(2.5,0.5)--cycle;
\draw[line width=1pt] (1.5,0) -- (2.5,0);
\draw[line width=1pt] (0,0) -- (-1,0);
\path (0.75,0) node [style=sergio]{\large $\mathsf{A}$};
\path (3.25,0) node [style=sergio]{\large $\mathsf{A}$};
\draw[line width=1pt] (4,0) -- (5,0);
\draw[line width=2pt, color=red] (0.75,0.5) -- (0.75,1);
\draw[line width=2pt, color=red] (0.75,-2.5) -- (0.75,-3);
\draw[line width=3pt, color=blue] (0.75,-1.5) -- (0.75,-0.5);
\path (0.5,1) node [style=sergio]{$p$};
\path (0.5,-1) node [style=sergio]{$a$};
\draw[line width=2pt, color=red] (3.25,-2.5) -- (3.25,-3);
\draw[line width=2pt, color=red] (3.25,0.5) -- (3.25,1);
\draw[line width=3pt, color=blue] (3.25,-1.5) -- (3.25,-0.5);
\path (3,1) node [style=sergio]{$p$};
\path (3,-1) node [style=sergio]{$a$};
\path (-1.5,-1) node [style=sergio]{$\cdots$};
\path (5.5,-1) node [style=sergio]{$\cdots$};
\path (-2.5,-1) node [style=sergio]{$\rho=$};
\filldraw[fill=white, draw=black, rounded corners] (2.5,-2.5)--(4,-2.5)--(4,-1.5)--(2.5,-1.5)--cycle;
\filldraw[fill=white, draw=black, rounded corners] (0,-2.5)--(1.5,-2.5)--(1.5,-1.5)--(0,-1.5)--cycle;
\path (0.5,-3) node [style=sergio]{$p$};
\path (3,-3) node [style=sergio]{$p$};
\path (3.25,-2) node [style=sergio]{\large $\mathsf{A}^*$};
\path (0.75,-2) node [style=sergio]{\large $\mathsf{A}^*$};
\draw[line width=1pt] (1.5,-2) -- (2.5,-2);
\draw[line width=1pt] (0,-2) -- (-1,-2);
\draw[line width=1pt] (4,-2) -- (5,-2);
\end{tikzpicture},
\label{LPDO}
\end{align}
where the inner index $a$ is termed the ``Kraus index'' with dimension $d_{\kappa}$, which signifies the classical mixture of distinct quantum states.

It is essential to note that each physical index is accompanied by a Kraus index, representing the environmental degree of freedom.
Hence, $\ket{\psi_{p\otimes a}}$ in Eq.~\eqref{1d MPS} serves as a local purification of the density matrix $\rho$ within the LPDO framework.
Analogous to the well-established connection between the locality of interaction and the efficient MPS representation, achieving local purification requires locality between system and environment, ensuring that the system spins only interact with adjacent ancillae.

\subsection{The weak injectivity condition}
\begin{figure}
\centering
\includegraphics[width=0.8\linewidth]{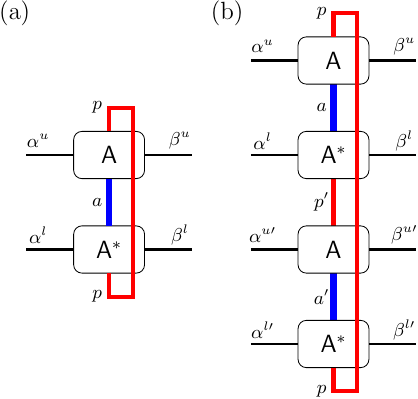}
\caption{Transfer matrices corresponding to (a) weak injectivity in Eq.~\eqref{weak transfer matrix} and (b) strong injectivity in Eq.~\eqref{strong transfer matrix} respectively.}
\label{injectivity}
\end{figure}

In the pure state scenario discussed in Section~\ref{sec.pure_SPT}, nontrivial SPT phases arise from the class of short-range correlated states, or injective MPSs.
These states, without symmetry constraints, can be transformed into trivial product states under FDLU circuits.
Incorporating symmetries, the injectivity of the MPS yields a unique mapping from the symmetry transformation in the physical space to that in the virtual space, facilitating the classification of SPT phases in pure states.
Therefore, the extension of the injectivity condition to LPDO becomes essential.

In our investigation, we identify two versions of the injectivity conditions linked with LPDOs, crucial for mitigating the presence of topological or long-range order phenomena that could potentially compromise the symmetry preservation inherent to ASPT phases.
Within this section, our initial focus is on the weak injectivity condition, defined for the purified MPS as described in Eq.~\eqref{1d MPS}.

\paragraph*{Definition} (Weak injectivity condition) 
An LPDO is deemed weakly injective provided that the corresponding purified MPS in Eq.~\eqref{1d MPS} exhibits injectivity, wherein its local tensor forms an injective map
\begin{align}
    A_{\alpha\beta}^{p,a}: (\mathbb{C}^D)^{\otimes 2}\rightarrow (\mathbb{C}^{d_p}\otimes \mathbb{C}^{d_{\kappa}})^{\otimes L}.
\end{align}

Returning to the purified MPS in Eq.~\eqref{1d MPS} and leveraging the resemblance between the weak injectivity of LPDO and injectivity of MPS, one can directly illustrate that for a weakly injective LPDO, the two-point correlation function of the mixed state $\rho$
\begin{align}
\begin{aligned}
C^{(1)}(i, j) &\equiv \mathrm{Tr}\left(\rho O_i O_j\right) - \mathrm{Tr}\left(\rho O_j\right)\mathrm{Tr}\left(\rho O_j\right)\\
= &\braket{\psi_{p\otimes a}|O_i^pO_j^p|\psi_{p\otimes a}}\\
&-\braket{\psi_{p\otimes a}|O_i^p|\psi_{p\otimes a}}\braket{\psi_{p\otimes a}|O_j^p|\psi_{p\otimes a}}\\ \sim&\ e^{-|i-j|/\xi}
\end{aligned}\label{Linear_corr}
\end{align}
decays exponentially with distance.
This necessitates that the transfer matrix
\begin{align}
    \mathbb{E} = \sum_{p, a}A^{p, a}_{\alpha^u\beta^u}{A}^{*p, a}_{\alpha^l\beta^l},  
    \label{weak transfer matrix}
\end{align}
depicted graphically in Fig.~\ref{injectivity}(a) possesses a non-degenerate spectrum, aligning with the conventional injectivity condition for the purified state.

Through the application of the weak injectivity condition, it becomes apparent that the LPDO formulated in this context can be linked bi-directionally to a trivial product state utilizing finite-depth local quantum channels.
This implies that LPDOs conforming to the weak injectivity condition signify density matrices with short-range entanglement and correlations.

Firstly, it is straightforward to establish that an LPDO can be derived from a trivial product state through the utilization of a finite-depth local quantum channel.
This quantum channel can be obtained by executing an FDLU circuit across both the physical and ancillary spaces, followed by the subsequent elimination of the ancillary degrees of freedom.
The weak injectivity property of an LPDO implies that its purification, denoted as $|\psi_{p\otimes a}\rangle$, can be adiabatically connected via FDLU circuits to a trivial product state within the composite Hilbert space that encompasses both physical and ancillary degrees of freedom.
This connection is succinctly expressed as:
\begin{align}
|\psi_{p\otimes a}\rangle=U\ket{0}_p\otimes\ket{0}_a.
\end{align}
Hence, we achieve a clear and explicit construction of the quantum channel that facilitates the trajectory from the trivial product state to an LPDO.
\begin{align}
\rho_{\mathrm{LPDO}}=\mathrm{Tr}_a\left(U\ket{0}\langle0|_p\otimes\ket{0}\langle0|_aU^\dag\right).\label{Unitary_channel}
\end{align}

Conversely, it is always feasible to convert a density matrix into a trivial product state applying a local amplitude damping channel~\cite{Nielsen2010}
\begin{align}
    \ket{0}\hspace{-1mm}\bra{0} = \left(\bigotimes_{i=1}^N\mathcal{E}_i\right)(\rho),
\end{align}
where the Kraus operators for $\mathcal{E}_i$ at each site $i$ are defined as follows
\begin{align}
    K_1 = \ket{0}\hspace{-1mm}\bra{0}, K_2 = \ket{0}\hspace{-1mm}\bra{1}.
\end{align}
In summary, weakly injective LPDOs describe density matrices entangled over short ranges, a fundamental prerequisite for decohered ASPTs.

\subsection{The strong injectivity condition}

In essence, the presence of a weakly injective LPDO precludes the emergence of a long-range order within linear correlation functions for the density matrix.
Consequently, this mechanism serves to avert conventional instances of spontaneous symmetry breaking within mixed states.
However, recent studies~\cite{Lee2023, Ma2023B, Sala2024, Lessa2024B} underscore the possibility of a more subtle manifestation of symmetry-breaking patterns within mixed states.
Specifically, it has been shown that it is possible to engineer mixed states exhibiting short-range entanglement, while spontaneously breaking a strong symmetry to a weak counterpart.
Such intricate patterns elude detection by conventional means, such as the linear two-point correlation function for the density matrix.
In order to comprehensively address symmetric mixed states, it becomes imperative to preclude the prospect of such strong-to-weak spontaneous symmetry-breaking (SW-SSB) orders.
Consequently, there arises the need for a stringent criterion, which we refer to as the strong injectivity condition.

It is convenient to introduce the strong injectivity condition within the double space formalism.
Let us assume that the spectrum decomposition of the density matrix is given by
\begin{align}
    \rho = \sum_k \lambda_k\ket{\psi_k}\hspace{-1mm}\bra{\psi_k}.   
\end{align}
The (unnormalized) double state for the mixed state is defined as
\begin{align}
    \sket{\rho} = \sum_k \lambda_k\ket{\psi_k}\otimes\ket{\psi_k^*},
\end{align}
which can be graphically depicted as
\begin{align}
\begin{tikzpicture}[scale=0.75]
\tikzstyle{sergio}=[rectangle,draw=none]
\filldraw[fill=white, draw=black, rounded corners] (0,-0.5)--(1.5,-0.5)--(1.5,0.5)--(0,0.5)--cycle;
\filldraw[fill=white, draw=black, rounded corners] (2.5,-0.5)--(4,-0.5)--(4,0.5)--(2.5,0.5)--cycle;
\draw[line width=1pt] (1.5,0) -- (2.5,0);
\draw[line width=1pt] (0,0) -- (-1,0);
\path (0.75,0) node [style=sergio]{\large $\mathsf{A}$};
\path (3.25,0) node [style=sergio]{\large $\mathsf{A}$};
\draw[line width=1pt] (4,0) -- (5,0);
\draw[line width=2pt, color=red] (0.75,0.5) -- (0.75,1);
\draw[line width=3pt, color=blue] (0.75,-1.5) -- (0.75,-0.5);
\path (0.5,1) node [style=sergio]{$p$};
\path (0.5,-1) node [style=sergio]{$a$};
\draw[line width=2pt, color=red] (3.25,0.5) -- (3.25,1);
\draw[line width=3pt, color=blue] (3.25,-1.5) -- (3.25,-0.5);
\path (3,1) node [style=sergio]{$p$};
\path (3,-1) node [style=sergio]{$a$};
\path (-1.5,-1) node [style=sergio]{$\cdots$};
\path (5.5,-1) node [style=sergio]{$\cdots$};
\path (-2.5,-1) node [style=sergio]{$\sket{\rho}=$};
\filldraw[fill=white, draw=black, rounded corners] (2.5,-2.5)--(4,-2.5)--(4,-1.5)--(2.5,-1.5)--cycle;
\filldraw[fill=white, draw=black, rounded corners] (0,-2.5)--(1.5,-2.5)--(1.5,-1.5)--(0,-1.5)--cycle;
\path (3.25,-2) node [style=sergio]{\large $\mathsf{A}^*$};
\path (0.75,-2) node [style=sergio]{\large $\mathsf{A}^*$};
\draw[line width=1pt] (1.5,-2) -- (2.5,-2);
\draw[line width=1pt] (0,-2) -- (-1,-2);
\draw[line width=1pt] (4,-2) -- (5,-2);
\draw[line width=2pt, color=red] (0.75,-2.5) -- (0.75,-3) -- (1.25,-3) -- (1.25,1);
\draw[line width=2pt, color=red] (3.25,-2.5) -- (3.25,-3) -- (3.75,-3) -- (3.75,1);
\path (1.5,1.1) node [style=sergio]{$p^{\prime}$};
\path (4,1.1) node [style=sergio]{$p^{\prime}$};
\end{tikzpicture}.
\label{double state}
\end{align}
Utilizing the concept of the double state, the strong injectivity condition for an LPDO is defined as follows:
\paragraph*{Definition} (Strong injectivity condition) 
An LPDO is deemed strongly injective if its corresponding double state $\sket{\rho}$ is an injective MPS, or equivalently, its local tensor given by
\begin{align}
    \sum_{a} A_{\alpha^u\beta^u}^{p, a}{A}_{\alpha^l\beta^l}^{*p^{\prime}, a}: (\mathbb{C}^D)^{\otimes 4}\rightarrow (\mathbb{C}^{d_p})^{\otimes 2L}
\end{align}
forms an injective map.

The strong injectivity condition can be equivalently stated as the non-degeneracy condition concerning the following transfer matrix (as shown in Fig.~\ref{injectivity}(b))
\begin{align}
    \mathbb{E}^{(2)} = \sum_{p, a, p^{\prime}, a^{\prime}}A^{p, a}_{\alpha^u\beta^u}{A}^{*p{'}, a}_{\alpha^l\beta^l}A^{p{'}, a{'}}_{\alpha^u{'}\beta^u{'}}{A}^{*p, a{'}}_{\alpha^l{'}\beta^l{'}}.
    \label{strong transfer matrix}
\end{align}
It implies the R\'enyi-2 correlation function, defined as
\begin{align}
    C^{(2)}(i, j)&\equiv \frac{\mathrm{Tr}\left(\rho O_i O_j\rho O_i O_j\right) - \mathrm{Tr}\left(\rho O_i \rho O_i\right)\mathrm{Tr}\left(\rho O_j\rho O_j\right)}{\mathrm{Tr}(\rho^2)}\nonumber\\
    = &\sbraket{\rho|O_iO_j\otimes O_i^*O_j^*|\rho}\nonumber\\
    &- \sbraket{\rho|O_i\otimes O_i^*|\rho}\sbraket{\rho|O_j\otimes O_j^*|\rho}\nonumber\\
    \sim&\ e^{-|i-j|/\xi'},
\label{Renyi2}
\end{align}
decays exponentially with distance.
In fact, while the R\'enyi-2 correlation function appears as an ordinary two-point correlation function in the double state $\sket{\rho}$, it manifests itself as a non-linear entity within the original density matrix representation.
Contrary to the weak injectivity condition, which imposes no restrictions on the values of such correlation functions, the absence of SW-SSB within the mixed state requires the exponential decay of $C^{(2)}$.
Consequently, the imposition of a strong injectivity condition becomes imperative to ensure this decay behavior.
In Appendix~\ref{Sec.SWSSB}, we provide an example of an LPDO that is weakly injective but not strongly injective.
This density matrix exhibits SW-SSB, i.e., carrying a long-range R\'enyi-2 correlation function.
On the other hand, the strong injectivity condition implies the weak one.
We now consider the contrapositive statement, i.e., an LPDO that is not weakly injective also fails to satisfy the strong injectivity condition.
To prove this, we note that an LPDO without weak injectivity allows for a nontrivial symmetry transformation $V$ on the virtual indices, leaving the physical and Kraus indices unaffected.
\begin{align}
\begin{tikzpicture}[scale=0.8]
\tikzstyle{sergio}=[rectangle,draw=none]
\filldraw[fill=white, draw=black, rounded corners] (-0.25,-0.5)--(1.25,-0.5)--(1.25,0.5)--(-0.25,0.5)--cycle;
\draw[line width=1pt] (1.25,0) -- (1.75,0);
\draw[line width=1pt] (-0.25,0) -- (-0.75,0);
\draw[line width=2pt, color=red] (0.5,0.5) -- (0.5,1);
\draw[line width=3pt, color=blue] (0.5,-0.5) -- (0.5,-1);
\path (0.5,0) node [style=sergio]{\large $\mathsf{A}$};
\path (0.85,-1) node [style=sergio]{$a$};
\path (0.85,1) node [style=sergio]{$p$};
\filldraw[fill=white, draw=black, rounded corners] (4.25,-0.5)--(5.75,-0.5)--(5.75,0.5)--(4.25,0.5)--cycle;
\draw[line width=1pt] (4.25,0) -- (2.75,0);
\filldraw[fill=white, draw=black] (3.5,0)circle (12pt);
\path (3.5,0) node [style=sergio]{\small $V^{-1}$};
\path (5,0) node [style=sergio]{\large $\mathsf{A}$};
\draw[line width=1pt] (5.75,0) -- (7.25,0);
\filldraw[fill=white, draw=black] (6.5,0)circle (12pt);
\path (6.5,0) node [style=sergio]{$V$};
\draw[line width=2pt, color=red] (5,0.5) -- (5,1);
\draw[line width=3pt, color=blue] (5,-0.5) -- (5,-1);
\path (5.35,-1) node [style=sergio]{$a$};
\path (5.35,1) node [style=sergio]{$p$};
\path (2.25,0) node [style=sergio]{$=$};
\end{tikzpicture},
\end{align}
from which we immediately reach the following equation
\begin{align}
\begin{tikzpicture}[scale=0.8]
\tikzstyle{sergio}=[rectangle,draw=none]
\filldraw[fill=white, draw=black, rounded corners] (-0.25,-1.5)--(1.25,-1.5)--(1.25,-0.5)--(-0.25,-0.5)--cycle;
\filldraw[fill=white, draw=black, rounded corners] (-0.25,-3)--(1.25,-3)--(1.25,-2)--(-0.25,-2)--cycle;
\draw[line width=1pt] (1.25,-1) -- (1.75,-1);
\draw[line width=1pt] (-0.25,-1) -- (-0.75,-1);
\draw[line width=1pt] (1.25,-2.5) -- (1.75,-2.5);
\draw[line width=1pt] (-0.25,-2.5) -- (-0.75,-2.5);
\draw[line width=2pt, color=red] (0.5,-0.5) -- (0.5,0);
\draw[line width=3pt, color=blue] (0.5,-1.5) -- (0.5,-2);
\draw[line width=2pt, color=red] (0.5,-3) -- (0.5,-3.5);
\path (0.5,-1) node [style=sergio]{\large $\mathsf{A}$};
\path (0.5,-2.5) node [style=sergio]{\large $\mathsf{A^*}$};
\path (2.5,-1.75) node [style=sergio]{$=$};
\filldraw[fill=white, draw=black, rounded corners] (4.75,-1.5)--(6.25,-1.5)--(6.25,-0.5)--(4.75,-0.5)--cycle;
\filldraw[fill=white, draw=black, rounded corners] (4.75,-3)--(6.25,-3)--(6.25,-2)--(4.75,-2)--cycle;
\draw[line width=1pt] (4.75,-1) -- (3.25,-1);
\filldraw[fill=white, draw=black] (4,-1)circle (12pt);
\path (4,-1) node [style=sergio]{\small$V^{-1}$};
\path (5.5,-1) node [style=sergio]{\large $\mathsf{A}$};
\draw[line width=1pt] (6.25,-1) -- (7.75,-1);
\filldraw[fill=white, draw=black] (7,-1)circle (12pt);
\path (7,-1) node [style=sergio]{$V$};
\draw[line width=1pt] (4.75,-2.5) -- (3.25,-2.5);
\filldraw[fill=white, draw=black] (4,-2.5)circle (12pt);
\path (4,-2.5) node [style=sergio]{\small$V^{*-1}$};
\path (5.5,-2.5) node [style=sergio]{\large $\mathsf{A^*}$};
\draw[line width=1pt] (6.25,-2.5) -- (7.75,-2.5);
\filldraw[fill=white, draw=black] (7,-2.5)circle (12pt);
\path (7,-2.5) node [style=sergio]{$V^*$};
\draw[line width=2pt, color=red] (5.5,-3) -- (5.5,-3.5);
\draw[line width=3pt, color=blue] (5.5,-1.5) -- (5.5,-2);
\draw[line width=2pt, color=red] (5.5,-0.5) -- (5.5,0);
\end{tikzpicture}.
\end{align}
Therefore, the LPDO in question does not satisfy the strong injectivity condition, thereby justifying our earlier argument.

In summary, by imposing the two aforementioned injectivity conditions, we effectively preclude any long-range order that could cause the breakdown of either strong or weak symmetries inherent within the LPDO \footnote{It is worth noting a subtlety here: the decay of the R\'enyi-2 correlators serves as a necessary condition for the absence of SW-SSB in mixed states. However, whether it is a sufficient condition is not yet clear in general~\cite{Lessa2024B}. However, as we will elucidate later, the combination of these two injectivity conditions allows us to derive accurate classification data for decohered ASPT phases, particularly in dimensions $(1+1)D$ and $(2+1)D$. Therefore, we conjecture that these two conditions suffice for the proposed classification framework.}.
This facilitates our analysis and classification of short-range correlated mixed states while conserving both strong and weak symmetries.

\subsection{Symmetries in the LPDO representations}
\label{Sec.LPDO}
With the weak and strong injectivity conditions established for LPDOs, we proceed to formulate the actions of the strong and weak symmetries on the density matrix.
To illustrate the distinctions between these two types of symmetries, we focus on the upper half of the LPDO as outlined in Eq.~\eqref{LPDO}.

The density matrix remains invariant under the implementation of a strong (exact) symmetry $U(k)$ on one side, characterized by the operation
\begin{align}
U(k)\rho=e^{i\theta}\rho.
\end{align}
In essence, the action of strong symmetry is independent of the Kraus index governing the classical mixture of various components.
In other words, the purified MPS $\ket{\psi_{p\otimes a}}$ in Eq.~\eqref{1d MPS} satisfies a global symmetry $U(k)_p\otimes I_a$ defined on the joint of physical and Kraus indices, rendering the transformation of the local tensor as
\begin{align}
\begin{tikzpicture}[scale=0.8]
\tikzstyle{sergio}=[rectangle,draw=none]
\filldraw[fill=white, draw=black, rounded corners] (-0.25,-0.5)--(1.25,-0.5)--(1.25,0.5)--(-0.25,0.5)--cycle;
\draw[line width=1pt] (1.25,0) -- (1.75,0);
\draw[line width=1pt] (-0.25,0) -- (-0.75,0);
\draw[line width=2pt, color=red] (0.5,0.5) -- (0.5,2);
\draw[line width=3pt, color=blue] (0.5,-0.5) -- (0.5,-1);
\path (0.5,0) node [style=sergio]{\large $\mathsf{A}$};
\path (0.85,-1) node [style=sergio]{$a$};
\path (0.85,2) node [style=sergio]{$p$};
\filldraw[fill=white, draw=black] (0.5,1.25)circle (10pt);
\path (0.5,1.25) node [style=sergio]{\small $U_k$};
\path (2.5,0) node [style=sergio]{$=e^{i\theta_k}$};
\filldraw[fill=white, draw=black, rounded corners] (4.75,-0.5)--(6.25,-0.5)--(6.25,0.5)--(4.75,0.5)--cycle;
\draw[line width=1pt] (4.75,0) -- (3.25,0);
\filldraw[fill=white, draw=black] (4,0)circle (10pt);
\path (4,0) node [style=sergio]{\scriptsize $V^{-1}_k$};
\path (5.5,0) node [style=sergio]{\large $\mathsf{A}$};
\draw[line width=1pt] (6.25,0) -- (7.75,0);
\filldraw[fill=white, draw=black] (7,0)circle (10pt);
\path (7,0) node [style=sergio]{\small $V_k$};
\draw[line width=2pt, color=red] (5.5,0.5) -- (5.5,1);
\draw[line width=3pt, color=blue] (5.5,-0.5) -- (5.5,-1);
\path (5.85,-1) node [style=sergio]{$a$};
\path (5.85,1) node [style=sergio]{$p$};
\end{tikzpicture}.
\label{1d exact}
\end{align}
In this context, $U(k)$ denotes the strong symmetry action operating on the physical index, while $V(k)$ represents the corresponding gauge transformation acting on the virtual index.
The weak injectivity condition ensures the uniqueness, up to a phase factor, of $V(k)$, given that the symmetry action on the Kraus index is restricted to remain as the identity.

Conversely, the weak (average) symmetry $U(g)$ requires application on both sides of a density matrix, expressed as
\begin{align}
U(g)\rho U(g)^\dag=\rho.
\end{align}
Equivalently, in terms of the purified MPS $\ket{\psi_{p\otimes a}}$ in Eq.~\eqref{1d MPS}, we assume the tensor remains invariant under the symmetry action $U(g)_p\otimes M(g)_a^{\dagger}$ on the joint of system and ancilla, i.e.,
\begin{align}
    U(g)_p\otimes M(g)_a^{\dagger}\ket{\psi_{p\otimes a}}\propto  \ket{\psi_{p\otimes a}},
\end{align}
where $M(g)_a$ is a unitary transformation satisfying that $M(g)_a^{\dag}M(g)_a=I$.
Here, naturally the symmetry action on the system and ancilla factorizes because they belong to different Hilbert spaces.
Furthermore, as previously assumed, the ancillary space has the same locality structure as the physical space, leading to the factorization of the symmetry action on the ancilla.
Similar to the pure-state case, the purified MPS $\ket{\psi_{p\otimes a}}$ (which is injective) satisfies the following local tensor condition~\cite{Schuch2011, Chen2011A}
\begin{align}
\begin{tikzpicture}[scale=0.8]
\tikzstyle{sergio}=[rectangle,draw=none]
\filldraw[fill=white, draw=black, rounded corners] (-0.25,-0.5)--(1.25,-0.5)--(1.25,0.5)--(-0.25,0.5)--cycle;
\draw[line width=1pt] (1.25,0) -- (1.75,0);
\draw[line width=1pt] (-0.25,0) -- (-0.75,0);
\draw[line width=2pt, color=red] (0,0.5) -- (0,2);
\draw[line width=3pt, color=blue] (5.75,1) -- (5.75,0.5);
\path (0.5,0) node [style=sergio]{\large $\mathsf{A}$};
\path (5.875,1.1875) node [style=sergio]{$a$};
\path (0.35,2) node [style=sergio]{$p$};
\filldraw[fill=white, draw=black] (0,1.25)circle (10pt);
\path (0,1.25) node [style=sergio]{\small $U_g$};
\path (2.5,0) node [style=sergio]{$=e^{i\theta_g}$};
\filldraw[fill=white, draw=black, rounded corners] (4.75,-0.5)--(6.25,-0.5)--(6.25,0.5)--(4.75,0.5)--cycle;
\draw[line width=1pt] (4.75,0) -- (3.25,0);
\filldraw[fill=white, draw=black] (4,0)circle (10pt);
\path (4,0) node [style=sergio]{\scriptsize $V^{-1}_g$};
\path (5.5,0) node [style=sergio]{\large $\mathsf{A}$};
\draw[line width=1pt] (6.25,0) -- (7.75,0);
\filldraw[fill=white, draw=black] (7,0)circle (10pt);
\path (7,0) node [style=sergio]{\small $V_g$};
\draw[line width=2pt, color=red] (5.25,0.5) -- (5.25,1);
\draw[line width=3pt, color=blue] (1,2) -- (1,0.5);
\path (1.375,2) node [style=sergio]{$a$};
\path (5.1625,1.1875) node [style=sergio]{$p$};
\filldraw[fill=white, draw=black] (1,1.25)circle (10pt);
\path (1,1.25) node [style=sergio]{\small $M_g^{\dagger}$};
\end{tikzpicture}.
\end{align}
With these considerations, the symmetry action on an LPDO is visually depicted as
\begin{align}
\begin{tikzpicture}[scale=0.8]
\tikzstyle{sergio}=[rectangle,draw=none]
\filldraw[fill=white, draw=black, rounded corners] (-0.25,-0.5)--(1.25,-0.5)--(1.25,0.5)--(-0.25,0.5)--cycle;
\draw[line width=1pt] (1.25,0) -- (1.75,0);
\draw[line width=1pt] (-0.25,0) -- (-0.75,0);
\draw[line width=2pt, color=red] (0.5,0.5) -- (0.5,2);
\draw[line width=3pt, color=blue] (0.5,-0.5) -- (0.5,-1);
\path (0.5,0) node [style=sergio]{\large $\mathsf{A}$};
\path (0.85,-1) node [style=sergio]{$a$};
\path (0.85,2) node [style=sergio]{$p$};
\filldraw[fill=white, draw=black] (0.5,1.25)circle (10pt);
\path (0.5,1.25) node [style=sergio]{\small $U_g$};
\path (2.5,0) node [style=sergio]{$=e^{i\theta_g}$};
\filldraw[fill=white, draw=black, rounded corners] (4.75,-0.5)--(6.25,-0.5)--(6.25,0.5)--(4.75,0.5)--cycle;
\draw[line width=1pt] (4.75,0) -- (3.25,0);
\filldraw[fill=white, draw=black] (4,0)circle (10pt);
\path (4,0) node [style=sergio]{\scriptsize $V^{-1}_g$};
\path (5.5,0) node [style=sergio]{\large $\mathsf{A}$};
\draw[line width=1pt] (6.25,0) -- (7.75,0);
\filldraw[fill=white, draw=black] (7,0)circle (10pt);
\path (7,0) node [style=sergio]{\small $V_g$};
\draw[line width=2pt, color=red] (5.5,0.5) -- (5.5,1);
\draw[line width=3pt, color=blue] (5.5,-0.5) -- (5.5,-2);
\path (5.85,-2) node [style=sergio]{$a$};
\path (5.85,1) node [style=sergio]{$p$};
\filldraw[fill=white, draw=black] (5.5,-1.25)circle (10pt);
\path (5.5,-1.25) node [style=sergio]{\small $M_g$};
\end{tikzpicture},
\label{1d average}
\end{align}
where we have utilized the unitarity of $M(g)$.
In contrast to the strong symmetry scenario, in the case of weak symmetry, an additional symmetry transformation $M(g)$ operates on the Kraus index.
Moreover, the application of weak symmetry operators on both sides must maintain the invariance of the density matrix, as depicted in the following figure
\begin{align}
\begin{tikzpicture}[scale=0.8]
\tikzstyle{sergio}=[rectangle,draw=none]
\filldraw[fill=white, draw=black, rounded corners] (-0.25,-1.5)--(1.25,-1.5)--(1.25,-0.5)--(-0.25,-0.5)--cycle;
\filldraw[fill=white, draw=black, rounded corners] (-0.25,-3)--(1.25,-3)--(1.25,-2)--(-0.25,-2)--cycle;
\draw[line width=1pt] (1.25,-1) -- (1.75,-1);
\draw[line width=1pt] (-0.25,-1) -- (-0.75,-1);
\draw[line width=1pt] (1.25,-2.5) -- (1.75,-2.5);
\draw[line width=1pt] (-0.25,-2.5) -- (-0.75,-2.5);
\draw[line width=2pt, color=red] (0.5,-0.5) -- (0.5,1);
\draw[line width=3pt, color=blue] (0.5,-1.5) -- (0.5,-2);
\draw[line width=2pt, color=red] (0.5,-3) -- (0.5,-4.5);
\path (0.5,-1) node [style=sergio]{\large $\mathsf{A}$};
\path (0.5,-2.5) node [style=sergio]{\large $\mathsf{A^*}$};
\filldraw[fill=white, draw=black] (0.5,0.25)circle (10pt);
\path (0.5,0.25) node [style=sergio]{\small $U_g$};
\filldraw[fill=white, draw=black] (0.5,-3.75)circle (10pt);
\path (0.5,-3.75) node [style=sergio]{\small $U_g^*$};
\path (2.5,-1.75) node [style=sergio]{$=$};
\filldraw[fill=white, draw=black, rounded corners] (4.75,-0.5)--(6.25,-0.5)--(6.25,0.5)--(4.75,0.5)--cycle;
\filldraw[fill=white, draw=black, rounded corners] (4.75,-4)--(6.25,-4)--(6.25,-3)--(4.75,-3)--cycle;
\draw[line width=1pt] (4.75,0) -- (3.25,0);
\filldraw[fill=white, draw=black] (4,0)circle (10pt);
\path (4,0) node [style=sergio]{\small $V^{-1}_g$};
\path (5.5,0) node [style=sergio]{\large $\mathsf{A}$};
\draw[line width=1pt] (6.25,0) -- (7.75,0);
\filldraw[fill=white, draw=black] (7,0)circle (10pt);
\path (7,0) node [style=sergio]{\small $V_g$};
\draw[line width=1pt] (4.75,-3.5) -- (3.25,-3.5);
\filldraw[fill=white, draw=black] (4,-3.5)circle (10pt);
\path (4.1,-3.5) node [style=sergio]{\small $V_g^{*-1}$};
\path (5.5,-3.5) node [style=sergio]{\large $\mathsf{A^*}$};
\draw[line width=1pt] (6.25,-3.5) -- (7.75,-3.5);
\filldraw[fill=white, draw=black] (7,-3.5)circle (10pt);
\path (7,-3.5) node [style=sergio]{\small $V^*_g$};
\draw[line width=2pt, color=red] (5.5,-4) -- (5.5,-4.5);
\draw[line width=3pt, color=blue] (5.5,-0.5) -- (5.5,-3);
\draw[line width=2pt, color=red] (5.5,0.5) -- (5.5,1);
\filldraw[fill=white, draw=black] (5.5,-1.25)circle (10pt);
\path (5.5,-1.25) node [style=sergio]{\small $M_g$};
\filldraw[fill=white, draw=black] (5.5,-2.25)circle (10pt);
\path (5.5,-2.25) node [style=sergio]{\small $M_g^*$};
\end{tikzpicture},
\end{align}
necessitating $M(g)$ to be a unitary operator to ensure the preservation of symmetry when applied to both sides of the density matrix, consistent with the above discussion.

Now, we argue that with combined weak and strong injectivity conditions, the gauge transformation $V(g)$ on the virtual index becomes uniquely determined (up to a phase) for a given $U(g)$, although $M(g)$ may not possess uniqueness.
To begin with, if the LPDO is weakly injective, then $V(g)$ is uniquely determined for fixed $U(g)$ and $M(g)$.
Let us assume there exist two sets of solutions for Eq.~\eqref{1d average} corresponding to the same $U(g)$, denoted as $\{M_1(g), V_1(g)\}$ and $\{M_2(g), V_2(g)\}$, where the representations $V_1(g)$ and $V_2(g)$ are linearly independent operators in the virtual space.
Next, consider the simultaneous implementation of $U(g)$ on the upper and lower sides of the LPDO, inducing a transformation $\overline{V}(g)$ supported on the joint space of the upper and lower indices.
With the strong injectivity condition ensured, such gauge transformation is also unique, leading to
\begin{align}
V_1(g)\otimes V_1^*(g) = \overline{V}(g) = V_2(g)\otimes V_2^*(g),
\end{align}
from which we deduce that $V_1(g)=V_2(g)$.
Consequently, the weak symmetry action on virtual indices is uniquely defined for an LPDO that satisfies both weak and strong injectivity conditions.

\subsection{ASPT protected by strong and weak symmetries}

In this section, we aim to demonstrate that a nontrivial ASPT phase necessitates protection from both strong and weak symmetries.

\subsubsection{Only strong symmetry $K$}
We first consider the scenario where only a strong symmetry group $K$ is present.
In this case, the classification mirrors that of pure-state SPT phases protected by the same symmetry group.
Similar to the pure state case, the symmetry action $U(k)$ on physical indices forms a linear representation of $K$ akin to Eq.~\eqref{linear rep}.
Subsequently, we consider the sequential implementation of two symmetry operations $U(k_2)$ and $U(k_1)$ ($k_1, k_2 \in K$) on the physical index, which is equivalent to the symmetry operation $U(k_1k_2)$.
According to Eq.~\eqref{1d exact}, these transformations yield $V(k_1)V(k_2)$ and $V(k_1k_2)$ on the right virtual index, respectively.
Since these resulting tensors must be identical, we deduce
\begin{align}
V(k_1)V(k_2) = \mu_2(k_1, k_2)V(k_1k_2), \quad \mu_2(k_1, k_2)\in U(1).
\end{align}
Furthermore, considering different orders to combine three transformations $V(k_1)$, $V(k_2)$, and $V(k_3)$ in the virtual index, we obtain the same consistency condition as that in Eq.~\eqref{2-cocycle}, implying that $\mu_2$ forms a 2-cocycle characterized by $\mu_2\in \H^2[K, U(1)]$.
Therefore, if only a strong symmetry is present, the classification of ASPT phases mirrors that of SPT phases in pure states.

However, it is crucial to note that the validity of the above classification is based on the constraint that the symmetry action $U(k)$ can only be applied to one side of the density matrix, and the phase structures on the upper and lower virtual indices are separate and well defined individually.
Continuing, we will now elucidate how the scenario changes drastically in the absence of strong symmetry and when only weak symmetry is present.

\subsubsection{Only weak symmetry $G$}
In the scenario where a mixed state is solely protected by a weak symmetry group $G$, the symmetry action on the physical index still forms a linear representation of $G$, expressed as
\begin{align}
U(g)U(h) = U(gh), \quad g, h \in G.
\end{align}
Now, let us apply both sides of the above equation to the physical index.
Following the symmetry transformation rule delineated in Eq.~\eqref{1d average}, the former induces the transformation $V(g)V(h)$ on the right virtual index and the transformation $M(g)M(h)$ on the Kraus index, while the latter results in $V(gh)$ and $M(gh)$ on these two indices. 
Given that these two tensors are completely identical, $V(g)$ and $M(g)$ must adhere to the following conditions
\begin{align}
V(g)V(h)=&\, \nu_2(g,h)V(gh), \quad\nu_2(g, h) \in U(1)\\
&M(g)M(h) = M(gh).
\end{align}
Here there could be a phase ambiguity for $V$ since $\nu_2$ on the right virtual index cancels with $\nu_2^{-1}$ on the left side, while the transformation on the Kraus index cannot differ a phase factor.
Therefore, $V(g)$ forms a projective representation of $G$ with the phase structure characterized by $\nu_2$, while $M(g)$ forms a linear representation of $G$ that facilitates permutation between different components within the density matrix.

However, due to the requirement that weak symmetry must be applied simultaneously to both the upper and lower sides of an LPDO to ensure its invariance, the virtual index on the lower side of the LPDO will exhibit a conjugate phase ambiguity of $\nu_2^{-1}$.
Specifically, when also sequentially applying $U(h)^*$ and $U(g)^*$ to the lower physical index, we have
\begin{align}
V(g)^*V(h)^*=\nu_2(g,h)^{-1}V(gh)^*.
\end{align}
Consequently, the cumulative phase ambiguity on virtual indices cancels out between the upper and lower virtual legs, suggesting the absence of nontrivial ASPT states in the presence of solely weak symmetry.
Another perspective on this is that the state in the double space $\sket{\rho}$ becomes trivial since there is no projective representation on the composed virtual leg.
As a result, the corresponding density matrix is also trivial.

\subsubsection{Both strong and weak symmetries}
Finally, let us illustrate that if the system is jointly protected by strong symmetry $K$ and weak symmetry $G$, distinct ASPT phases can emerge, characterized by the phase structure of the entire group $\widetilde{G}$.
We would like to emphasize that, in this context, we only consider a simple case where $\widetilde{G}=K\times G$ and treat $\widetilde{G}$ as a unified entity to engage in abstract discussions regarding its phase structure.
However, in subsequent sections, we will discuss a general group extension $\widetilde{G}$ from $K$ and $G$ and explore the intricate internal structures of $\widetilde{G}$, particularly the complex interplay between $K$ and $G$.

The group element $\tilde{g}\in \widetilde{G}$ is defined as $\tilde{g}=(k, g)$ with $k\in K$ and $g\in G$.
We require that the symmetry action $\widetilde{U}(\tilde{g})$ on the physical index forms a linear representation of $\widetilde{G}$, i.e,
\begin{align}
    \widetilde{U}(\tilde{g})\widetilde{U}(\tilde{g}^{\prime}) = \widetilde{U}(\tilde{g}\tilde{g}^{\prime}).\label{linear rep entire}
\end{align}
The symmetry action of $\widetilde{U}(\tilde{g})$ on the physical index is depicted as
\begin{align}
\begin{tikzpicture}[scale=0.8]
\tikzstyle{sergio}=[rectangle,draw=none]
\filldraw[fill=white, draw=black, rounded corners] (-0.25,-0.5)--(1.25,-0.5)--(1.25,0.5)--(-0.25,0.5)--cycle;
\draw[line width=1pt] (1.25,0) -- (1.75,0);
\draw[line width=1pt] (-0.25,0) -- (-0.75,0);
\draw[line width=2pt, color=red] (0.5,0.5) -- (0.5,2);
\draw[line width=3pt, color=blue] (0.5,-0.5) -- (0.5,-1);
\path (0.5,0) node [style=sergio]{\large $\mathsf{A}$};
\path (0.85,-1) node [style=sergio]{$a$};
\path (0.85,2) node [style=sergio]{$p$};
\filldraw[fill=white, draw=black] (0.5,1.25)circle (10pt);
\path (0.5,1.25) node [style=sergio]{\small $\widetilde{U}_{\tilde{g}}$};
\path (2.5,0) node [style=sergio]{$=e^{i\tilde{\theta}_{\tilde{g}}}$};
\filldraw[fill=white, draw=black, rounded corners] (4.75,-0.5)--(6.25,-0.5)--(6.25,0.5)--(4.75,0.5)--cycle;
\draw[line width=1pt] (4.75,0) -- (3.25,0);
\filldraw[fill=white, draw=black] (4,0)circle (10pt);
\path (4,0) node [style=sergio]{\scriptsize $\widetilde{V}^{-1}_{\tilde{g}}$};
\path (5.5,0) node [style=sergio]{\large $\mathsf{A}$};
\draw[line width=1pt] (6.25,0) -- (7.75,0);
\filldraw[fill=white, draw=black] (7,0)circle (10pt);
\path (7,0) node [style=sergio]{\small $\widetilde{V}_{\tilde{g}}$};
\draw[line width=2pt, color=red] (5.5,0.5) -- (5.5,1);
\draw[line width=3pt, color=blue] (5.5,-0.5) -- (5.5,-2);
\path (5.85,-2) node [style=sergio]{$a$};
\path (5.85,1) node [style=sergio]{$p$};
\filldraw[fill=white, draw=black] (5.5,-1.25)circle (10pt);
\path (5.5,-1.25) node [style=sergio]{\small $\widetilde{M}_{\tilde{g}}$};
\end{tikzpicture},
\label{1d total}
\end{align}
where $\widetilde{V}(\tilde{g})$ and $\widetilde{M}(\tilde{g})$ are the gauge transformations on the virtual and Kraus indices respectively.
Similar to the previous argument for weak symmetry, $\widetilde{V}(\tilde{g})$ and $\widetilde{M}(\tilde{g})$ satisfy the following conditions
\begin{align}
\widetilde{V}(\tilde{g})\widetilde{V}(\tilde{g}^{\prime})=&\, \tilde{\nu}_2(\tilde{g},\tilde{g}^{\prime})\widetilde{V}(\tilde{g}\tilde{g}^{\prime}), \quad\tilde{\nu}_2(\tilde{g}, \tilde{g}^{\prime}) \in U(1)\\
&\widetilde{M}(\tilde{g})\widetilde{M}(\tilde{g}^{\prime}) = \widetilde{M}(\tilde{g}\tilde{g}^{\prime}).
\label{proj rep}
\end{align}
indicating that $\widetilde{V}(\tilde{g})$ forms a projective representation of $\widetilde{G}$ characterized by $\tilde{\nu}_2\in U(1)$ and $\widetilde{M}(\tilde{g})$ forms a linear representation of $\widetilde{G}$.

Consider a strong symmetry operator $\widetilde{U}[(k, 1)] (k\in K)$ acting on the upper physical index, followed by two weak symmetry operators $\widetilde{U}[(1, h)]$ and $\widetilde{U}[(1, g)] (g, h\in G)$ acting on the physical indices on both sides sequentially.
These operations are equivalent to the symmetry operation $\widetilde{U}[(k, gh)]$ on the upper physical index and only to the operation $\widetilde{U}[(1, gh)]$ on the lower physical index.
On the upper side of the LPDO, $\widetilde{V}(\tilde{g})$ satisfies
\begin{align}
\begin{aligned}
&\widetilde{V}[(1, g)]\widetilde{V}[(1, h)]\widetilde{V}[(k, 1)]\\
=&\tilde{\nu}_2[(1, g), (1, h)]\tilde{\nu}_2[(1, gh),(k, 1)]\widetilde{V}[(k, gh)].
\label{upper}
\end{aligned}
\end{align}
Similarly, $\widetilde{V}(\tilde{g})$ on the lower side of LPDO satisfies
\begin{align}
\widetilde{V}[(1, g)]^{*}\widetilde{V}[(1, h)]^{*}=\tilde{\nu}_2^{-1}[(1, g),(1, h)]\widetilde{V}[(1, gh)]^{*}.
\label{lower}
\end{align}
From Eqs.~\eqref{upper} and~\eqref{lower}, we observe a nontrivial total phase ambiguity $\tilde{\nu}_2[(1, gh),(k, 1)]$ remaining uncanceled on virtual indices.
This suggests the potential existence of a nontrivial ASPT phase characterized by the phase structure involving both the strong and the weak symmetries.

In summary, a nontrivial ASPT order requires protection from both strong symmetry $K$ and weak symmetry $G$.
As we encounter a combination of weak and strong symmetry groups (potentially constituting a nontrivial extension structure), we will delve into a formalism in the following sections that treat the two groups in slightly distinct manners.
This approach facilitates the manifestation of the decorated domain wall picture for ASPTs.

\section{Warmup: Classification of $(1+1)D$ ASPT with strong fermion parity symmetry}
As a preliminary example, we will consider systems characterized by strong $K=\mathbb{Z}_2^f$ fermion parity symmetry and weak bosonic $G=G_b$ symmetry.
In other words, we assume that all bosonic symmetries are broken down to weak symmetry due to decoherence, while only the fermion parity symmetry remains strong.
In the following section, we will extend the discussion to generalize the strong symmetry to an arbitrary abelian symmetry group $K$.
Notably, though physically relevant, the fermionic case has certain technical subtleties due to fermionic statistics.
Therefore, readers who are not familiar with fermionic tensor networks and topological states can directly turn to Sec.~\ref {Sec:Bosonic} for general discussions of bosonic ASPT phases.

\subsection{$(1+1)D$ fermionic SPT phases: Supervector spaces and twisted boundary conditions}
To begin, we briefly revisit the structure of fermionic MPS (fMPS) and the classification of SPT phases protected by symmetry groups that incorporate fermion parity symmetry $\mathbb{Z}_2^f$ in $(1+1)D$~\cite{Bultinck2017A, Wang2018, Kapustin2018, Wang2020}.
Subsequently, we will employ analogous techniques to address fermionic LPDO (fLPDO).

The complete symmetry group of a fermionic system comprises a bosonic symmetry $G_b$ and the fermion parity symmetry $\mathbb{Z}_2^f$, featuring a central extension of $\mathbb{Z}_2^f$ by $G_b$ with the subsequent short exact sequence
\begin{align}
1\rightarrow\mathbb{Z}_2^f\rightarrow G_f\rightarrow G_b\rightarrow1.
\label{fermion parity extension}
\end{align}
Such a group structure is characterized by the factor system $\omega_2\in\mathcal{H}^2(G_b,\mathbb{Z}_2^f)$, subject to the 2-cocycle condition
\begin{align}
\omega_2(g, h)+\omega_2(gh, l) = \omega_2(g, hl)+ \omega_2(h, l)~(\mathrm{mod}~2).
\end{align}
A nontrivial $\omega_2$ means a nontrivial group extension of $\mathbb{Z}_2^f$ by $G_b$.

In the following classification, we will address fermion parity and bosonic symmetry separately.
Initially, let us focus on the fermion parity symmetry.
To achieve this, we need to extend the vector space used in the definition of MPS to a graded supervector space $V$ with a direct sum form
\begin{align}
V=V^0\oplus V^1,
\label{super vector space}
\end{align}
where the vectors in $V^0/V^1$ exhibit even/odd fermion parity.
The fermion parity of a vector $|i\rangle$ is denoted by $|i|\in{0,1}$.
We can define a fermion parity symmetry operator $P_f$ whose action on the vector $|i\rangle$ is given by
\begin{align}
    P_f|i\rangle=(-1)^{|i|}|i\rangle,
\end{align}
or equivalently
\begin{align}
    P_f = \left[\begin{array}{cc}
    I & 0\\
    0 & -I
    \end{array}\right]
\end{align}
for the vector space in Eq.~\eqref{super vector space}.
The direct product of the vectors $|i\rangle$ and $|j\rangle$ possesses fermion parity $|i|+|j|~(\mathrm{mod}~2)$.
We denote the graded tensor product as
\begin{align}
|i\rangle\otimes_g|j\rangle\in V\otimes_gV,
\label{graded product}
\end{align}
where the label $g$ highlights the tensor product's graded structure.
Due to the fermionic nature of this graded structure, we can define a fermion reordering operator as
\begin{align}
\begin{aligned}
\mathcal{F}:~&V\otimes_gW & \rightarrow &~~~W\otimes_gV\\
&|i\rangle\otimes_g|j\rangle & \mapsto &~~~(-1)^{|i||j|}|j\rangle\otimes_g|i\rangle.
\end{aligned}
\label{fermion reordering}
\end{align}

With these ingredients, we are ready to define fMPS.
The formal definition of fMPS has the same form as that in Eq.~\eqref{MPS2}, however, the vector spaces of the physical index and virtual indices are replaced by the supervectors with the graded structure.
A given local tensor (as in Eq.~\eqref{MPS2}) should preserve the fermion parity symmetry, namely $|i|=(|\beta|-|\alpha|)~\mod 2$. 
From the fermion anti-commutation relation in Eq.~\eqref{fermion reordering}, different contracting orders of an fMPS may exhibit additional minus signs. 
Fortunately, they will at most result in a global phase of the final state, provided that each local tensor has a well-defined total parity.
We refer the reader to Ref.~\cite{Bultinck2017A} for more details. 

With parity-conserving local tensors alone, one can only produce an even-fermion parity state (the total fermion parity is labeled by $n_0\in \{0,1\}$) under a periodic boundary condition (PBC).
To construct fMPS with odd fermion parity, it is convenient to introduce an additional odd fermion parity tensor $Y$ without a physical index, which serves as a fermion parity twist operator.
The $Y$ tensor can be inserted into the virtual leg at the right end of the system, as shown below
\begin{align}
\begin{tikzpicture}[scale=0.75]
\tikzstyle{sergio}=[rectangle,draw=none]
\filldraw[fill=white, draw=black, rounded corners] (0,-2.5)--(1.5,-2.5)--(1.5,-1.5)--(0,-1.5)--cycle;
\draw[line width=2pt, color=red] (-1.75,-1.5) -- (-1.75,-1);
\draw[line width=1pt] (1.5,-2) -- (2.5,-2);
\draw[line width=1pt] (0,-2) -- (-1,-2);
\path (0.75,-2) node [style=sergio]{\large $\mathsf{Y}$};
\filldraw[fill=white, draw=black, rounded corners] (-2.5,-2.5)--(-1,-2.5)--(-1,-1.5)--(-2.5,-1.5)--cycle;
\draw[line width=1pt] (-2.5,-2) -- (-3.5,-2);
\path (-1.75,-2) node [style=sergio]{\large $\mathsf{A}$};
\end{tikzpicture}.
\end{align}
The $Y$ tensor has a general form, 
\begin{align}
Y=\sum\limits_{\alpha,\beta}Y_{\alpha\beta}|\alpha)\otimes_g(\beta|,\label{Y_def}
\end{align}
where $Y_{\alpha\beta}=0$ if the total fermion parity carried by $|\alpha)$ and $|\beta)$ is even.
This is enforced to guarantee that $Y$ alters the parity on the virtual index.
In essence, the $Y$ operator anticommutes with fermion parity, namely
\begin{align}
    YP_f=-P_fY.\label{Y_Pf}
\end{align}
The simplest form of the $Y$ tensor can be constructed as~\cite{Bultinck2017A}
\begin{align}
    Y = \left[\begin{array}{cc}
    0 & I\\
    -I & 0
    \end{array}\right].\label{Y_construct}
\end{align}

We can now present the general fMPS for $(1+1)D$ short-range entangled fermionic states
\begin{align}
|\psi_f\rangle=\sum\limits_{i_j}\mathrm{Tr}\left(A^{i_1}\cdots A^{i_N}Y^{n_0}\right)|i_1\cdots i_N\rangle,\label{attachment_fMPS}
\end{align}
where the exponent $n_0\in\mathbb{Z}_2$ determines the fermion parity of the state.
This construction yields wavefunctions that are eigenstates of the fermion parity symmetry.

Next, we move on to the consideration of the full symmetry $G_f$ of the fMPS.
The group elements can be denoted by the form $\tilde{g}=P_f^{n_g}g$, where $n_g\in \Z_2$ and $g\in G_b$, and the group product rule is defined as
\begin{align}
\left(P_f^{n_g}g\right)\cdot\left(P_f^{n_h}h\right)=P_f^{n_g+n_h+\omega_2(g,h)}gh.
\end{align}
The symmetry action on the physical index is described by
\begin{align}
\widetilde{U}\left(P_f^{n_g}g\right):=P_f^{n_g}U(g),
\label{tilde U}
\end{align}
where $U(g)$ represents the transformation associated with the bosonic symmetry.
With this definition, if we require the total symmetry action $\widetilde{U}$ to form a linear representation of $G_f$ as Eq.~\eqref{linear rep entire}, the transformation of the bosonic symmetry on the physical index needs to satisfy~\cite{Bultinck2017A, Wang2020}
\begin{align}
U(g)U(h)=P_f^{\omega_2(g,h)}U(gh).
\label{bosonic symmetry}
\end{align}

Similar to the bosonic case, the symmetry action on the physical index leads to a gauge transformation on the virtual indices
\begin{align}
\begin{tikzpicture}[scale=0.8]
\tikzstyle{sergio}=[rectangle,draw=none]
\filldraw[fill=white, draw=black, rounded corners] (-0.25,-0.5)--(1.25,-0.5)--(1.25,0.5)--(-0.25,0.5)--cycle;
\draw[line width=1pt] (1.25,0) -- (1.75,0);
\draw[line width=1pt] (-0.25,0) -- (-0.75,0);
\draw[line width=2pt, color=red] (0.5,0.5) -- (0.5,2);
\path (0.5,0) node [style=sergio]{\large $\mathsf{A}$};
\filldraw[fill=white, draw=black] (0.5,1.25)circle (10pt);
\path (0.5,1.25) node [style=sergio]{\small $\widetilde{U}_{\tilde{g}}$};
\path (2.5,0) node [style=sergio]{$=e^{i\theta_g}$};
\filldraw[fill=white, draw=black, rounded corners] (4.75,-0.5)--(6.25,-0.5)--(6.25,0.5)--(4.75,0.5)--cycle;
\draw[line width=1pt] (4.75,0) -- (3.25,0);
\filldraw[fill=white, draw=black] (4,0)circle (10pt);
\path (4,0) node [style=sergio]{\scriptsize $\widetilde{V}^{-1}_{\tilde{g}}$};
\path (5.5,0) node [style=sergio]{\large $\mathsf{A}$};
\draw[line width=1pt] (6.25,0) -- (7.75,0);
\filldraw[fill=white, draw=black] (7,0)circle (10pt);
\path (7,0) node [style=sergio]{\small $\widetilde{V}_{\tilde{g}}$};
\draw[line width=2pt, color=red] (5.5,0.5) -- (5.5,1);
\end{tikzpicture},
\label{1d fMPS symmetry}
\end{align}
where $\widetilde{V}(\tilde{g})$, taking the form
\begin{align}
\widetilde{V}(\tilde{g}) = P_f^{n_g}V(g),
\end{align}
is expected to be a projective representation of $G_f$.
As the projective representation concerns both the fermion parity and the bosonic symmetry, the data for the projective representation are a bit more involved than in the bosonic case. 

The next step is to discuss the interplay between bosonic symmetry $G_b$ and fermion parity $\Z_2^f$.
We apply two bosonic symmetry operators $U(h)$ and $U(g)$ sequentially. 
With Eq.~\eqref{bosonic symmetry}, one can show the multiplication rule for the gauge transformation on the virtual indices now has the following form,
\begin{align}
V(g)V(h)=\nu_2(g,h)P_f^{\omega_2(g,h)}V(gh),
\label{virtual proj rep_fMPS}
\end{align}
where $\nu_2\in U(1)$ is a possible phase ambiguity.
A $V(g)$ operator acting on the virtual index can potentially change the fermion parity on that virtual leg.
This is labeled by a $\mathbb{Z}_2$ index $n_1(g)$ defined through the commutation relation between $V(g)$ and $P_f$, namely
\begin{align}
    P_f V(g) = (-1)^{n_1(g)}V(g)P_f.\label{n1}
\end{align}
Specifically, when $n_1(g)=1$, $V(g)$ switches the fermion parity on the virtual leg.

In addition, for an MPS with odd fermion parity ($n_0=1$), applying $U(g)$ to physical indices of all sites simultaneously will induce a pair of $V(g)$ and $V(g)^{-1}$ on each virtual index except at the right end of the MPS where $Y$ is attached.
For the former case, $V(g)$ and $V(g)^{-1}$ will be canceled out, while we are left with $V(g)YV(g)^{-1}$ on the last virtual leg, which should be equivalent to the original state with only $Y$ on that leg. 
Therefore, $Y$ and $V(g)$ should commute with each other up to a phase.
Since the $Y$ tensor is also a fermionic operator that alters the fermion charge on the virtual index, it has the following commutation relation with $V(g)$,
\begin{align}
    YV(g) = (-1)^{n_1(g)}V(g)Y,
\end{align}
which can be explicitly derived from Eq.~\eqref{fermion reordering}.
Therefore, $V(g)$ carries two possible topological invariants: one is the fermion parity $n_1(g)$ in Eq.~\eqref{n1}, and the other is the phase ambiguity $\nu_2(g, h)$ from the sequential application of different symmetry operators on the physical index that has been defined in Eq.~\eqref{virtual proj rep_fMPS}.
We will discuss the consistency conditions for these two indices respectively in the following. 

As for the self-consistent conditions for the fermion parity $n_1(g)$ of the representation $V(g)$, we consider the simultaneous implementation of $P_f$ from the left to each side of Eq.~\eqref{virtual proj rep_fMPS}, where the left-hand side becomes 
\begin{align}
    \begin{aligned}
    \textrm{L.H.S.} &= P_f V(g)V(h) = (-1)^{n_1(g)+n_1(h)}V(g)V(h)P_f\\
    &= (-1)^{n_1(g)+n_1(h)}\nu_2(g,h)P_f^{\omega_2(g,h)}V(gh) P_f,
    \end{aligned}\label{LHS}
\end{align}
while the right-hand side is transformed into
\begin{align}
    \begin{aligned}
        \textrm{R.H.S.} & = P_f\nu_2(g,h)P_f^{\omega_2(g,h)}V(gh)\\
        &= \nu_2(g,h)(-1)^{n_1(gh)}P_f^{\omega_2(g,h)}V(gh)P_f.
    \end{aligned}\label{RHS}
\end{align}
The above relations require that the fermion parities of $V(g)V(h)$ and $V(gh)$ should be equal, i.e.,
\begin{align}
n_1(g)+n_1(h)-n_1(gh)=0~(\mathrm{mod}~2),\label{1-cocycle_fMPS}
\end{align}
which implies that $n_1$ should be a 1-cocycle with $\Z_2$ coefficient, classified by $\H^1(G_b, \Z_2)$.

In addition, taking the attachment $Y^{n_0}$ into account gives the following relation 
\begin{align}
\begin{aligned}
&V(g)V(h)Y^{n_0}=(-1)^{n_0[n_1(g)+n_1(h)]}Y^{n_0}V(g)V(h)\\
=&(-1)^{n_0[n_1(g)+n_1(h)]}\nu_2(g,h)Y^{n_0}P_f^{\omega_2(g,h)}V(gh),
\end{aligned}
\end{align}
which can also be evaluated as follows
\begin{align}
\begin{aligned}
&V(g)V(h)Y^{n_0}=\nu_2(g,h)P_f^{\omega_2(g,h)}V(gh)Y^{n_0}\\
=&(-1)^{n_0[n_1(gh)+\omega_2(g,h)]}\nu_2(g,h)Y^{n_0}P_f^{\omega_2(g,h)}V(gh),
\end{aligned}
\end{align}
where the factor $(-1)^{n_0\omega_2(g, h)}$ comes from the commutation between $P_f^{\omega_2(g, h)}$ and $Y^{n_0}$.
The above relations indicate that $n_1$ should also be a 1-cocycle with $\Z_2$ coefficient and a twisted cocycle condition, namely
\begin{align}
n_0[n_1(g)+n_1(h)-n_1(gh)-\omega_2(g,h)] = 0~(\mathrm{mod}~2).
\label{twisted 1-cocycle}
\end{align}
For a nontrivial $n_0$, this twisted 1-cocycle condition has solutions if and only if $n_0\omega_2(g, h)$ is a 2-coboundary in $\H^2 [G_b, \Z_2]$.
In addition, since both $n_0$ and $\omega_2$ take values in $\Z_2$, we reach the following requirement
\begin{align}
    n_0\omega_2(g, h) = 0~(\mathrm{mod}~2).\label{cup}
\end{align}

Finally, it is time to derive the associativity condition for three bosonic symmetry actions $V(g)$, $V(h)$, and $V(l)$ on virtual indices, which can be composed with the following two different orders
\begin{align}
\begin{aligned}
    &V(g)V(h)V(l) = \nu_2 (g, h)P_f^{\omega_2(g, h)}V(gh)V(l)\\
    =&\nu_2(g, h)\nu_2(gh, l)P_f^{\omega_2(g, h)+\omega_2(gh, l)}V(ghl),
\end{aligned}
\end{align}
and
\begin{align}
    \begin{aligned}
        &V(g)V(h)V(l) = V(g)\nu_2(h, l)P_f^{\omega_2(h, l)}V(hl)\\
        =& \nu_2(h, l)(-1)^{n_1(g)\omega_2(h, l)}P_f^{\omega_2(h, l)}V(g)V(hl)\\
        =& \nu_2(h, l)\nu_2(g, hl)(-1)^{n_1(g)\omega_2(h, l)}P_f^{\omega_2(h, l)+\omega_2(g, hl)}V(ghl).
    \end{aligned}
\end{align}
Therefore, $\nu_2$ satisfies a twisted 2-cocycle condition of the bosonic symmetry group, namely
\begin{align}
    \frac{\nu_2(g,h)\nu_2(gh,l)}{\nu_2(g,hl)\nu_2(h,l)}=(-1)^{n_1(g)\omega_2(h, l)}\equiv \mathcal{O}_3(g,h,l),\label{twisted 2-cocycle fMPS}
\end{align}
which is classified by $\nu_2\in\H^2[G_b,U(1)]$ with an additional constraint that $\mathcal{O}_3(g,h,l)$ should be a 3-coboundary in $\H^3[G_b, U(1)]$ to guarantee the above equation to have solutions.

We summarize the consistency conditions for constructing an fMPS representation of $(1+1)D$ SPT state.
There are three topological invariants labeling different phases:
\begin{enumerate}[1.]
\item $n_0\in \H^0[G_b, h^2(\Z_2^f)]=\Z_2$: Fermion parity of fMPS encoded by attachment $Y^{n_0}$ in Eq.~\eqref{attachment_fMPS} with the condition in Eq.~\eqref{cup}.
Here, $h^2(\Z_2^f)=\Z_2$ is the classification of $(1+1)D$ invertible topological phases.
\item $n_1\in\H^1[G_b,h^1(\Z_2^f)]$: The fermion parity of the symmetry action on virtual indices $V(g)$ with the 1-cocycle condition in Eq.~\eqref{1-cocycle_fMPS}, or simply $\mathrm{d}n_1=0$.
\item $\nu_2\in\H^2[G_b,U(1)]$: The phase ambiguity of $V(g)$ with the 2-cocycle condition in Eq.~\eqref{twisted 2-cocycle fMPS}, or simply $\mathrm{d}\nu_2=\mathcal{O}_3$, and the condition of $\mathcal{O}_3$ to be a 3-coboundary in $\H^3[G_b, U(1)]$.
\end{enumerate}

The physical meaning of each topological invariant is as follows: $n_0=1$ indicates that the corresponding fMPS with PBC has an odd fermion parity, which means that the system hosts a Kitaev Majorana chain. $n_1(g)=1$ indicates that the action of $U(g)$ in a finite region leaves one complex fermion on the virtual index at each end, which is exactly the complex fermion decoration on the symmetry defects of $G_b$. A nontrivial $\nu_2$ labels a bosonic SPT phase solely protected by bosonic symmetry $G_b$. These classification data and obstruction functions are identical to the results obtained from spectrum sequence methods~\cite{Wang2018, Wang2020}.

\subsection{$(1+1)D$ ASPT with strong fermion parity}
 
Inspired by the fMPS classification of fermionic SPT states, in this section, we will construct fermionic LPDO (fLPDO) to describe fermionic ASPTs.
The formal definition of the fLPDO has the identical form of Eq.~\eqref{LPDO}, however, all the vector spaces are again replaced by the supervector space with fermionic graded structure in Eq.~\eqref{super vector space}.
The fermion parity conservation for the local tensor is written as $|p|=(|\beta|-|\alpha|) \mod 2$ and $|a|=0$, i.e., the Kraus index is always in even fermion parity.
The physical reason lies in that the fermion parity is always kept as a strong symmetry, and thus all components of the decomposition in $\rho$ should exhibit the same global fermion parity, which we still refer to as $n_0$.

Similar to the pure state case, to construct states with different fermion parities $n_0$, one needs to attach the symmetry twist operator $Y^{n_0}$, an odd fermion parity tensor without a physical index. Graphically, it is represented as
\begin{align}
\begin{tikzpicture}[scale=0.75]
\tikzstyle{sergio}=[rectangle,draw=none]
\filldraw[fill=white, draw=black, rounded corners] (0,-0.5)--(1.5,-0.5)--(1.5,0.5)--(0,0.5)--cycle;
\filldraw[fill=white, draw=black, rounded corners] (0,-2.5)--(1.5,-2.5)--(1.5,-1.5)--(0,-1.5)--cycle;
\draw[line width=2pt, color=red] (-1.75,0.5) -- (-1.75,1);
\draw[line width=2pt, color=red] (-1.75,-2.5) -- (-1.75,-3);
\draw[line width=3pt, color=blue] (-1.75,-1.5) -- (-1.75,-0.5);
\draw[line width=1pt] (1.5,0) -- (2.5,0);
\draw[line width=1pt] (1.5,-2) -- (2.5,-2);
\draw[line width=1pt] (0,0) -- (-1,0);
\draw[line width=1pt] (0,-2) -- (-1,-2);
\path (0.75,0) node [style=sergio]{\large $\mathsf{Y}$};
\path (0.75,-2) node [style=sergio]{\large $\mathsf{Y^*}$};
\filldraw[fill=white, draw=black, rounded corners] (-2.5,-0.5)--(-1,-0.5)--(-1,0.5)--(-2.5,0.5)--cycle;
\filldraw[fill=white, draw=black, rounded corners] (-2.5,-2.5)--(-1,-2.5)--(-1,-1.5)--(-2.5,-1.5)--cycle;
\draw[line width=1pt] (-2.5,0) -- (-3.5,0);
\draw[line width=1pt] (-2.5,-2) -- (-3.5,-2);
\path (-1.75,0) node [style=sergio]{\large $\mathsf{A}$};
\path (-1.75,-2) node [style=sergio]{\large $\mathsf{A^*}$};
\label{attachment}
\end{tikzpicture},
\end{align}
where $Y$ is defined in the same way as Eqs.~\eqref{Y_def} and~\eqref{Y_Pf}.

Next, we consider the total symmetry group $G_f$, which is a central extension of weak bosonic symmetry $G_b$ and strong fermion parity $\Z_2^f$, with a group structure characterized by $\omega_2\in \H^2[G_b, \Z_2^f]$.
We will discuss the consistency conditions, especially their relations and differences from those in the pure state case.
The actions of bosonic symmetry on physical and virtual indices have the same form as those in Eqs.~\eqref{bosonic symmetry} and~\eqref{virtual proj rep_fMPS}.
However, as we have illustrated in Sec.~\ref{Sec.LPDO}, the ASPT density matrix is invariant under weak symmetry if and only if we act on the symmetry operators on both sides of the density matrix.
Therefore, we must consider the simultaneous actions of $U(h)$ and $U(g)$ on the physical indices of upper and lower that leave $V(g)V(h)$ on the right virtual index on the upper side and $V^{*}(g)V^{*}(h)$ on the right virtual index on the lower side.
On each of them, we get
\begin{align}
\begin{aligned}
V(g)V(h)&=\nu_2(g,h)P_f^{\omega_2(g,h)}V(gh)\\
V(g)^*V(h)^*&=\nu_2^{-1}(g,h)P_f^{\omega_2(g,h)}V(gh)^*,
\end{aligned}
\end{align}
then we see that the total phase ambiguity at the right virtual indices (including both upper and lower) is canceled.
Consequently, the phase ambiguity of the weak symmetry group $G_b$ itself is not a topological invariant, which is consistent with the conclusion we have already reached: there is no nontrivial ASPT state solely protected by weak symmetry.

Alternatively, we will demonstrate that as the fermion parity $P_f$ acts as a strong symmetry and can be applied on only one side of the LPDO, the parity of $V(g)$, denoted as $n_1(g)$ and defined equivalently to Eq.~\eqref{n1}, continues to serve as a nontrivial topological invariant.
This characteristic distinguishes various ASPT phases that are jointly protected by $\Z_2^f$ and $G_b$.
To verify this, we consider the sequential implementation of $U(h)$, $U(g)$ on both sides of the density matrix, then only apply $P_f$ on the upper side.
Similar to the discussions in Eqs.~\eqref{LHS} and~\eqref{RHS}, on the right virtual index of the upper side, we have
\begin{align}
    \begin{aligned}
    P_f V(g)V(h) &= (-1)^{n_1(g)+n_1(h)}\nu_2(g,h)P_f^{\omega_2(g,h)}V(gh) P_f\\
    &= (-1)^{n_1(gh)}\nu_2(g,h)P_f^{\omega_2(g,h)}V(gh)P_f,
    \end{aligned}
\end{align}
while on the right virtual index of the lower side, we simply obtain
\begin{align}
    V(g)^*V(h)^* = \nu_2(g,h)^{-1}P_f^{\omega_2(g,h)}V(gh)^{*}
\end{align}
It is noted that the phase ambiguity is not completely canceled out when grouping the upper and lower sides together, hence the following self-consistent condition
\begin{align}
    n_1(g)+n_1(h)-n_1(gh)=0~(\mathrm{mod}~2)\label{1-cocycle_fLPDO}
\end{align}
still holds.
The physical interpretation of this result is rooted in the inability to exchange the fermion parity charges carried by $V(g)V(h)$ and $V(g)^*V(h)^*$.
This limitation arises from the fermion parity acting as a strong symmetry, where the charges of strong symmetry or the associated phase structures on the upper and lower sides must be conserved individually.
Due to the same argument, the twisted 1-cocycle condition in Eq.~\eqref{twisted 1-cocycle} also holds, leading to the following condition for the total fermion parity $n_0$
\begin{align}
    n_0\omega_2(g, h)= 0~(\mathrm{mod}~2).\label{cup_fLPDO}
\end{align}
On the contrary, the condition for solutions of Eq.~\eqref{twisted 2-cocycle fMPS} to exist no longer needs to be satisfied as the phase ambiguity of $\nu_2(g, h)$ must be canceled out on virtual indices when applying weak symmetries simultaneously on both sides of a density matrix.

Therefore, the topological invariants of a $(1+1)D$ fLPDO are
\begin{enumerate}[1.]
\item $n_0\in\H^0[G_b, h^2(\Z_2^f)]=\Z_2$: The fermion parity of the fLPDO encoded by the attachment $Y^{n_0}$ with the condition in Eq.~\eqref{cup_fLPDO};
\item $n_1\in\H^1[G_b,h^1(\Z_2^f)]$: The fermion parity of the symmetry action on virtual indices $V(g)$ with the 1-cocycle condition in Eq.~\eqref{1-cocycle_fLPDO}, or $\mathrm{d}n_1=0$, but without the coboundary constraint for $n_1$ here in contrast to the pure-state SPT phase.
\end{enumerate}
We see that the fLPDO construction gives an accurate classification of the $(1+1)D$ ASPT phases with a strong fermion parity symmetry and a weak $G_b$ symmetry~\cite{Ma2023B} but with a much more concise interpretation through the tensor network formalism.

A crucial aspect of this construction is that the decorated domain wall structure of the ASPT states is inherently encoded in the fLPDOs.
Consider the truncated symmetry operator $U_R(g)=\prod_{i\in R}U_i(g)$, where $R$ denotes a connected finite region of a $(1+1)D$ fLPDO.
On the right-hand side, there will be $V(g)$ on the upper virtual index and $V(g)^*$ on the lower virtual index, each carrying a fermion parity $n_1(g)$.
Moreover, since the fermion parity acts as the charge of a strong symmetry, the fermion parities of $V(g)$ and $V(g)^*$ are individually well-defined, as charges cannot tunnel.
In essence, a truncated symmetry operator $U_R(g)$ introduces two symmetry defects on the boundary of $R$, each carrying a nontrivial charge of $\Z_2^f$ labeled by $n_1(g)$.

\section{$(1+1)D$ ASPT phases with general group extension}\label{Sec:Bosonic}
In the previous section, we adopt $K = \Z_2^f$ and $G=G_b$ as an illustrative example.
Generalization to arbitrary strong and weak abelian symmetry groups is the main focus of this section.
Specifically, the total symmetry group $\widetilde{G}$ arises as an extension of an abelian group $K$ with another group $G$, whose group structure is specified by the factor system $\omega_2\in \H^2(G, K)$ of the following short exact sequence
\begin{align}
1\rightarrow K\rightarrow\widetilde{G}\rightarrow G\rightarrow1.
\label{general extension}
\end{align}
The consistency condition of the factor system reads as
\begin{align}
    \omega_2(g, h)\cdot\omega_2(gh, l) = \omega_2(g, hl)\cdot\omega_2(h, l),
\end{align}
where $\omega_2$ takes the value of the group element in $K$.
Since a finite abelian group $K$ has the general form of $\prod\limits_{j=1}^m\Z_{n_j}$, we denote the group element $k\in K$ as $(k_1, \cdots, k_m)$ with $k_j\in \Z_{n_j}$, and similarly $\omega_2=(\omega_2^1, \cdots, \omega_2^m)$.

\subsection{$(1+1)D$ SPT phases with general group extension}
Here, we generalize the treatment in fermionic MPS to discuss the SPT classification in $(1+1)D$ pure states under a general group extension in Eq.~\eqref{general extension}.
For the symmetry group $K$, we directly encode the $K$-charge in the graded structure of the MPS and consider the twist of the $K$ symmetry by attaching a corresponding $Y$ block.
Implementing the $G$ symmetry demands that each subspace on the physical index of a given $K$-charge be a linear representation of the symmetry $G$. 

A $K$-charge is a one-dimensional linear representation of the Abelian group $K$, which is classified by 
\begin{align}
\H^1[K,U(1)]=K=\prod\limits_{j=1}^m\Z_{n_j}.
\label{K charge}
\end{align}
Therefore, the linear representation of $K$ can be labeled by an $m$-component vector $(k_1,\cdots,k_m)$, where $k_j\in\Z_{n_j}$ for $j=1,\cdots,m$ and is referred to as the $K$-charge.

Now we consider the SPT phases jointly protected by $K$ and $G$.
The group element in $\widetilde{G}$ is $\tilde{g}=(k, g)$, where $k\in K, g\in G$, with the following product rule
\begin{align}
    (k_1, g)\cdot (k_2, h) = [k_1k_2\omega_2(g, h), gh].\label{group structure}
\end{align}
A symmetry operator $\widetilde{U}(\Tilde{g})$ acted on the physical index can be expanded as
\begin{align}
\widetilde{U}(\Tilde{g})=\hat{O}(k)U(g),\label{U_tilde}
\end{align}
which should be a linear representation of $\Tilde{G}$ satisfying Eq.~\eqref{linear rep entire}.
Combined with Eq.~\eqref{group structure}, the symmetry actions on the physical index satisfy that
\begin{align}
    \hat{O}(k_1)U(g)\hat{O}(k_2)U(h) = \hat{O}[k_1k_2\omega_2(g, h)]U(gh),
\end{align}
from which we can obtain the following multiplication rules for $\hat{O}(k)$ and $U(g)$.
Firstly, we take $g=h=e$ (the identity group element) and reach that the symmetry action of $\hat{O}(k)$ on the physical index itself should form a linear representation, i.e.,
\begin{align}
    \hat{O}(k_1)\hat{O}(k_2) = \hat{O}(k_1k_2).\label{O1O2}
\end{align}
Meanwhile, by taking $k_1=k_2=e$, it can be easily verified that $U(g)$, the symmetry action associated with the $G$ group, needs to satisfy the following twisted group multiplication rules
\begin{align}
U(g)U(h)=\hat{O}[\omega_2(g,h)]U(gh).\label{Ugh}
\end{align}
Finally, we find that the application of $\hat{O}(k)$ and $U(g)$ should be commutative, i.e.,
\begin{align}
     U(g)\hat{O}(k) = \hat{O}(k)U(g),\label{OkUg}
\end{align}
where we have assumed $k_1=h=e$.

Similar to the fermion parity case, the application of $\widetilde{U}(\tilde{g})$ on the physical index leaves $\widetilde{V}(\tilde{g})$ at the right virtual index, as shown in Eq.~\eqref{1d fMPS symmetry}, which takes the form of
\begin{align}
    \widetilde{V}(\tilde{g})=\hat{O}^{\prime}(k)V(g),
\end{align}
where the combination formulas of $\hat{O}^{\prime}(k)$ and $V(g)$ are deduced as follows.
By sequentially applying two symmetry operators $\hat{O}(k_1)$ and $\hat{O}(k_2)$ to the physical index, we obtain the operator $\hat{O}^{\prime}(k_1)\hat{O}^{\prime}(k_2)$ on the virtual index.
This should be equivalent to the application of $\hat{O}(k_1k_2)$ that results in $\hat{O}^{\prime}(k_1k_2)$ on the virtual index.
Therefore, we can deduce the following relation for the operator $\hat{O}^{\prime}(k)$
\begin{align}
    \hat{O}^{\prime}(k_1)\hat{O}^{\prime}(k_2) = \mu_2(k_1,k_2)\hat{O}^{\prime}(k_1k_2), \quad \mu_2\in U(1).\label{Oprime}
\end{align}
Similarly, applying each side of Eq.~\eqref{Ugh} to the physical index, we are left with $V(g)V(h)$ and $\hat{O}^{\prime}[\omega_2(g,h)]V(gh)$ on the virtual index, respectively, where the resulting tensors should be equivalent, i.e., 
\begin{align}
V(g)V(h)=\nu_2(g,h)\hat{O}^{\prime}[\omega_2(g,h)]V(gh),~ \nu_2\in U(1).
\label{virtual proj rep}
\end{align}

In this case, there are two phase structures $\mu_2(k_1, k_2)$ and $\nu_2(g, h)$ that characterize the actions of $K$ and $G$ respectively.
Roughly speaking, they describe the SPT phases solely protected by $K$ or $G$.
The consistency condition for $\mu_2(k_1,k_2)$ can be easily derived since the symmetry action on the virtual index in Eq.~\eqref{Oprime} has the same form as Eq.~\eqref{Vk}.
Specifically, the following two ways to fuse the symmetry operations
\begin{align}
\begin{aligned}
\hat{O}^{\prime}(k_1)\hat{O}^{\prime}(k_2)\hat{O}^{\prime}(k_3)&=\mu_2(k_1,k_2)\hat{O}^{\prime}(k_1k_2)\hat{O}^{\prime}(k_3)\\
&=\mu_2(k_1,k_2)\mu_2(k_1k_2,k_3)\hat{O}^{\prime}(k_1k_2k_3),
\end{aligned}
\end{align}
and
\begin{align}
\begin{aligned}
\hat{O}^{\prime}(k_1)\hat{O}^{\prime}(k_2)\hat{O}^{\prime}(k_3)&=\hat{O}^{\prime}(k_1)\mu(k_2,k_3)\hat{O}^{\prime}(k_2k_3)\\
&=\mu_2(k_1,k_2k_3)\mu_2(k_2,k_3)\hat{O}^{\prime}(k_1k_2k_3)
\end{aligned}
\end{align}
must be consistent, from which we obtain the consistency equation for $\mu_2$,
\begin{align}
\frac{\mu_2(k_1,k_2)\mu_2(k_1k_2,k_3)}{\mu_2(k_1,k_2k_3)\mu_2(k_2,k_3)}=1.\label{2-cocycle K}
\end{align}
It implies that $\mu_2(k_1,k_2)$ is a 2-cocycle with $U(1)$ coefficient, and $\hat{O}^{\prime}(k)$ can be a projective representation of the group $K$ classified by the group 2-cohomology $\H^2[K, U(1)]$.
Therefore, $\mu_2(k_1, k_2)$ exactly characterizes the classification of $(1+1)D$ SPT phases when only group $K$ exists~\cite{Chen2011A}.

Before deriving the consistency condition for $\nu_2(g, h)$, we first focus on the interplay between $K$ and $G$, which is reflected by the nonzero $K$-charge $n_1(g)=[n_1^1(g),\cdots,n_1^m(g)]\in\H^1[K,U(1)]$ carried by each $V(g)$.
To formally define the $K$-charge $n_1(g)$, we consider the application of $U(g)\hat{O}(k)$ and $\hat{O}(k)U(g)$ on the physical index, and the resulting transformations on the virtual index should be equivalent up to a phase, i.e.,
\begin{align}
    \hat{O}^{\prime}(k)V(g) = e^{i\phi(k, g)}V(g)\hat{O}^{\prime}(k).\label{OprimeV}
\end{align}
As the above relation must hold for all $k=(k_1, \cdots, k_m)$, we can choose a specific case where only $k_j = 1$ (generator of $\Z_{n_j}$) while all the other elements are zero, whose symmetry action is denoted as $\hat{O}^{\prime}_j$.
Therefore, the requirement in Eq.~\eqref{OprimeV} means that
\begin{align}
    \hat{O}^{\prime}_jV(g)=\alpha_j(g)V(g)\hat{O}^{\prime}_j,
\end{align}
for $j=1, \cdots,m$.
Recall that $\hat{O}^{\prime}(k)$ forms a projective representation of group $K$, the above set $\{\hat{O}^{\prime}_j\}$ is able to generate the symmetry action for any $k$ on the virtual index in the following way
\begin{align}
    \hat{O}^{\prime}(k) = \beta(k) \prod\limits_{j=1}^m\hat{O}^{\prime k_j}_j,
\end{align}
where $\beta(k)$ is a global phase factor defined for each $k$, which in principle can be derived from $\mu_2$ in Eq.~\eqref{Oprime}, but its explicit form is irrelevant in this case.
Specifically, $\hat{O}^{\prime n_j}_j\propto \hat{O}^{\prime}(e) \propto I$, where $e$ is the group identity element in $K$, resulting in the following condition for $\alpha_j$
\begin{align}
    \alpha_j(g)^{n_j} = 1.
\end{align}
Therefore, we can define $\alpha_j(g) = e^{i2\pi n_1^j(g)/n_j}$ with $n_1^j(g)\in \Z_{n_j}$ for $j=1, \cdots, m$, which we refer to as ``the $K$-charge carried by $V(g)$''.
In summary, the commutation relation between the symmetry actions of $K$ and $G$ on the virtual index reads as
\begin{align}
    \hat{O}^{\prime}(k)V(g)=V(g)\hat{O}^{\prime}(k)\prod\limits_{j=1}^me^{i2\pi k_j\cdot n_1^j(g)/n_j}.\label{K-charge V}
\end{align}

Now we are ready to consider the consistency condition for the $K$-charge of $V(g)$ from the mixed anomaly of $K$ and $G$.
Consider left-multiplying Eq.~\eqref{virtual proj rep} with $\hat{O}^{\prime}(k)$, from which we obtain
\begin{align}
    \begin{aligned}
    &\textrm{L.H.S.} = \hat{O}^{\prime}(k) V(g)V(h)\\
    =& \prod\limits_{j=1}^me^{i{2\pi}{k_j}[n_1^j(g)+n_1^j(h)]/n_j}V(g)V(h)\hat{O}^{\prime}(k)
    \end{aligned}\label{LHS_G}
\end{align}
and
\begin{align}
    \begin{aligned}
        &\textrm{R.H.S.} = \nu_2(g,h)\hat{O}^{\prime}(k)\hat{O}^{\prime}[\omega_2(g,h)]V(gh)\\
        =& \frac{\mu_2[k, \omega_2(g,h)]}{\mu_2[\omega_2(g,h), k]}\prod\limits_{j=1}^me^{i{2\pi}{k_j}n_1^j(gh)/n_j}V(g)V(h)\hat{O}^{\prime}(k),
    \end{aligned}\label{RHS_G}
\end{align}
respectively.
The above constraint requires that 
\begin{align}
    n_1^j(g)+n_1^j(h)-n_1^j(gh) &= \mathcal{O}_2^j(g,h)~(\mathrm{mod}~n_j)\label{1-cocycle}
\end{align}
for $\forall j=1,\cdots,m$, where the obstruction function $\mathcal{O}_2^j$ is defined as
\begin{align}
    \mathcal{O}_2^j(g, h) \equiv \frac{n_j}{2\pi}\arg{\frac{\mu_2[1_j, \omega_2(g,h)]}{\mu_2[\omega_2(g,h), 1_j]}}\label{Obstruction2}
\end{align}
with $1_j$ abbreviating the group element of $K$ whose $j$-th element $k_j=1$ and all the other are zero.
It implies that $n_1(g)$ satisfies a twisted 1-cocycle condition with coefficient $\H^1[K,U(1)]$, i.e., $n_1\in\H^1\left[G,\H^1[K,U(1)]\right]$, which has solutions if and only if $\mathcal{O}_2^j$ is a 2-coboundary in $\H^2(G, \Z_{n_j})$.
In addition, we have
\begin{align}
    \frac{\mu_2[k, \omega_2(g,h)]}{\mu_2[\omega_2(g,h), k]} &= \prod\limits_{j=1}^m\left(\frac{\mu_2[1_j, \omega_2(g,h)]}{\mu_2[\omega_2(g,h), 1_j]}\right)^{k_j}.\label{mu_k_prop}
\end{align} 

Finally, let us derive the consistency condition for the phase factor system $\nu_2$ characterizing the projective representation $V(g)$ on virtual indices.
Sequentially applying the symmetry operators $U(l)$, $U(h)$, and $U(g)$ on the physical index ($g, h, l\in G$) leaves $V(g)V(h)V(l)$ on the right virtual index.
On the one hand, we have
\begin{align}
\begin{aligned}
&V(g)V(h)V(l)=\nu_2(g,h)\hat{O}^{\prime}[\omega_2(g,h)]V(gh)V(l)\\
=&\nu_2(g,h)\nu_2(gh,l)\mu_2[\omega_2(g, h), \omega_2(gh, l)]\\
&\hat{O}^{\prime}[\omega_2(g,h)\cdot\omega_2(gh,l)]V(ghl);
\end{aligned}
\label{3 sequential 1}
\end{align}
on the other hand, 
\begin{align}
\begin{aligned}
\label{3 sequential 2}
&V(g)V(h)V(l)=V(g)\nu_2(h,l)\hat{O}^{\prime}[\omega_2(h,l)]V(hl)\\
=&\prod\limits_{j=1}^me^{i{2\pi}\omega_2^j(h,l)n_1^j(g)/n_j}\nu_2(h,l)\hat{O}^{\prime}[\omega_2(h,l)]V(g)V(hl)\\
=&\prod\limits_{j=1}^me^{i{2\pi}\omega_2^j(h,l)n_1^j(g)/n_j}\nu_2(h,l)\nu_2(g,hl)\\
&\mu_2[\omega_2(h,l), \omega_2(g, hl)]\hat{O}^{\prime}[\omega_2(h,l)\cdot\omega_2(g,hl)]V(ghl).
\end{aligned}
\end{align}
The above two formulas imply the following twisted 2-cocycle condition for the phase structure characterizing the SPT pure state protected by $G$
\begin{align}
\frac{\nu_2(g,h)\nu_2(gh,l)}{\nu_2(g,hl)\nu_2(h,l)}=\mathcal{O}_3(g,h,l)
\label{twisted 2-cocycle}
\end{align}
which has solutions if and only if the obstruction function
\begin{align}
\mathcal{O}_3(g,h,l)\equiv\prod\limits_{j=1}^me^{i{2\pi}\omega_2^j(h,l)n_1^j(g)/n_j}\frac{\mu_2[\omega_2(g, hl), \omega_2(h,l)]}{\mu_2[\omega_2(g, h), \omega_2(gh, l)]}\label{Obstruction3}
\end{align}
is a 3-coboundary in $\H^3[G, U(1)]$.

We summarize the consistency conditions for the construction of an MPS for $(1+1)D$ $\widetilde{G}$-symmetric SPT state.
There are three topological invariants to label different phases:
\begin{enumerate}[1.]
\item $\mu_2\in\H^2[K,U(1)]$: different SPT phases solely protected by the symmetry $K$ with the 2-cocycle condition in Eq.~\eqref{2-cocycle K}, or simply $\mathrm{d}\mu_2=0$.
\item $n_1\in \H^1\left[G,\H^1[K,U(1)]\right]$: the $K$-charge carried by the projective representation $V(g)$ on virtual indices, labeling different SPT phases jointly protected by $K$ and $G$, with the twisted 1-cocycle condition in Eq.~\eqref{1-cocycle}, or simply $\mathrm{d}n_1=\mathcal{O}_2$, together with the condition that $\mathcal{O}_2^j$ in Eq.~\eqref{Obstruction2} is a 2-coboundary in $\H^2[G, \Z_{n_j}]$ and the constraint in Eq.~\eqref{mu_k_prop}.
\item $\nu_2\in \H^2[G,U(1)]$:  different SPT states protected by $G$ exclusively with the twisted 2-cocycle condition in Eq.~\eqref{twisted 2-cocycle}, or simply $\mathrm{d}\nu_2 = \mathcal{O}_3$, and the condition that $\mathcal{O}_3$ in Eq.~\eqref{Obstruction3} is a 3-coboundary in $\H^3[G, U(1)]$.
\end{enumerate}
The obstructions in Eq.~\eqref{Obstruction2} and~\eqref{Obstruction3} mean that for nontrivial $\omega_2$ some $n_1$ and $\mu_2$ may not be well defined as long as these conditions are not satisfied, from which some nontrivial intrinsic ASPT phases may emerge to be introduced later.

\subsection{Classification of ASPT in $(1+1)D$ systems}
In this section, we provide the classification data for ASPTs protected by a strong symmetry $K$ and a weak symmetry $G$, allowing a possible non-trivial extension characterized by $\omega_2 \in \H^2(G, K)$.  

In this case, the symmetry actions for the total group on the physical and virtual indices still take the forms of $\widetilde{U}(\Tilde{g})=\hat{O}(k)U(g)$ and $\widetilde{V}(\Tilde{g})=\hat{O}^{\prime}(k)V(g)$ ($\Tilde{g}=(k,g)$, where $k\in K$ and $g\in G$), respectively.
Similar to the fLPDO case, the phase ambiguity $\mu_2(k_1, k_2)$ for the strong symmetry $K$, as defined in Eq.~\eqref{Oprime}, should be conserved at each side of the density matrix since the applications of $U(k)$ on different sides are independent.
Therefore, the 2-cocycle condition generated from the consistency of fusing three symmetry actions on the virtual index $\hat{O}^{\prime}(k_1)\hat{O}^{\prime}(k_2)\hat{O}^{\prime}(k_3)$ remains, namely
\begin{align}
\frac{\mu_2(k_1,k_2)\mu_2(k_1k_2,k_3)}{\mu_2(k_1,k_2k_3)\mu_2(k_2,k_3)}=1,\label{2-cocycle K LPDO}
\end{align}
characterizing different ASPT phases exclusively protected by the strong symmetry $K$.

As for the weak symmetry $G$, the corresponding symmetry actions on the physical and virtual index satisfy Eqs.~\eqref{Ugh} and~\eqref{virtual proj rep}.
However, different from the strong symmetry $K$ that can be applied only on one side, a weak symmetry operator $U(g)$ must be simultaneously applied to both sides of the density matrix to guarantee its invariance under symmetry transformation.
Applying two symmetry operators $U(h)$ and $U(g)$ sequentially to the physical indices on both sides, we obtain $V(g)V(h)$ and $V(g)^*V(h)^*$ on the virtual index of the upper and lower sides, respectively.
According to Eq.~\eqref{virtual proj rep}, we have the following relation of the operator $V(g)$ acting on the virtual indices as
\begin{align}
\begin{aligned}
V(g)V(h)&=\nu_2(g,h)\hat{O}^{\prime}[\omega_2(g,h)]V(gh),\\
V(g)^{*}V(h)^{*}&=\nu_2(g,h)^{-1}\hat{O}^{\prime}[\omega_2(g,h)]^{*}V(gh)^{*},
\end{aligned}
\end{align}
where $\nu_2(g,h)$ is a $U(1)$ phase. 
Similar to the discussion of the fermion parity case, the total phase ambiguity on the virtual indices is canceled from the upper and lower sides, indicating that $\nu_2$ is not a topological invariant characterizing distinct ASPT phases.
In other words, if there is only weak symmetry, we cannot have any nontrivial ASPT density matrix. 
As a consequence, we do not have to worry about the coboundary condition for $\hat{O}_3$ to promise a solution for the twisted cocycle condition of $\nu_2$ in Eq.~\eqref{twisted 2-cocycle}.

Finally, let us consider the $K$-charge $n_1(g)=[n_1^1(g),\cdots,n_1^m(g)]\in\H^1[K,U(1)]$ carried by each $V(g)$ defined in Eq.~\eqref{K-charge V} to explore the interplay between strong and weak symmetries.
As $K$ is a strong symmetry, the $K$-charge of $V(g)$ cannot tunnel between the upper and lower spaces and must be conserved on each side.
Consider the sequential application of two weak symmetry operators $U(h)$ and $U(g)$ on an LPDO, followed by the application of a strong symmetry operator $\hat{O}(k)$ only on the upper side, which (similar to Eqs.~\eqref{LHS_G} and~\eqref{RHS_G}) leaves on the virtual space the following transformation
\begin{align}
\begin{aligned}\label{ovv}
 &\hat{O}^{\prime}(k) V(g)V(h)\\
 =&\prod\limits_{j=1}^me^{i{2\pi}{k_j}[n_1^j(g)+n_1^j(h)]/n_j}V(g)V(h)\hat{O}^{\prime}(k)\\
 =& \frac{\mu_2[k, \omega_2(g,h)]}{\mu_2[\omega_2(g,h), k]}\prod\limits_{j=1}^me^{i{2\pi}{k_j}n_1^j(gh)/n_j}V(g)V(h)\hat{O}^{\prime}(k),
\end{aligned}
\end{align}
on the upper side. The two lines are the result of two ways of combining and permuting the three operators. On the lower side, the transformation is simply $V(g)^{*}V(h)^{*}$.
We see that the total phase ambiguity after grouping the upper and lower virtual indices is not canceled.
The consistency condition for the two lines in Eq.~\eqref{ovv} indicates the following condition
\begin{align}
    n_1^j(g)+n_1^j(h)-n_1^j(gh) &= \mathcal{O}_2^j(g,h)~(\mathrm{mod}~n_j)\label{1-cocycle LPDO}
\end{align}
with
\begin{align}
    \mathcal{O}_2^j(g, h) \equiv \frac{n_j}{2\pi}\arg{\frac{\mu_2[1_j, \omega_2(g,h)]}{\mu_2[\omega_2(g,h), 1_j]}}\label{Obstruction2 LPDO}
\end{align}
and
\begin{align}
    \frac{\mu_2[k, \omega_2(g,h)]}{\mu_2[\omega_2(g,h), k]} = \prod\limits_{j=1}^m\left(\frac{\mu_2[1_j, \omega_2(g,h)]}{\mu_2[\omega_2(g,h), 1_j]}\right)^{k_j}.\label{mu_k_prop LPDO}
\end{align}

We summarize the consistency conditions for constructing an LPDO of $(1+1)D$ $\widetilde{G}$-symmetric ASPT state.
There are two topological invariants to label different phases:
\begin{enumerate}[1.]
\item $\mu_2\in \H^2[K,U(1)]$: the ASPT phases solely protected by the strong symmetry $K$ with the 2-cocycle condition in Eq.~\eqref{2-cocycle K LPDO}, or simply $\mathrm{d}\mu_2=0$.
\item $n_1\in\H^1\left[G,\H^1[K,U(1)]\right]$: the ASPT phases jointly protected by the strong and weak symmetries, with a 1-cocycle condition in Eq.~\eqref{1-cocycle LPDO}, or simply $\mathrm{d}n_1=\mathcal{O}_2$, together with the condition that $\mathcal{O}_2^j$ in Eq.~\eqref{Obstruction2 LPDO} is a 2-coboundary in $\H^2[G, \Z_{n_j}]$ and the constraint in Eq.~\eqref{mu_k_prop LPDO}.
\end{enumerate}
There is no coboundary condition for $\hat{O}_3$ as that for pure-state SPT.

At first glance, it might seem that the classification of ASPT phases protected by $\widetilde{G}$ is simpler than that of pure-state SPT phases protected by the same symmetry group, where both $K$ and $G$ are strong, because the latter has one more topological invariant.
However, we will see that this is not always the case, as the twisted 2-cocycle condition for $\nu_2$ in Eq.~\eqref{twisted 2-cocycle} imposes more restrictions on $n_1$ and $\omega_2$, i.e., $\hat{O}_3$ should be a 3-coboundary in $\H^3[G, U(1)]$.
To be more specific, for a nontrivial group extension labeled by $\omega_2\neq 0$, certain $n_1(g)$ may not be well defined if Eq.~\eqref{twisted 2-cocycle} may have no solution.
In contrast, for ASPT phases where $G$ is a weak symmetry, the constraint of Eq.\eqref{twisted 2-cocycle} no longer holds.
Therefore, such a choice of nontrivial $n_1$, which may disrupt the coboundary condition, is legitimate.
In this case, the ASPT phase has no pure-state SPT counterpart and is referred to as an intrinsic ASPT phase.
We note that a necessary condition for the emergence of intrinsic ASPT phases is a nontrivial group extension characterized by $\omega_2$ in the short exact sequence in Eq.~\eqref{general extension}.

In the following sections, we will provide two examples of explicit LPDO constructions for ASPTs whose weak and strong symmetries belong to the same group $K=\Z_2$ and $G=\Z_2$ while their group extensions are different.

\subsection{Example: decohered Haldane phase in $(1+1)D$ systems}
When $K=\Z_2$ and $G=\Z_2$ have trivial group extension, the ASPT phase is a descendant of the pure state SPT protected by $\Z_2\times \Z_2$ symmetry, namely the cluster state.
We begin with the $(1+1)D$ cluster state whose Hamiltonian is
\begin{align}
    H=-\sum_i^N\left(\sigma_i^z \tau_{i+1 / 2}^x\sigma_{i+1}^z +\tau_{i-1 / 2}^z \sigma_i^x \tau_{i+1 / 2}^z\right),
\end{align}
where $N$ is the number of sites.
In this model, there are two spin-$\frac{1}{2}$ degrees of freedom $\sigma_i$ and $\tau_{i+1/2}$ at the site $i$, each carrying a linear representation of one of $\Z_2$'s.
The ground state of this model follows the decorated domain wall construction~\cite{Chen2014}, i.e.,
\begin{align}
\begin{aligned}
    \ket{\psi} &= \frac{1}{2^{N/2}}\sum_{\{\sigma_i\}}\ket{\cdots\uparrow_\sigma\rightarrow_{\tau}\uparrow_{\sigma}\rightarrow_{\tau}\uparrow_{\sigma}\leftarrow_{\tau}\downarrow_{\sigma}\leftarrow_{\tau}\uparrow_{\sigma}\cdots}\\
    &\equiv \frac{1}{2^{N/2}}\sum_{\{\sigma_i\}}\ket{\psi_{\{\sigma_i\}}},
\end{aligned}\label{cluster}
\end{align}
where $\tau$-spin excitations are positioned at the domain walls of $\sigma$ spins.
$\prod_i \sigma_i^x$ and $\prod_i \tau_{i+1/2}^x$ define two global $\Z_2$ symmetries of the system.
Specifically, $\prod_i\sigma_i^x$ map two states with the same domain wall configurations but opposite $\sigma$ spins to each other, while $\prod_i\tau_{i+1/2}^x$ stabilizes each state $\ket{\psi_{\{\sigma_i\}}}$ with eigenvalue $1$ under PBC, i.e.,
\begin{align}
    \prod_i \tau_{i+1/2}^x\ket{\psi_{\{\sigma_i\}}} = \ket{\psi_{\{\sigma_i\}}}. 
\end{align}

This state can be represented by an injective MPS with $D=2$ (not normalized)
\begin{align}
\begin{tikzpicture}[scale=0.8]
\tikzstyle{sergio}=[rectangle,draw=none]
\filldraw[fill=white, draw=black, rounded corners] (-2,-0.5)--(-0.5,-0.5)--(-0.5,0.5)--(-2,0.5)--cycle;
\draw[line width=1pt] (-0.5,0) -- (0,0);
\draw[line width=1pt] (-2,0) -- (-2.5,0);
\draw[line width=2pt, color=red] (-1.6,0.5) -- (-1.6,1);
\draw[line width=2pt, color=red] (-0.9,0.5) -- (-0.9,1);
\path (-1.25,0) node [style=sergio]{\large $\mathsf{A}$};
\path (-1.4,0.8) node [style=sergio]{$\sigma$};
\path (-0.7,0.8) node [style=sergio]{$\tau$};
\path (-1.6,1.4) node [style=sergio]{$\ket{\uparrow}$};
\path (-0.8,1.4) node [style=sergio]{$\ket{\rightarrow}$};
\path (-2.5,0.3) node [style=sergio]{$0$};
\path (-0,0.3) node [style=sergio]{$0$};
\path (0.5,0) node [style=sergio]{$=1$};
\filldraw[fill=white, draw=black, rounded corners] (2,-0.5)--(3.5,-0.5)--(3.5,0.5)--(2,0.5)--cycle;
\draw[line width=1pt] (3.5,0) -- (4,0);
\draw[line width=1pt] (2,0) -- (1.5,0);
\draw[line width=2pt, color=red] (2.4,0.5) -- (2.4,1);
\draw[line width=2pt, color=red] (3.1,0.5) -- (3.1,1);
\path (2.75,0) node [style=sergio]{\large $\mathsf{A}$};
\path (2.6,0.8) node [style=sergio]{$\sigma$};
\path (3.3,0.8) node [style=sergio]{$\tau$};
\path (2.4,1.4) node [style=sergio]{$\ket{\uparrow}$};
\path (3.2,1.4) node [style=sergio]{$\ket{\leftarrow}$};
\path (1.5,0.3) node [style=sergio]{$0$};
\path (4,0.3) node [style=sergio]{$1$};
\path (4.5,0) node [style=sergio]{$=1$};
\tikzstyle{sergio}=[rectangle,draw=none]
\filldraw[fill=white, draw=black, rounded corners] (-2,-3.5)--(-0.5,-3.5)--(-0.5,-2.5)--(-2,-2.5)--cycle;
\draw[line width=1pt] (-0.5,-3) -- (0,-3);
\draw[line width=1pt] (-2,-3) -- (-2.5,-3);
\draw[line width=2pt, color=red] (-1.6,-2.5) -- (-1.6,-2);
\draw[line width=2pt, color=red] (-0.9,-2.5) -- (-0.9,-2);
\path (-1.25,-3) node [style=sergio]{\large $\mathsf{A}$};
\path (-1.4,-2.2) node [style=sergio]{$\sigma$};
\path (-0.7,-2.2) node [style=sergio]{$\tau$};
\path (-1.6,-1.6) node [style=sergio]{$\ket{\downarrow}$};
\path (-0.8,-1.6) node [style=sergio]{$\ket{\leftarrow}$};
\path (-2.5,-2.7) node [style=sergio]{$1$};
\path (-0,-2.7) node [style=sergio]{$0$};
\path (0.5,-3) node [style=sergio]{$=1$};
\filldraw[fill=white, draw=black, rounded corners] (2,-3.5)--(3.5,-3.5)--(3.5,-2.5)--(2,-2.5)--cycle;
\draw[line width=1pt] (3.5,-3) -- (4,-3);
\draw[line width=1pt] (2,-3) -- (1.5,-3);
\draw[line width=2pt, color=red] (2.4,-2.5) -- (2.4,-2);
\draw[line width=2pt, color=red] (3.1,-2.5) -- (3.1,-2);
\path (2.75,-3) node [style=sergio]{\large $\mathsf{A}$};
\path (2.6,-2.2) node [style=sergio]{$\sigma$};
\path (3.3,-2.2) node [style=sergio]{$\tau$};
\path (2.4,-1.6) node [style=sergio]{$\ket{\downarrow}$};
\path (3.2,-1.6) node [style=sergio]{$\ket{\rightarrow}$};
\path (1.5,-2.7) node [style=sergio]{$1$};
\path (4,-2.7) node [style=sergio]{$1$};
\path (4.5,-3) node [style=sergio]{$=1$};
\end{tikzpicture},\label{MPS_cluster}
\end{align}
whose symmetry action can be depicted as
\begin{align}
\begin{tikzpicture}[scale=0.8]
\tikzstyle{sergio}=[rectangle,draw=none]
\filldraw[fill=white, draw=black, rounded corners] (0.25,-0.5)--(1.75,-0.5)--(1.75,0.5)--(0.25,0.5)--cycle;
\draw[line width=1pt] (1.75,0) -- (2.25,0);
\draw[line width=1pt] (0.25,0) -- (-0.25,0);
\draw[line width=2pt, color=red] (0.65,0.5) -- (0.65,2);
\draw[line width=2pt, color=red] (1.35,0.5) -- (1.35,2);
\path (1,0) node [style=sergio]{\large $\mathsf{A}$};
\path (2.75,0) node [style=sergio]{$=$};
\filldraw[fill=white, draw=black] (0.65,1.25)circle (10pt);
\path (0.65,1.25) node [style=sergio]{$\sigma_x$};
\filldraw[fill=white, draw=black, rounded corners] (4.75,-0.5)--(6.25,-0.5)--(6.25,0.5)--(4.75,0.5)--cycle;
\draw[line width=1pt] (4.75,0) -- (3.25,0);
\filldraw[fill=white, draw=black] (4,0)circle (10pt);
\path (4,0) node [style=sergio]{$X$};
\path (5.5,0) node [style=sergio]{\large $\mathsf{A}$};
\draw[line width=1pt] (6.25,0) -- (7.75,0);
\filldraw[fill=white, draw=black] (7,0)circle (10pt);
\path (7,0) node [style=sergio]{$X$};
\draw[line width=2pt, color=red] (5.15,0.5) -- (5.15,1);
\draw[line width=2pt, color=red] (5.85,0.5) -- (5.85,1);
\end{tikzpicture},
\end{align}
and
\begin{align}
\begin{tikzpicture}[scale=0.8]
\tikzstyle{sergio}=[rectangle,draw=none]
\filldraw[fill=white, draw=black, rounded corners] (0.25,-0.5)--(1.75,-0.5)--(1.75,0.5)--(0.25,0.5)--cycle;
\draw[line width=1pt] (1.75,0) -- (2.25,0);
\draw[line width=1pt] (0.25,0) -- (-0.25,0);
\draw[line width=2pt, color=red] (0.65,0.5) -- (0.65,2);
\draw[line width=2pt, color=red] (1.35,0.5) -- (1.35,2);
\path (1,0) node [style=sergio]{\large $\mathsf{A}$};
\path (2.75,0) node [style=sergio]{$=$};
\filldraw[fill=white, draw=black] (1.35,1.25)circle (10pt);
\path (1.35,1.25) node [style=sergio]{$\tau_x$};
\filldraw[fill=white, draw=black, rounded corners] (4.75,-0.5)--(6.25,-0.5)--(6.25,0.5)--(4.75,0.5)--cycle;
\draw[line width=1pt] (4.75,0) -- (3.25,0);
\filldraw[fill=white, draw=black] (4,0)circle (10pt);
\path (4,0) node [style=sergio]{$Z$};
\path (5.5,0) node [style=sergio]{\large $\mathsf{A}$};
\draw[line width=1pt] (6.25,0) -- (7.75,0);
\filldraw[fill=white, draw=black] (7,0)circle (10pt);
\path (7,0) node [style=sergio]{$Z$};
\draw[line width=2pt, color=red] (5.15,0.5) -- (5.15,1);
\draw[line width=2pt, color=red] (5.85,0.5) -- (5.85,1);
\end{tikzpicture},
\end{align}
which demonstrates that the $(1+1)D$ cluster state belongs to a nontrivial SPT phase characterized by the projective representation on virtual indices.

In the following, we assume that one of the $\Z_2$ symmetries, namely $G = \{1, g\}$ with $U(g) = \prod_i \sigma_i^x$, becomes weak, while the other $K = \{1, k\}$ with $\hat{O}(k) = \prod_i \tau_{i+1/2}^x$ remains strong.
We consider the following mixed state
\begin{align}
    \rho = \frac{1}{2^N}\sum_{i}\ket{\psi_{\{\sigma_i\}}}\hspace{-1mm}\bra{\psi_{\{\sigma_i\}}},
    \label{Z4}
\end{align}
which is the incoherent superposition of different components in the cluster state.
From the above discussion, it can easily be verified that $\prod_i \tau_{i+1/2}^x$ is a strong symmetry, while $ \prod_i \sigma_i^x$ acts as a weak symmetry as it exchanges states with opposite $\sigma$ configurations.
In a sense, Eq.~\eqref{Z4} is a fixed-point density matrix for the ASPT phase, as the weak degrees of freedom are completely classical in this density matrix. 
This mixed state can be realized by the following disordered system
\begin{align}
    &H_0=-\sum_i^N\sigma_i^z \tau_{i+1 / 2}^x\sigma_{i+1}^z,\\
    &H_{\rm disorder} = -\sum_i^N h_i \sigma_i^z 
\end{align}
for $h_i=\pm 1$ randomly chosen for each site.
The symmetry property for the disordered Hamiltonians read as
\begin{align}
    &[H_0, \hat{O}(k)] = 0, \quad [H_0, U(g)] = 0,\\
    &[H_{\rm disorder}, \hat{O}(k)] = 0, \quad \{H_{\rm disorder}, U(g)\} = 0,
\end{align}
i.e., the weak symmetry can commute with the disordered term with an extra phase.

Now we propose a general construction of the LPDO representation for such a mixed state, i.e., we start from the MPS representation $\mathsf{A_0}$ for the pure state $\ket{\psi} = \sum_k\sqrt{\lambda_k}\ket{\psi_k}$, then the corresponding mixed state $\rho = \sum_k\lambda_k\ket{\psi_k}\bra{\psi_k}$ can be expressed by the following LPDO tensor
\begin{align}
\begin{tikzpicture}[scale=0.8]
\tikzstyle{sergio}=[rectangle,draw=none]
\filldraw[fill=white, draw=black, rounded corners] (-2,-0.5)--(-0.5,-0.5)--(-0.5,0.5)--(-2,0.5)--cycle;
\draw[line width=1pt] (-0.5,0) -- (0,0);
\draw[line width=1pt] (-2,0) -- (-2.5,0);
\draw[line width=2pt, color=red] (-1.25,0.5) -- (-1.25,1);
\path (-1.25,0) node [style=sergio]{\large $\mathsf{A}$};
\path (-1.1,1.2) node [style=sergio]{$p$};
\path (-2.5,0.3) node [style=sergio]{$\alpha$};
\path (-0,0.3) node [style=sergio]{$\beta$};
\path (0.5,0) node [style=sergio]{$=$};
\filldraw[fill=white, draw=black, rounded corners] (1.5,-0.5)--(3,-0.5)--(3,0.5)--(1.5,0.5)--cycle;
\draw[line width=1pt] (3,0) -- (3.5,0);
\draw[line width=1pt] (1.5,0) -- (1,0);
\draw[line width=2pt, color=red] (2.25,0.5) -- (2.25,1);
\path (2.25,0) node [style=sergio]{\large $\mathsf{A_0}$};
\path (2.4,1.2) node [style=sergio]{$p$};
\path (1,0.3) node [style=sergio]{$\alpha$};
\path (3.5,0.3) node [style=sergio]{$\beta$};
\draw[line width=3pt, color=blue] (-1.6,-1) -- (-1.6,-0.5);
\draw[line width=3pt, color=blue] (-0.9,-1) -- (-0.9,-0.5);
\path (-1.8,-1.2) node [style=sergio]{$a_l$};
\path (-0.7,-1.2) node [style=sergio]{$a_r$};
\path (4.6,0) node [style=sergio]{\large $\delta_{\alpha, a_l}\delta_{\beta, a_r}$};
\end{tikzpicture},
\label{general LPDO}
\end{align}

With this construction, the LPDO representation for Eq.~\eqref{Z4} reads as
\begin{align}
\begin{tikzpicture}[scale=0.8]
\tikzstyle{sergio}=[rectangle,draw=none]
\filldraw[fill=white, draw=black, rounded corners] (-2,-0.5)--(-0.5,-0.5)--(-0.5,0.5)--(-2,0.5)--cycle;
\draw[line width=1pt] (-0.5,0) -- (0,0);
\draw[line width=1pt] (-2,0) -- (-2.5,0);
\draw[line width=2pt, color=red] (-1.6,0.5) -- (-1.6,1);
\draw[line width=2pt, color=red] (-0.9,0.5) -- (-0.9,1);
\path (-1.25,0) node [style=sergio]{\large $\mathsf{A}$};
\path (-1.4,0.8) node [style=sergio]{$\sigma$};
\path (-0.7,0.8) node [style=sergio]{$\tau$};
\path (-1.6,1.4) node [style=sergio]{$\ket{\uparrow}$};
\path (-0.8,1.4) node [style=sergio]{$\ket{\rightarrow}$};
\path (-2.5,0.3) node [style=sergio]{$0$};
\path (-0,0.3) node [style=sergio]{$0$};
\draw[line width=3pt, color=blue] (-1.6,-1) -- (-1.6,-0.5);
\draw[line width=3pt, color=blue] (-0.9,-1) -- (-0.9,-0.5);
\path (-1.6,-1.3) node [style=sergio]{$0$};
\path (-0.9,-1.3) node [style=sergio]{$0$};
\path (0.5,0) node [style=sergio]{$=1$};
\filldraw[fill=white, draw=black, rounded corners] (2,-0.5)--(3.5,-0.5)--(3.5,0.5)--(2,0.5)--cycle;
\draw[line width=1pt] (3.5,0) -- (4,0);
\draw[line width=1pt] (2,0) -- (1.5,0);
\draw[line width=2pt, color=red] (2.4,0.5) -- (2.4,1);
\draw[line width=2pt, color=red] (3.1,0.5) -- (3.1,1);
\path (2.75,0) node [style=sergio]{\large $\mathsf{A}$};
\path (2.6,0.8) node [style=sergio]{$\sigma$};
\path (3.3,0.8) node [style=sergio]{$\tau$};
\path (2.4,1.4) node [style=sergio]{$\ket{\uparrow}$};
\path (3.2,1.4) node [style=sergio]{$\ket{\leftarrow}$};
\path (1.5,0.3) node [style=sergio]{$0$};
\path (4,0.3) node [style=sergio]{$1$};
\draw[line width=3pt, color=blue] (2.4,-1) -- (2.4,-0.5);
\draw[line width=3pt, color=blue] (3.1,-1) -- (3.1,-0.5);
\path (2.4,-1.3) node [style=sergio]{$0$};
\path (3.1,-1.3) node [style=sergio]{$1$};
\path (4.5,0) node [style=sergio]{$=1$};
\tikzstyle{sergio}=[rectangle,draw=none]
\filldraw[fill=white, draw=black, rounded corners] (-2,-4)--(-0.5,-4)--(-0.5,-3)--(-2,-3)--cycle;
\draw[line width=1pt] (-0.5,-3.5) -- (0,-3.5);
\draw[line width=1pt] (-2,-3.5) -- (-2.5,-3.5);
\draw[line width=2pt, color=red] (-1.6,-3) -- (-1.6,-2.5);
\draw[line width=2pt, color=red] (-0.9,-3) -- (-0.9,-2.5);
\path (-1.25,-3.5) node [style=sergio]{\large $\mathsf{A}$};
\path (-1.4,-2.7) node [style=sergio]{$\sigma$};
\path (-0.7,-2.7) node [style=sergio]{$\tau$};
\path (-1.6,-2.1) node [style=sergio]{$\ket{\downarrow}$};
\path (-0.8,-2.1) node [style=sergio]{$\ket{\leftarrow}$};
\path (-2.5,-3.2) node [style=sergio]{$1$};
\path (-0,-3.2) node [style=sergio]{$0$};
\draw[line width=3pt, color=blue] (-1.6,-4.5) -- (-1.6,-4);
\draw[line width=3pt, color=blue] (-0.9,-4.5) -- (-0.9,-4);
\path (-1.6,-4.8) node [style=sergio]{$1$};
\path (-0.9,-4.8) node [style=sergio]{$0$};
\path (0.5,-3.5) node [style=sergio]{$=1$};
\filldraw[fill=white, draw=black, rounded corners] (2,-4)--(3.5,-4)--(3.5,-3)--(2,-3)--cycle;
\draw[line width=1pt] (3.5,-3.5) -- (4,-3.5);
\draw[line width=1pt] (2,-3.5) -- (1.5,-3.5);
\draw[line width=2pt, color=red] (2.4,-3) -- (2.4,-2.5);
\draw[line width=2pt, color=red] (3.1,-3) -- (3.1,-2.5);
\path (2.75,-3.5) node [style=sergio]{\large $\mathsf{A}$};
\path (2.6,-2.7) node [style=sergio]{$\sigma$};
\path (3.3,-2.7) node [style=sergio]{$\tau$};
\path (2.4,-2.1) node [style=sergio]{$\ket{\downarrow}$};
\path (3.2,-2.1) node [style=sergio]{$\ket{\rightarrow}$};
\path (1.5,-3.2) node [style=sergio]{$1$};
\path (4,-3.2) node [style=sergio]{$1$};
\draw[line width=3pt, color=blue] (2.4,-4.5) -- (2.4,-4);
\draw[line width=3pt, color=blue] (3.1,-4.5) -- (3.1,-4);
\path (2.4,-4.8) node [style=sergio]{$1$};
\path (3.1,-4.8) node [style=sergio]{$1$};
\path (4.5,-3.5) node [style=sergio]{$=1$};
\end{tikzpicture},
\label{1d tensor element}
\end{align}
with the following symmetry actions
\begin{align}
\begin{tikzpicture}[scale=0.8]
\tikzstyle{sergio}=[rectangle,draw=none]
\filldraw[fill=white, draw=black, rounded corners] (0.25,-0.5)--(1.75,-0.5)--(1.75,0.5)--(0.25,0.5)--cycle;
\draw[line width=1pt] (1.75,0) -- (2.25,0);
\draw[line width=1pt] (0.25,0) -- (-0.25,0);
\draw[line width=2pt, color=red] (0.65,0.5) -- (0.65,2);
\draw[line width=2pt, color=red] (1.35,0.5) -- (1.35,2);
\draw[line width=3pt, color=blue] (0.65,-1) -- (0.65,-0.5);
\draw[line width=3pt, color=blue] (1.35,-1) -- (1.35,-0.5);
\path (1,0) node [style=sergio]{\large $\mathsf{A}$};
\path (2.75,0) node [style=sergio]{$=$};
\filldraw[fill=white, draw=black] (0.65,1.25)circle (9pt);
\path (0.65,1.25) node [style=sergio]{$\sigma_x$};
\filldraw[fill=white, draw=black, rounded corners] (4.75,-0.5)--(6.25,-0.5)--(6.25,0.5)--(4.75,0.5)--cycle;
\draw[line width=1pt] (4.75,0) -- (3.25,0);
\filldraw[fill=white, draw=black] (4,0)circle (9pt);
\path (4,0) node [style=sergio]{$X$};
\path (5.5,0) node [style=sergio]{\large $\mathsf{A}$};
\draw[line width=1pt] (6.25,0) -- (7.75,0);
\filldraw[fill=white, draw=black] (7,0)circle (9pt);
\path (7,0) node [style=sergio]{$X$};
\draw[line width=2pt, color=red] (5.15,0.5) -- (5.15,1);
\draw[line width=2pt, color=red] (5.85,0.5) -- (5.85,1);
\draw[line width=3pt, color=blue] (5.15,-2) -- (5.15,-0.5);
\draw[line width=3pt, color=blue] (5.85,-2) -- (5.85,-0.5);
\filldraw[fill=white, draw=black] (5.15,-1.25)circle (9pt);
\path (5.15,-1.25) node [style=sergio]{$X$};
\filldraw[fill=white, draw=black] (5.85,-1.25)circle (9pt);
\path (5.85,-1.25) node [style=sergio]{$X$};
\end{tikzpicture},\label{G=Z2}
\end{align}
and
\begin{align}
\begin{tikzpicture}[scale=0.8]
\tikzstyle{sergio}=[rectangle,draw=none]
\filldraw[fill=white, draw=black, rounded corners] (0.25,-0.5)--(1.75,-0.5)--(1.75,0.5)--(0.25,0.5)--cycle;
\draw[line width=1pt] (1.75,0) -- (2.25,0);
\draw[line width=1pt] (0.25,0) -- (-0.25,0);
\draw[line width=2pt, color=red] (0.65,0.5) -- (0.65,2);
\draw[line width=2pt, color=red] (1.35,0.5) -- (1.35,2);
\draw[line width=3pt, color=blue] (0.65,-1) -- (0.65,-0.5);
\draw[line width=3pt, color=blue] (1.35,-1) -- (1.35,-0.5);
\path (1,0) node [style=sergio]{\large $\mathsf{A}$};
\path (2.75,0) node [style=sergio]{$=$};
\filldraw[fill=white, draw=black] (1.35,1.25)circle (10pt);
\path (1.35,1.25) node [style=sergio]{$\tau_x$};
\filldraw[fill=white, draw=black, rounded corners] (4.75,-0.5)--(6.25,-0.5)--(6.25,0.5)--(4.75,0.5)--cycle;
\draw[line width=1pt] (4.75,0) -- (3.25,0);
\filldraw[fill=white, draw=black] (4,0)circle (10pt);
\path (4,0) node [style=sergio]{$Z$};
\path (5.5,0) node [style=sergio]{\large $\mathsf{A}$};
\draw[line width=1pt] (6.25,0) -- (7.75,0);
\filldraw[fill=white, draw=black] (7,0)circle (10pt);
\path (7,0) node [style=sergio]{$Z$};
\draw[line width=2pt, color=red] (5.15,0.5) -- (5.15,1);
\draw[line width=2pt, color=red] (5.85,0.5) -- (5.85,1);
\draw[line width=3pt, color=blue] (5.15,-1) -- (5.15,-0.5);
\draw[line width=3pt, color=blue] (5.85,-1) -- (5.85,-0.5);
\end{tikzpicture}.\label{K=Z2}
\end{align}
Here $Z$ on virtual indices in Eq.~\eqref{K=Z2} just plays the role of counting the $K$-charge, i.e., $\hat{O}^{\prime}(k)$ in Eq.~\eqref{K-charge V}, while $X$ on virtual indices in Eq.~\eqref{G=Z2} carries a nontrivial $K$-charge $n_1=1$, labeling the nontrivial ASPT phase jointly protected by $K=\Z_2$ and $G=\Z_2$.
As the pure-state SPT counterpart of this ASPT phase, i.e., with the same symmetry structure $\widetilde{G} = K\times G$ but both $K$ and $G$ are strong, is just the well-known Haldane phase, we denote our constructed ASPT phase as the decohered Haldane phase. 

Moreover, we demonstrate that every ASPT phase with a pure-state SPT counterpart under the same symmetry group $\widetilde{G}$ can be constructed using the LPDO formalism outlined in Eq.~\eqref{general LPDO}, starting from an MPS representing the corresponding SPT phase.
This construction ensures that the resulting ASPT density matrix retains exactly the same topological invariants as those of the original MPS.

In the case of the strong symmetry action $U(k)$, the transformations on the MPS virtual indices adopt a diagonal form, merely attaching certain phases to different virtual subspaces.
These transformations are directly translated to the LPDO virtual indices.
As there is no need for operations on the Kraus indices in Eq.~\eqref{general LPDO} since not only they should not be permuted, but also no additional phases need to be added to the tensor, the representation adheres faithfully to the strong symmetry action described in Eq.~\eqref{1d exact}.
On the contrary, weak symmetry transformations acting on the virtual indices of the MPS, denoted as $V(g)$, may yield a nontrivial $K$-charge.
Consequently, $V(g)^{-1}$ and $V(g)$ should be mapped to both the virtual and Kraus indices in Eq.~\eqref{general LPDO}.
Additionally, an optional transformation $M$ may be applied to the Kraus indices to rectify any undesired phase induced by $V(g)^{-1}$ and $V(g)$, consistent with the weak symmetry transformation described in Eq.~\eqref{1d average}.

It is noteworthy that in both scenarios, the projective representation on the virtual indices of the LPDO remains identical to that of the original MPS.
This ensures the preservation of the same set of topological invariants, namely $n_0\in \H^2[K, U(1)]$ and $n_1\in \H^1\left[G,\H^1[K,U(1)]\right]$.
Therefore, our LPDO framework in Eq.~\eqref{general LPDO} can be effectively used to construct any ASPT phase that possesses a pure-state SPT counterpart.

\subsection{Intrinsic ASPT in $(1+1)D$ systems}
In the above case of the decohered Haldane phase, the total symmetry group is the direct product of strong and weak symmetries, i.e., $\widetilde{G} = K\times G$.
In other words, the group structure $\omega_2$ is trivial, which means that the right-hand side of Eq.~\eqref{twisted 2-cocycle} is equal to $1$ and the twisted 2-cocycle condition degrades to the conventional one shown in Eq.~\eqref{2-cocycle}, consistent with the fact that the decohered Haldane phase has a pure-state SPT correspondence. 

On the other hand, our LPDO construction is also suitable for describing the intrinsic ASPT state that does not have a pure-state SPT counterpart~\cite{Ma2023B}.
To construct an intrinsic ASPT phase, we need to consider a nontrivial group extension labeled by $\omega_2$.
We can still adopt $K=\Z_2$ and $G=\Z_2$, but with a nontrivial extension to the $\Z_4$ symmetry, i.e.,
\begin{align}
    1\rightarrow \Z_2\rightarrow \Z_4\rightarrow \Z_2 \rightarrow 1.
\end{align}
It is known that there is no nontrivial SPT phase for $(1+1)D$ pure states protected by $\Z_4$ since $\H^2[\Z_4, U(1)] = 1$, while we can construct an intrinsic ASPT phase jointly protected by strong $K=\Z_2$ and weak $G=\Z_2$ symmetries.

Since the decorated domain wall structure of this intrinsic ASPT phase mirrors that of the direct product symmetry class, the fixed-point density matrix can assume precisely the same form as the density matrix expressed in Eq.~\eqref{Z4}. 
However, it is important to note that the symmetry action must differ in this context.
Specifically, the strong symmetry action $\hat{O}(k) = \prod_i \tau_{i+1/2}^x$ remains unchanged, while the weak symmetry becomes $U(g) = \prod_i \sigma_{i}^xe^{i\frac{\pi}{4}(1-\tau_{i+1/2}^x)}$.
It can be easily verified that
\begin{align}
    U(g)U(g) = \hat{O}(k)U(1),
\end{align}
indicating the group structure is $\Z_4$ and the group extension is nontrivial.

Now the implementation of symmetry action $U(g)$ leaves each component in Eq.~\eqref{Z4} an additional phase, 
\begin{align}
    \prod_i \sigma_{i}^xe^{i\frac{\pi}{4}(1-\tau_{i+1/2}^x)}\ket{\psi_{\{\sigma_i\}}} = e^{i\frac{\pi}{2}N_{\textrm{DW}}}\ket{\psi_{\{\sigma_i\}}},
\end{align}
where $N_{\textrm{DW}}$ is the number of domain walls in configurations $\{\sigma_i\}$.
It means that $U(g)$ is not a symmetry action for the original cluster state in Eq.~\eqref{cluster} any longer.
However, the density matrix in Eq.~\eqref{Z4} should remain unchanged under the simultaneous implementation of $U(g)$ on both sides, i.e.,
\begin{align}
    U(g)\rho U(g)^{\dagger} = \rho,
\end{align}
where the additional phase will be canceled out. 
Therefore, $U(g)$ defined above is indeed a weak symmetry for the density matrix $\rho$.

Now we will see how the two symmetries are consistent within an LPDO representation and the projective representation of the two symmetries is constructed.
We can utilize the same tensor given in the last section while defining new symmetry operations.
For simplicity, we denote $R= e^{i\frac{\pi}{4}(1-\tau^x)}$.
The symmetry action of $K$ in Eq.~\eqref{K=Z2} remains unchanged, while the new symmetry transformation of $G$ on the local tensor can be derived as
\begin{align}
\begin{tikzpicture}[scale=0.8]
\tikzstyle{sergio}=[rectangle,draw=none]
\filldraw[fill=white, draw=black, rounded corners] (0.25,-0.5)--(1.75,-0.5)--(1.75,0.5)--(0.25,0.5)--cycle;
\draw[line width=1pt] (1.75,0) -- (2.25,0);
\draw[line width=1pt] (0.25,0) -- (-0.25,0);
\draw[line width=2pt, color=red] (0.65,0.5) -- (0.65,2);
\draw[line width=2pt, color=red] (1.35,0.5) -- (1.35,2);
\draw[line width=3pt, color=blue] (0.65,-1) -- (0.65,-0.5);
\draw[line width=3pt, color=blue] (1.35,-1) -- (1.35,-0.5);
\path (1,0) node [style=sergio]{\large $\mathsf{A}$};
\path (2.75,0) node [style=sergio]{$=$};
\filldraw[fill=white, draw=black] (0.65,1.25)circle (9pt);
\path (0.65,1.25) node [style=sergio]{$\sigma_x$};
\filldraw[fill=white, draw=black] (1.35,1.25)circle (9pt);
\path (1.35,1.25) node [style=sergio]{$R$};
\filldraw[fill=white, draw=black, rounded corners] (4.75,-0.5)--(6.25,-0.5)--(6.25,0.5)--(4.75,0.5)--cycle;
\draw[line width=1pt] (4.75,0) -- (3.25,0);
\filldraw[fill=white, draw=black] (4,0)circle (9pt);
\path (4,0) node [style=sergio]{$X$};
\path (5.5,0) node [style=sergio]{\large $\mathsf{A}$};
\draw[line width=1pt] (6.25,0) -- (7.75,0);
\filldraw[fill=white, draw=black] (7,0)circle (9pt);
\path (7,0) node [style=sergio]{$X$};
\draw[line width=2pt, color=red] (5.15,0.5) -- (5.15,1);
\draw[line width=2pt, color=red] (5.85,0.5) -- (5.85,1);
\draw[line width=3pt, color=blue] (5.15,-3) -- (5.15,-0.5);
\draw[line width=3pt, color=blue] (5.85,-3) -- (5.85,-0.5);
\filldraw[fill=white, draw=black] (5.15,-1.25)circle (9pt);
\path (5.15,-1.25) node [style=sergio]{$X$};
\filldraw[fill=white, draw=black] (5.85,-1.25)circle (9pt);
\path (5.85,-1.25) node [style=sergio]{$X$};
\filldraw[fill=white, draw=black, rounded corners] (4.75,-2.6)--(6.25,-2.6)--(6.25,-1.9)--(4.75,-1.9)--cycle;
\path (5.5,-2.25) node [style=sergio]{$M$};
\end{tikzpicture},\label{new G=Z2}
\end{align}
with an additional transformation on the Kraus index being $M=e^{i\frac{\pi}{4}(1-ZZ)}$ (viewed from down to up). 

Similar to the previous case, $X$ on virtual indices in Eq.~\eqref{new G=Z2} carries a nontrivial $K$-charge $n_1=1$, labeling the nontrivial ASPT phase jointly protected by $K=\Z_2$ and $G=\Z_2$, while the additional $M$ does not affect the topological invariants as the related phase structure should be canceled out under simultaneous implementation of weak symmetry on both sides.
We highlight that in this case, Eq.~\eqref{twisted 2-cocycle} for pure-state SPT becomes
\begin{align}
    1 = \frac{\nu_2(g,g)\nu_2(1,g)}{\nu_2(g,1)\nu_2(g,g)}=\prod\limits_{j=1}^me^{i{2\pi}\omega_2^j(g,g)n_1^j(g)/n_j} = -1,
\end{align}
which obviously has no solution.
Therefore, we have constructed an intrinsic ASPT phase that does not correspond to any pure-state SPT phase, which has the same fixed point as the decohered Haldane phase but with different symmetry actions and group structures.

\subsection{Topological invariants under symmetric quantum channels}
Here, we demonstrate that the above topological numbers $\mu_2$ and $n_1$ are invariant under a local quantum channel $\mathcal{E}(\rho) = \sum_{\kappa}E_{\kappa}\rho E_{\kappa}^{\dagger}$ that preserves both the strong and weak symmetry of the mixed state.
An arbitrary local channel can be constructed by implementing a finite-depth local unitary on a larger system (the system in consideration combined with ancillary degrees of freedom in a product state) and tracing out the ancilla, i.e.,
\begin{align}
    \mathcal{E}\left(\rho_p\right)=\mathrm{Tr}_a\left[U\left(\rho_p\otimes\ket{0}\langle0|_a\right)U^\dag\right].
\end{align}
The unitary evolution in the system-ancilla composite in the above equation can be depicted as
\begin{align}
\begin{tikzpicture}[scale=0.8]
\tikzstyle{sergio}=[rectangle,draw=none]
\draw[line width=1pt] (-4.5,-1) -- (-6.5,-1);
\filldraw[fill=white, draw=black, rounded corners] (-7.5,-1.5)--(-6.5,-1.5)--(-6.5,-0.5)--(-7.5,-0.5)--cycle;
\filldraw[fill=white, draw=black, rounded corners] (-6,-1.5)--(-5,-1.5)--(-5,-0.5)--(-6,-0.5)--cycle;
\filldraw[fill=white, draw=black, rounded corners] (-4.5,-1.5)--(-3.5,-1.5)--(-3.5,-0.5)--(-4.5,-0.5)--cycle;
\draw[line width=2pt, color=red] (-7.25,-0.5) -- (-7.25,0);
\draw[line width=2pt, color=red] (-7.25,-2) -- (-7.25,-1.5);
\draw[line width=2pt, color=red] (-4.25,0) -- (-4.25,-0.5);
\draw[line width=2pt, color=red] (-4.25,-1.5) -- (-4.25,-2);
\draw[line width=2pt, color=red] (-5.75,-1.5) -- (-5.75,-2);
\draw[line width=2pt, color=red] (-5.75,-0.5) -- (-5.75,0);
\path (-7,-1) node [style=sergio]{\large $E_1$};
\path (-5.5,-1) node [style=sergio]{\large $E_2$};
\path (-4,-1) node [style=sergio]{\large $E_3$};
\path (-3,-1) node [style=sergio]{$=$};
\draw[line width=1pt] (0.5,-1) -- (-1.5,-1);
\filldraw[fill=white, draw=black, rounded corners] (-2.5,-1.5)--(-1.5,-1.5)--(-1.5,-0.5)--(-2.5,-0.5)--cycle;
\filldraw[fill=white, draw=black, rounded corners] (-1,-1.5)--(0,-1.5)--(0,-0.5)--(-1,-0.5)--cycle;
\filldraw[fill=white, draw=black, rounded corners] (0.5,-1.5)--(1.5,-1.5)--(1.5,-0.5)--(0.5,-0.5)--cycle;
\draw[line width=2pt, color=red] (-2,-0.5) -- (-2,0);
\draw[line width=2pt, color=red] (-2,-2) -- (-2,-1.5);
\draw[line width=2pt, color=red] (1,-1.5) -- (1,-2);
\draw[line width=2pt, color=red] (-0.5,-1.5) -- (-0.5,-2);
\draw[line width=2pt, color=red] (1,0) -- (1,-0.5);
\draw[line width=2pt, color=red] (-0.5,-0.5) -- (-0.5,0);
\path (-2,-1) node [style=sergio]{\large $E_1$};
\path (-0.5,-1) node [style=sergio]{\large $E_2$};
\path (1,-1) node [style=sergio]{\large $E_3$};
\draw[line width=3pt, color=blue] (-6.75,-0.5) -- (-6.75,0) -- (-6.25,0) -- (-6.25,-2);
\draw[line width=3pt, color=blue] (-5.25,-0.5) -- (-5.25,0) -- (-4.75,0) -- (-4.75,-2);
\draw[line width=3pt, color=blue] (-3.75,-0.5) -- (-3.75,0) -- (-3.25,0) -- (-3.25,-2);
\draw[line width=3pt, color=blue] (-1.25,-2) -- (-1.25,-1.25) -- (-1.5,-1.25);
\draw[line width=3pt, color=blue] (0.25,-2) -- (0.25,-1.25) -- (0,-1.25);
\draw[line width=3pt, color=blue] (1.75,-2) -- (1.75,-1.25) -- (1.5,-1.25);
\draw[line width=3pt, color=blue] (-6.75,-2) -- (-6.75,-1.5);
\draw[line width=3pt, color=blue] (-5.25,-2) -- (-5.25,-1.5);
\draw[line width=3pt, color=blue] (-3.75,-2) -- (-3.75,-1.5);
\filldraw[fill=white, draw=black] (-6.75,-2.25)circle (10pt);
\path (-6.75,-2.25) node [style=sergio]{\small $\ket{0}$};
\filldraw[fill=white, draw=black] (-5.25,-2.25)circle (10pt);
\path (-5.25,-2.25) node [style=sergio]{\small $\ket{0}$};
\filldraw[fill=white, draw=black] (-3.75,-2.25)circle (10pt);
\path (-3.75,-2.25) node [style=sergio]{\small $\ket{0}$};
\end{tikzpicture},
\end{align}
where the blue legs for the ancillary degrees of freedom are bent down, representing the fact that they are traced out eventually with the other side.
These Kraus operators for the local quantum channel are generalizations of local unitary evolutions in the MPS case shown in Eq.~\eqref{unitarydecomp}.

The symmetry conditions for the Kraus operators $E_{\kappa}$ in the quantum channel are the following.
For strong symmetry, each Kraus operator commutes with symmetry, namely
\begin{align}
\begin{tikzpicture}[scale=0.8]
\tikzstyle{sergio}=[rectangle,draw=none]
\draw[line width=1pt] (-4.5,-1) -- (-6.5,-1);
\filldraw[fill=white, draw=black, rounded corners] (-7.5,-1.5)--(-6.5,-1.5)--(-6.5,-0.5)--(-7.5,-0.5)--cycle;
\filldraw[fill=white, draw=black, rounded corners] (-6,-1.5)--(-5,-1.5)--(-5,-0.5)--(-6,-0.5)--cycle;
\filldraw[fill=white, draw=black, rounded corners] (-4.5,-1.5)--(-3.5,-1.5)--(-3.5,-0.5)--(-4.5,-0.5)--cycle;
\draw[line width=2pt, color=red] (-7,-0.5) -- (-7,1);
\draw[line width=2pt, color=red] (-7,-3) -- (-7,-1.5);
\draw[line width=2pt, color=red] (-4,-1.5) -- (-4,-3);
\draw[line width=2pt, color=red] (-5.5,-1.5) -- (-5.5,-3);
\draw[line width=2pt, color=red] (-4,1) -- (-4,-0.5);
\draw[line width=2pt, color=red] (-5.5,-0.5) -- (-5.5,1);
\path (-7,-1) node [style=sergio]{\large $E_1$};
\path (-5.5,-1) node [style=sergio]{\large $E_2$};
\path (-4,-1) node [style=sergio]{\large $E_3$};
\path (-3,-1) node [style=sergio]{$=$};
\filldraw[fill=white, draw=black] (-7,0.25)circle (10pt);
\path (-7,0.25) node [style=sergio]{\small $\hat{O}_k$};
\filldraw[fill=white, draw=black] (-5.5,0.25)circle (10pt);
\path (-5.5,0.25) node [style=sergio]{\small $\hat{O}_k$};
\filldraw[fill=white, draw=black] (-4,0.25)circle (10pt);
\path (-4,0.25) node [style=sergio]{\small $\hat{O}_k$};
\filldraw[fill=white, draw=black] (-7,-2.25)circle (10pt);
\path (-7,-2.25) node [style=sergio]{\small $\hat{O}_k^*$};
\filldraw[fill=white, draw=black] (-5.5,-2.25)circle (10pt);
\path (-5.5,-2.25) node [style=sergio]{\small $\hat{O}_k^*$};
\filldraw[fill=white, draw=black] (-4,-2.25)circle (10pt);
\path (-4,-2.25) node [style=sergio]{\small $\hat{O}_k^*$};
\draw[line width=1pt] (0.5,-1) -- (-1.5,-1);
\filldraw[fill=white, draw=black, rounded corners] (-2.5,-1.5)--(-1.5,-1.5)--(-1.5,-0.5)--(-2.5,-0.5)--cycle;
\filldraw[fill=white, draw=black, rounded corners] (-1,-1.5)--(0,-1.5)--(0,-0.5)--(-1,-0.5)--cycle;
\filldraw[fill=white, draw=black, rounded corners] (0.5,-1.5)--(1.5,-1.5)--(1.5,-0.5)--(0.5,-0.5)--cycle;
\draw[line width=2pt, color=red] (-2,-0.5) -- (-2,0);
\draw[line width=2pt, color=red] (-2,-2) -- (-2,-1.5);
\draw[line width=2pt, color=red] (1,-1.5) -- (1,-2);
\draw[line width=2pt, color=red] (-0.5,-1.5) -- (-0.5,-2);
\draw[line width=2pt, color=red] (1,0) -- (1,-0.5);
\draw[line width=2pt, color=red] (-0.5,-0.5) -- (-0.5,0);
\path (-2,-1) node [style=sergio]{\large $E_1$};
\path (-0.5,-1) node [style=sergio]{\large $E_2$};
\path (1,-1) node [style=sergio]{\large $E_3$};
\draw[line width=3pt, color=blue] (-6.25,-2) -- (-6.25,-1.25) -- (-6.5,-1.25);
\draw[line width=3pt, color=blue] (-4.75,-2) -- (-4.75,-1.25) -- (-5,-1.25);
\draw[line width=3pt, color=blue] (-3.25,-2) -- (-3.25,-1.25) -- (-3.5,-1.25);
\draw[line width=3pt, color=blue] (-1.25,-2) -- (-1.25,-1.25) -- (-1.5,-1.25);
\draw[line width=3pt, color=blue] (0.25,-2) -- (0.25,-1.25) -- (0,-1.25);
\draw[line width=3pt, color=blue] (1.75,-2) -- (1.75,-1.25) -- (1.5,-1.25);
\end{tikzpicture}.
\end{align}
The weak symmetry corresponds to the symmetry action $U(g)_p\otimes M(g)_a$ on the joint of system and ancilla for the unitary operator on the larger Hilbert space.
It can rotate the set of Kraus operators while keeping the total channel invariant, as demonstrated in the following equation
\begin{align}
\begin{tikzpicture}[scale=0.8]
\tikzstyle{sergio}=[rectangle,draw=none]
\draw[line width=1pt] (-4.5,-1) -- (-6.5,-1);
\filldraw[fill=white, draw=black, rounded corners] (-7.5,-1.5)--(-6.5,-1.5)--(-6.5,-0.5)--(-7.5,-0.5)--cycle;
\filldraw[fill=white, draw=black, rounded corners] (-6,-1.5)--(-5,-1.5)--(-5,-0.5)--(-6,-0.5)--cycle;
\filldraw[fill=white, draw=black, rounded corners] (-4.5,-1.5)--(-3.5,-1.5)--(-3.5,-0.5)--(-4.5,-0.5)--cycle;
\draw[line width=2pt, color=red] (-7,-0.5) -- (-7,1);
\draw[line width=2pt, color=red] (-7,-3) -- (-7,-1.5);
\draw[line width=2pt, color=red] (-4,-1.5) -- (-4,-3);
\draw[line width=2pt, color=red] (-5.5,-1.5) -- (-5.5,-3);
\draw[line width=2pt, color=red] (-4,1) -- (-4,-0.5);
\draw[line width=2pt, color=red] (-5.5,-0.5) -- (-5.5,1);
\path (-7,-1) node [style=sergio]{\large $E_1$};
\path (-5.5,-1) node [style=sergio]{\large $E_2$};
\path (-4,-1) node [style=sergio]{\large $E_3$};
\path (-3,-1) node [style=sergio]{$=$};
\filldraw[fill=white, draw=black] (-7,0.25)circle (10pt);
\path (-7,0.25) node [style=sergio]{\small $U_g$};
\filldraw[fill=white, draw=black] (-5.5,0.25)circle (10pt);
\path (-5.5,0.25) node [style=sergio]{\small $U_g$};
\filldraw[fill=white, draw=black] (-4,0.25)circle (10pt);
\path (-4,0.25) node [style=sergio]{\small $U_g$};
\filldraw[fill=white, draw=black] (-7,-2.25)circle (10pt);
\path (-7,-2.25) node [style=sergio]{\small $U_g^*$};
\filldraw[fill=white, draw=black] (-5.5,-2.25)circle (10pt);
\path (-5.5,-2.25) node [style=sergio]{\small $U_g^*$};
\filldraw[fill=white, draw=black] (-4,-2.25)circle (10pt);
\path (-4,-2.25) node [style=sergio]{\small $U_g^*$};
\draw[line width=1pt] (0.5,-1) -- (-1.5,-1);
\filldraw[fill=white, draw=black, rounded corners] (-2.5,-1.5)--(-1.5,-1.5)--(-1.5,-0.5)--(-2.5,-0.5)--cycle;
\filldraw[fill=white, draw=black, rounded corners] (-1,-1.5)--(0,-1.5)--(0,-0.5)--(-1,-0.5)--cycle;
\filldraw[fill=white, draw=black, rounded corners] (0.5,-1.5)--(1.5,-1.5)--(1.5,-0.5)--(0.5,-0.5)--cycle;
\draw[line width=2pt, color=red] (-2,-0.5) -- (-2,0);
\draw[line width=2pt, color=red] (-2,-2) -- (-2,-1.5);
\draw[line width=2pt, color=red] (1,-1.5) -- (1,-2);
\draw[line width=2pt, color=red] (-0.5,-1.5) -- (-0.5,-2);
\draw[line width=2pt, color=red] (1,0) -- (1,-0.5);
\draw[line width=2pt, color=red] (-0.5,-0.5) -- (-0.5,0);
\path (-2,-1) node [style=sergio]{\large $E_1$};
\path (-0.5,-1) node [style=sergio]{\large $E_2$};
\path (1,-1) node [style=sergio]{\large $E_3$};
\draw[line width=3pt, color=blue] (-6.25,-2) -- (-6.25,-1.25) -- (-6.5,-1.25);
\draw[line width=3pt, color=blue] (-4.75,-2) -- (-4.75,-1.25) -- (-5,-1.25);
\draw[line width=3pt, color=blue] (-3.25,-2) -- (-3.25,-1.25) -- (-3.5,-1.25);
\draw[line width=3pt, color=blue] (-1.25,-3) -- (-1.25,-1.25) -- (-1.5,-1.25);
\draw[line width=3pt, color=blue] (0.25,-3) -- (0.25,-1.25) -- (0,-1.25);
\draw[line width=3pt, color=blue] (1.75,-3) -- (1.75,-1.25) -- (1.5,-1.25);
\filldraw[fill=white, draw=black] (-1.25,-2.25)circle (10pt);
\path (-1.25,-2.25) node [style=sergio]{\small $M_g^1$};
\filldraw[fill=white, draw=black] (0.25,-2.25)circle (10pt);
\path (0.25,-2.25) node [style=sergio]{\small $M_g^2$};
\filldraw[fill=white, draw=black] (1.75,-2.25)circle (10pt);
\path (1.75,-2.25) node [style=sergio]{\small $M_g^3$};
\end{tikzpicture}.
\end{align}
Here, similar to the discussion on the density matrix, we assume that the symmetry action on the Kraus index is also factorized onto each site because the ancillary degrees of freedom have the same locality structure as the system itself. 
As a result, the strong and weak symmetry actions on the local tensors of the Kraus operator are

\begin{align}
\begin{tikzpicture}[scale=0.8]
\tikzstyle{sergio}=[rectangle,draw=none]
\draw[line width=1pt] (-3.5,-1) -- (-5.5,-1);
\filldraw[fill=white, draw=black, rounded corners] (-5,-1.5)--(-4,-1.5)--(-4,-0.5)--(-5,-0.5)--cycle;
\draw[line width=2pt, color=red] (-4.5,-1.5) -- (-4.5,-3);
\draw[line width=2pt, color=red] (-4.5,-0.5) -- (-4.5,1);
\path (-4.5,-1) node [style=sergio]{\large $E_2$};
\path (-3,-1) node [style=sergio]{$=$};
\filldraw[fill=white, draw=black] (-4.5,0.25)circle (10pt);
\path (-4.5,0.25) node [style=sergio]{\small $\hat{O}_k$};
\filldraw[fill=white, draw=black] (-4.5,-2.25)circle (10pt);
\path (-4.5,-2.25) node [style=sergio]{\small $\hat{O}_k^*$};
\draw[line width=1pt] (1.5,-1) -- (-2.5,-1);
\filldraw[fill=white, draw=black, rounded corners] (-1,-1.5)--(0,-1.5)--(0,-0.5)--(-1,-0.5)--cycle;
\draw[line width=2pt, color=red] (-0.5,-1.5) -- (-0.5,-2);
\draw[line width=2pt, color=red] (-0.5,-0.5) -- (-0.5,0);
\path (-0.5,-1) node [style=sergio]{\large $E_2$};
\draw[line width=3pt, color=blue] (-3.75,-2) -- (-3.75,-1.25) -- (-4,-1.25);
\draw[line width=3pt, color=blue] (0.25,-2) -- (0.25,-1.25) -- (0,-1.25);
\filldraw[fill=white, draw=black] (-1.75,-1)circle (10pt);
\path (-1.75,-1) node [style=sergio]{\small $\hat{O}_{1k}^{\prime-1}$};
\filldraw[fill=white, draw=black] (0.75,-1)circle (10pt);
\path (0.75,-1) node [style=sergio]{\small $\hat{O}_{2k}^{\prime}$};
\end{tikzpicture}
\end{align}
and 
\begin{align}
\begin{tikzpicture}[scale=0.8]
\tikzstyle{sergio}=[rectangle,draw=none]
\draw[line width=1pt] (-3.5,3.5) -- (-5.5,3.5);
\filldraw[fill=white, draw=black, rounded corners] (-5,3)--(-4,3)--(-4,4)--(-5,4)--cycle;
\draw[line width=2pt, color=red] (-4.5,3) -- (-4.5,1.5);
\draw[line width=2pt, color=red] (-4.5,4) -- (-4.5,5.5);
\path (-4.5,3.5) node [style=sergio]{\large $E_2$};
\path (-3,3.5) node [style=sergio]{$=$};
\filldraw[fill=white, draw=black] (-4.5,4.75)circle (10pt);
\path (-4.5,4.75) node [style=sergio]{\small $U_g$};
\filldraw[fill=white, draw=black] (-4.5,2.25)circle (10pt);
\path (-4.5,2.25) node [style=sergio]{\small $U_g^*$};
\draw[line width=1pt] (1.5,3.5) -- (-2.5,3.5);
\filldraw[fill=white, draw=black, rounded corners] (-1,3)--(0,3)--(0,4)--(-1,4)--cycle;
\draw[line width=2pt, color=red] (-0.5,3) -- (-0.5,2.5);
\draw[line width=2pt, color=red] (-0.5,4) -- (-0.5,4.5);
\path (-0.5,3.5) node [style=sergio]{\large $E_2$};
\draw[line width=3pt, color=blue] (-3.75,2.5) -- (-3.75,3.25) -- (-4,3.25);
\draw[line width=3pt, color=blue] (0.25,1.5) -- (0.25,3.25) -- (0,3.25);
\filldraw[fill=white, draw=black] (0.25,2.25)circle (10pt);
\path (0.25,2.25) node [style=sergio]{\small $M_g^2$};
\filldraw[fill=white, draw=black] (-1.75,3.5)circle (10pt);
\path (-1.75,3.5) node [style=sergio]{\small $V_{1g}^{-1}$};
\filldraw[fill=white, draw=black] (0.75,3.5)circle (10pt);
\path (0.75,3.5) node [style=sergio]{\small $V_{2g}$};
\end{tikzpicture}.
\end{align}

We now examine the phase structure between different $\hat{O}^{\prime}(k_1)$ and $\hat{O}^{\prime}(k_2)$ as defined in Eq.~\eqref{Oprime}, as well as the mixed anomaly between $\hat{O}^{\prime}(k)$ and $V(g)$ as defined in Eq.~\eqref{K-charge V}.
These phase structures define the cocycle data carried by the virtual legs of the tensor $E_i$.
We will argue that, to consistently implement the symmetries on the tensor $E_i$'s, these phase structures have to be trivial, which means that the effect of these local quantum channels cannot change the cocycle data in the original LPDO. 

For strong symmetry, much like in the pure-state case, the phase structures from the symmetry action (as defined in Eq. \eqref{Oprime}) on the two virtual indices of a bulk tensor must be canceled to maintain consistency.
Extending this reasoning to the boundary tensor, where only one virtual leg is present, one observes -- analogously to the MPS case -- that this phase factor must be virtual.
Consequently, for all virtual legs of the local quantum channel, the phase factors associated with strong symmetry are trivial.
In other words, the virtual legs exhibit a linear representation under strong symmetry.

As for the mixed anomaly between strong and weak symmetries, we will look at the phase structure that is defined in Eq. ~\eqref{K-charge V}.
Since the symmetry action on the Kraus index for a strong symmetry is identity $I$, which automatically commutes with the weak symmetry action $M_g$, the Kraus index cannot contribute any additional phase factor.
Therefore, the phase factors between strong and weak symmetry actions on the virtual indices should still be opposite for the two virtual legs from the same bulk tensor.
Moreover, the boundary condition again requires that all of these phases be trivial.
As a result, there is no mixed anomaly of strong and weak symmetry in the virtual legs of the local quantum channels.
Therefore, the implementation of a symmetric local channel preserves the two topological invariants $\mu_2$ and $n_1$ of the density matrix, suggesting that the resulting state belongs to the same ASPT phase as the original.

\subsection{Physical realization of ASPT phases in $(1+1)D$ systems}
Given the LPDO representation for a mixed state belonging to a specific ASPT phase, realizing this LPDO in decohered or disordered systems poses an essential challenge.
From an experimental perspective, we focus on finite systems with $N$ sites under the open boundary condition (OBC).

\subsubsection{Decohered systems}
To prepare an LPDO in a decohered system, we need to design quantum dynamics that drive an initial product state to the target density matrix.
Two techniques can be used to accomplish this. The first method to generate the desired LPDO involves directly preparing the purified MPS shown in Eq.~\ref{1d MPS} and tracing out the environmental degree of freedom, i.e., measuring them without needing to record the measurement outcome.
An arbitrary short-range correlated MPS of $N$ sites can be prepared by a quantum circuit of depth $O\left[\log{(N/\varepsilon)}\right]$~\cite{Malz2024} with an error bounded by $\varepsilon$.
The basic idea is to group adjacent $q$ sites together and approximate the resulting MPS by the fixed-point tensor under the renormalization group (RG).
The original MPS is then prepared by implementing a quantum circuit that realizes the inverse RG flow (inverse of disentanglers) on the fixed-point MPS.
Specifically, if the purified MPS of the provided LPDO is already a fixed point under RG flow, only a constant circuit depth is required.
Since the LPDO considered here satisfies both strong and weak injectivity conditions, the purified MPS is injective, making this method applicable.

\begin{figure}
    \centering
    \includegraphics[width=0.9\linewidth]{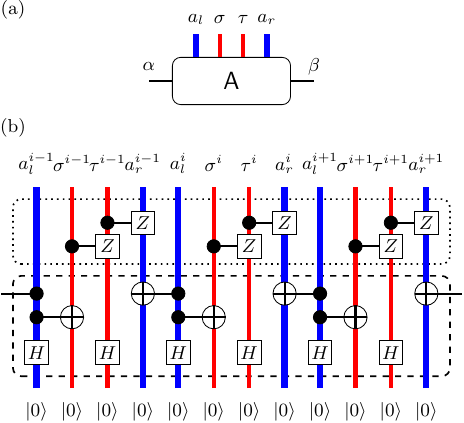}
    \caption{Quantum circuit to prepare the purified MPS.}
    \label{Fig: QC}
\end{figure}

As an illustrative example, we explicitly construct the quantum circuit used to prepare the purified MPS for the example of $(1+1)D$ ASPT phases discussed in the previous section, whose LPDO tensor element is presented in Eq.~\eqref{1d tensor element}.
As shown in Fig.~\ref{Fig: QC}(a), each site $i$ has four physical indices $\sigma^i$, $\tau^i$, $a_{l}^{i}$, and $a_{r}^{i}$, as well as two virtual indices $\alpha^i$ and $\beta^i$.
From the LPDO construction formalism in Eq.~\eqref{general LPDO}, we note that the relation $a_{r}^{i-1} = \beta^{i-1} = \alpha^{i} = a_{l}^{i}$ always holds in the wavefunction superposition.
Moreover, the original cluster MPS construction in Eq.~\eqref{MPS_cluster} ensures that $\alpha^i = \sigma^i$, which is inherited by the LPDO in Eq.~\eqref{1d tensor element}.
Therefore, for each component in the final wavefunction, we have $a_{r}^{i-1} = a_{l}^{i} = \sigma^{i}$, which forms a local ferromagnetic pattern.
Finally, we construct the quantum circuit shown in Fig.~\ref{Fig: QC}(b) to prepare the purified MPS.
The first part of the circuit (dashed frame) generates local GHZ states for $a_{r}^{i-1}$, $a_{l}^{i}$, and $\sigma^{i}$~\cite{Schoen2005, Chen2022}, while the second part (dotted frame) decorates the domain wall of $\sigma$ spins with $\tau$ excitations~\cite{Chen2014, Cao2023}.

The above method uses closed-system dynamics to prepare the purified MPS for a given LPDO.
Alternatively, we explore the possibility of using open-system dynamics that directly target the provided mixed state.
A recent work~\cite{Liu2025} proposed an analytical construction of a parent Lindbladian for fixed-point matrix product density operator (MPDO) under mixed-state RG~\cite{Cirac2017}, such that the MPDO can be realized as the steady state of this local Lindbladian.
Therefore, we can use a three-step process to prepare the provided LPDO.
\begin{enumerate}
    \item Contract the Kraus indices of LPDO and perform mixed-state RG to obtain the fixed-point MPDO $\rho \stackrel{\rm RG}{\longrightarrow} \rho_{\rm fixed}$~\cite{Cirac2017}.
    \item Construct the local Lindbladian $\mathcal{L}$ such that $\rho_{\rm fixed}$ is its steady state, i.e., $\rho_{\rm fixed} = \lim_{t\rightarrow\infty} e^{\mathcal{L}t}(\rho_0)$~\cite{Liu2025}.
    \item Construct another local Lindbladian $\mathcal{L}^{\prime}$ to drive the fixed-point MPDO back to the original LPDO with duration $t^{\prime}$, i.e., $\rho =  e^{\mathcal{L}^{\prime}t^{\prime}}(\rho_{\rm fixed})$.
\end{enumerate}
An effective model for simulating the Lindbladian evolution on a digital quantum computer is the multipartite collision model~\cite{Cattaneo2021, Cattaneo2023}, which introduces ancilla qubits to simulate the coupling between the system and environment.
However, while it is believed that the fixed-point MPDO characterizes the mixed-state phases of matter~\cite{RuizdeAlarcon2024}, designing $\mathcal{L}^{\prime}$ in step 3 remains a challenging task in general.
Some recent works have made some progress in this direction.
For example, if the correlation length of conditional mutual information remains finite during the RG process (which could be viewed as a local quantum channel) in step 1, there exists a quasi-local Lindbladian that reverses this RG flow~\cite{Sang2025}.
However, a construction for more general cases is still lacking.
Therefore, in the current stage, the pure-state method remains more straightforward and easy to implement than the mixed-state approach.

\subsubsection{Disordered systems}
In this section, we discuss the construction of disordered systems for a given density matrix represented by an LPDO.
Let us first formally define the problem at hand.
Our goal is to find a disordered system, represented by an ensemble of Hamiltonians $\{H_i\}$ with corresponding probability distribution $\{p_i\}$, such that the given density matrix can be decomposed as $\rho = \sum_i p_i\ket{\psi_i}\hspace{-0.5mm}\bra{\psi_i}$, where $\ket{\psi_i}$ is the gapped ground state of $H_i$ for each $i$.

In the previous sections, we treat Kraus indices as ancillae to describe the interaction between the system and environment.
From a classical mixture perspective, an LPDO also represents a statistical ensemble with all the Kraus indices constituting the label of states, which can be naturally realized in disordered systems.
As the ensemble dimension $d_{\kappa}^N$ scales exponentially with the system size $N$, explicitly writing down all Hamiltonians and their corresponding probabilities becomes impractical. 
Fortunately, this is not an issue in real experiments, where one only needs to efficiently sample from the probability distribution $\{p_i\}$ and construct the corresponding Hamiltonian $H_i$ for each sample.
The probability for a fixed Kraus configuration $\bm{a} = \{a_1, \cdots. a_N\}$ is given by
\begin{align}
    p_{\bm{a}} = \text{Tr}_{\bm{p}}[\ket{\psi_{\bm{p}\otimes \bm{a}}}\bra{\psi_{\bm{p}\otimes \bm{a}}}],
\end{align}
where $\bm{p} = \{p_1, \cdots, p_N\}$ denotes the physical configuration to be traced out.
This probability can also be expressed as the product of a series of conditional probabilities
\begin{align}
    p(a_1, \cdots, a_N) = p(a_1)p(a_2|a_1)\cdots p(a_N|a_1, \cdots a_{N-1}),
\end{align}
which leads to a site-by-site sampling scheme from this probability distribution for a given LPDO.
The procedure is illustrated as follows
\begin{enumerate}[1.]
\item Transform the provided LPDO into its right canonical form~\cite{Guo2024D}, satisfying that
\begin{align}
\begin{tikzpicture}[scale=0.8]
\tikzstyle{sergio}=[rectangle,draw=none]
\filldraw[fill=white, draw=black, rounded corners] (-0.25,-0.5)--(1.25,-0.5)--(1.25,0.5)--(-0.25,0.5)--cycle;
\filldraw[fill=white, draw=black, rounded corners] (-0.25,-2.5)--(1.25,-2.5)--(1.25,-1.5)--(-0.25,-1.5)--cycle;
\path (0.5,0) node [style=sergio]{\large $\mathsf{A}$};
\path (0.5,-2) node [style=sergio]{\large $\mathsf{A}^*$};
\draw[line width=2pt, color=red] (0.5,0.5) -- (0.5,1) -- (1,1) -- (1,-3) -- (0.5,-3) -- (0.5,-2.5);
\draw[line width=3pt, color=blue] (0.5,-1.5) -- (0.5,-0.5);
\path (0.25,1) node [style=sergio]{$p$};
\path (0.25,-1) node [style=sergio]{$a$};
\path (0.25,-3) node [style=sergio]{$p$};
\draw[line width=1pt] (-0.25,-2) -- (-1,-2);
\draw[line width=1pt] (1.25,0) -- (2,0) -- (2,-2) -- (1.25,-2);
\draw[line width=1pt] (-0.25,0) -- (-1,0);
\path (2.75,-1) node [style=sergio]{\large$=$};
\draw[line width=1pt] (3.5,0) -- (4.5,0) -- (4.5,-2) -- (3.5,-2);
\end{tikzpicture}.
\end{align}
\item Sample the Kraus index of the first site according to the probability distribution
\begin{align}
\begin{tikzpicture}[scale=0.8]
\tikzstyle{sergio}=[rectangle,draw=none]
\filldraw[fill=white, draw=black, rounded corners] (-0.25,-0.5)--(1.25,-0.5)--(1.25,0.5)--(-0.25,0.5)--cycle;
\filldraw[fill=white, draw=black, rounded corners] (-0.25,-2.5)--(1.25,-2.5)--(1.25,-1.5)--(-0.25,-1.5)--cycle;
\path (0.5,0) node [style=sergio]{\large $\mathsf{A_1}$};
\path (0.5,-2) node [style=sergio]{\large $\mathsf{A^*_1}$};
\draw[line width=2pt, color=red] (0.5,0.5) -- (0.5,1) -- (1,1) -- (1,-3) -- (0.5,-3) -- (0.5,-2.5);
\draw[line width=3pt, color=blue] (0.5,-1.5) -- (0.5,-1.1875);
\draw[line width=3pt, color=blue] (0.5,-0.8125) -- (0.5,-0.5);
\path (0.25,1) node [style=sergio]{$p$};
\path (0.25,-1) node [style=sergio]{$a_1$};
\path (0.25,-3) node [style=sergio]{$p$};
\path (-1.375,-1) node [style=sergio]{$p(a_1)\,\,=$};
\draw[line width=1pt] (1.25,0) -- (2,0) -- (2,-2) -- (1.25,-2);
\end{tikzpicture}.
\end{align}
\item For $k$ from $1$ to $N$, iteratively implement this step.
Suppose the sampling results for the first $k$ sites are $\{a_1, \cdots a_k\}$, the Kraus index of the $(k+1)$th site is then sampled from the conditional probability distribution
\begin{align}
    p(a_{k+1}|a_1, \cdots, a_k) = \frac{p(a_1, \cdots, a_{k+1})}{p(a_1, \cdots, a_k)},
\end{align}
where
\begin{align}
\begin{tikzpicture}[scale=0.8]
\tikzstyle{sergio}=[rectangle,draw=none]
\draw[line width=1pt] (1,0) -- (4,0) -- (4,-2) -- (1,-2);
\draw[line width=1pt] (-0.75,0) -- (0,0);
\draw[line width=1pt] (-0.75,-2) -- (0,-2);
\filldraw[fill=white, draw=black, rounded corners] (-2.25,-0.5)--(-0.75,-0.5)--(-0.75,0.5)--(-2.25,0.5)--cycle;
\filldraw[fill=white, draw=black, rounded corners] (-2.25,-2.5)--(-0.75,-2.5)--(-0.75,-1.5)--(-2.25,-1.5)--cycle;
\path (-1.5,0) node [style=sergio]{\large $\mathsf{A_{1}}$};
\path (-1.5,-2) node [style=sergio]{\large $\mathsf{A^{*}_{1}}$};
\draw[line width=2pt, color=red] (-1.5,0.5) -- (-1.5,1) -- (-1,1) -- (-1,-3) -- (-1.5,-3) -- (-1.5,-2.5);
\draw[line width=3pt, color=blue] (-1.5,-1.5) -- (-1.5,-1.1875);
\draw[line width=3pt, color=blue] (-1.5,-0.8125) -- (-1.5,-0.5);
\path (-1.75,1) node [style=sergio]{$p$};
\path (-1.75,-1) node [style=sergio]{$a_{1}$};
\path (-1.75,-3) node [style=sergio]{$p$};
\path (0.56,-1) node [style=sergio]{\large$\cdots$};
\filldraw[fill=white, draw=black, rounded corners] (1.75,-0.5)--(3.25,-0.5)--(3.25,0.5)--(1.75,0.5)--cycle;
\filldraw[fill=white, draw=black, rounded corners] (1.75,-2.5)--(3.25,-2.5)--(3.25,-1.5)--(1.75,-1.5)--cycle;
\path (2.5,0) node [style=sergio]{\large $\mathsf{A_{k+1}}$};
\path (2.5,-2) node [style=sergio]{\large $\mathsf{A^{*}_{k+1}}$};
\draw[line width=2pt, color=red] (2.5,0.5) -- (2.5,1) -- (3,1) -- (3,-3) -- (2.5,-3) -- (2.5,-2.5);
\draw[line width=3pt, color=blue] (2.5,-1.5) -- (2.5,-1.1875);
\draw[line width=3pt, color=blue] (2.5,-0.8125) -- (2.5,-0.5);
\path (2.25,1) node [style=sergio]{$p$};
\path (2.25,-1) node [style=sergio]{$a_{k+1}$};
\path (2.25,-3) node [style=sergio]{$p$};
\path (-4,-1) node [style=sergio]{$p(a_1,\cdots a_{k+1})\,\,=$};
\end{tikzpicture}.
\end{align}
\item Construct the parent Hamiltonian~\cite{Perez2007} for the reduced MPS, which only includes free physical indices and fixed $\bm{a} = {a_1, \cdots, a_N}$ based on each sampling outcome.
\end{enumerate}
In summary, in addition to the preservation of positivity and Hermiticity by construction, another advantage of our LPDO formalism over the MPDO is the physical picture under the Kraus index.
It represents environmental degrees of freedom in a decohered system and labels the statistical ensemble in a disordered system, facilitating the physical realization of an LPDO in both formalisms.

\section{Generalization to $(2+1)D$ systems}
Previous studies in the literature focusing on the structure or application of LPDO are mainly restricted to $(1+1)D$ systems.
In this section, we first generalize the definition and construction of LPDO to two spatial dimensions, then provide the classification of ASPT phases protected by strong fermion parity symmetry.

\subsection{Construction of LPDO in $(2+1)D$ systems}
In $(2+1)D$, a pure quantum state with area law entanglement on an arbitrary lattice $\Gamma$ can be represented by a projected entangled pair state (PEPS)~\cite{Schuch2007, Perez2008, Schuch2010}, whose bulk-edge correspondence provides a natural formalism for studying the emergence of topological order from the structure of local tensors~\cite{Cirac2011, Schuch2013, Yang2014, Yang2015}.

A PEPS defined by a rank-$5$ tensor with one physical index and four virtual indices is
\begin{align}
\begin{tikzpicture}[scale=0.8]
\tikzstyle{sergio}=[rectangle,draw=none]
\draw[line width=1pt] (2,2) -- (5,2);
\draw[line width=1pt] (3,1) -- (4,3);
\filldraw[fill=white, draw=black, rounded corners] (2.5,1.5)--(4,1.5)--(4.5,2.5)--(3,2.5)--cycle;
\draw[line width=2pt, color=red] (3.5,2) -- (3.5,3);
\path (2,2.25) node [style=sergio]{$\tau_1$};
\path (4.25,3) node [style=sergio]{$\tau_2$};
\path (5,2.25) node [style=sergio]{$\tau_3$};
\path (3.4,1) node [style=sergio]{$\tau_4$};
\path (3.85,2) node [style=sergio]{\large $\mathsf{A}$};
\path (3.25,3) node [style=sergio]{$\ket{i}$};
\path (-1.25,2) node [style=sergio]{\small $\mathsf{A}=\sum\limits_{i}\sum\limits_{\{\tau_j\}}A^{i}_{\{\tau_j\}}|i\rangle\bigotimes\limits_{j=1}^4(\tau_j|=$};
\end{tikzpicture},
\end{align}
where $i$ is a $d$-dimensional physical index, and $\{\tau_j\}$ are the virtual indices with dimension $D$. 

An area-law entangled mixed state $\rho$ can generally be represented by a projected entangled pair density operator (PEPDO).
On a specific lattice site $i$, a PEPDO is defined as a rank-6 tensor with two physical indices and four virtual indices
\begin{align}
\begin{tikzpicture}[scale=0.7]
\tikzstyle{sergio}=[rectangle,draw=none]
\draw[line width=1pt] (2,2) -- (5,2);
\draw[line width=1pt] (3,1) -- (4,3);
\draw[line width=2pt, color=red] (3.5,2) -- (3.5,1);
\filldraw[fill=white, draw=black, rounded corners] (2.5,1.5)--(4,1.5)--(4.5,2.5)--(3,2.5)--cycle;
\draw[line width=2pt, color=red] (3.5,2) -- (3.5,3);
\path (2.25,2.25) node [style=sergio]{$\tau_1$};
\path (4.25,3) node [style=sergio]{$\tau_2$};
\path (4.75,2.25) node [style=sergio]{$\tau_3$};
\path (2.7,1) node [style=sergio]{$\tau_4$};
\path (3.85,2) node [style=sergio]{\large $\mathsf{A}$};
\path (3.4,3.5) node [style=sergio]{$\ket{i}$};
\path (3.4,0.5) node [style=sergio]{$\langle i'|$};
\path (-2,2) node [style=sergio]{\small $\mathsf{A}=\sum\limits_{i, i'}\sum\limits_{\{\tau_j\}}A^{i,i'}_{\{\tau_j\}}|i\rangle\langle i'|\bigotimes\limits_{j=1}^4(\tau_j|=$};
\end{tikzpicture}.
\label{PEPDO}
\end{align}

Similar to the $(1+1)D$ cases, there exists a large class of PEPDO that also admits a locally purified form, which we refer to as $(2+1)D$ LPDO. 
It is obtained by starting from a pure state in the combined Hilbert space of physical and ancillary degrees of freedom, $|\psi_{p\otimes a}\rangle\in\H_p\otimes\H_a$, with the following PEPS form with bond dimension $D$
\begin{align}
\begin{tikzpicture}[scale=0.7]
\tikzstyle{sergio}=[rectangle,draw=none]
\draw[line width=1pt] (2,2) -- (7.5,2);
\draw[line width=1pt] (1,0) -- (6.5,0);
\draw[line width=1pt] (2,-1) -- (4,3);
\draw[line width=1pt] (4.5,-1) -- (6.5,3);
\filldraw[fill=white, draw=black, rounded corners] (2.5,1.5)--(4,1.5)--(4.5,2.5)--(3,2.5)--cycle;
\filldraw[fill=white, draw=black, rounded corners] (1.5,-0.5)--(3,-0.5)--(3.5,0.5)--(2,0.5)--cycle;
\filldraw[fill=white, draw=black, rounded corners] (5,1.5)--(6.5,1.5)--(7,2.5)--(5.5,2.5)--cycle;
\filldraw[fill=white, draw=black, rounded corners] (4,-0.5)--(5.5,-0.5)--(6,0.5)--(4.5,0.5)--cycle;
\draw[line width=2pt, color=red] (3.25,2) -- (3.25,3);
\draw[line width=3pt, color=blue] (3.75,2) -- (3.75,3);
\draw[line width=2pt, color=red] (2.25,0) -- (2.25,1);
\draw[line width=3pt, color=blue] (2.75,0) -- (2.75,1);
\path (3.25,3.25) node [style=sergio]{\small $p$};
\path (2.25,1.25) node [style=sergio]{\small $p$};
\path (2.75,1.25) node [style=sergio]{\small $a$};
\path (3.75,3.25) node [style=sergio]{\small $a$};
\path (-0.5,1) node [style=sergio]{\small $|\psi_{p\otimes a}\rangle=$};
\draw[line width=2pt, color=red] (5.75,2) -- (5.75,3);
\draw[line width=3pt, color=blue] (6.25,2) -- (6.25,3);
\path (5.75,3.25) node [style=sergio]{\small $p$};
\path (6.25,3.25) node [style=sergio]{\small $a$};
\draw[line width=2pt, color=red] (4.75,0) -- (4.75,1);
\draw[line width=3pt, color=blue] (5.25,0) -- (5.25,1);
\path (4.75,1.25) node [style=sergio]{\small $p$};
\path (5.25,1.25) node [style=sergio]{\small $a$};
\path (1,1) node [style=sergio]{\small $\cdots$};
\path (7.5,1) node [style=sergio]{\small $\cdots$};
\path (3.5,-1.25) node [style=sergio]{\small $\cdots$};
\path (5,3.25) node [style=sergio]{\small $\cdots$};
\end{tikzpicture},
\label{2d PEPS}
\end{align}
and then tracing out the ancilla Hilbert space.
The corresponding density matrix $\rho$ is
\begin{align}
\rho=\mathrm{Tr}_a\left(|\psi_{p\otimes a}\rangle\langle\psi_{p\otimes a}|\right).
\end{align}
The local tensor of LPDO on a lattice site has the following graphical representation
\begin{align}
\begin{tikzpicture}[scale=0.7]
\tikzstyle{sergio}=[rectangle,draw=none]
\draw[line width=1pt] (2,2) -- (5,2);
\draw[line width=1pt] (2,0) -- (5,0);
\draw[line width=1pt] (3,-1) -- (4,1);
\draw[line width=1pt] (3,1) -- (4,3);
\draw[line width=2pt, color=red] (3.5,0) -- (3.5,-1.25);
\filldraw[fill=white, draw=black, rounded corners] (2.5,-0.5)--(4,-0.5)--(4.5,0.5)--(3,0.5)--cycle;
\draw[line width=3pt, color=blue] (3.5,2) -- (3.5,0);
\filldraw[fill=white, draw=black, rounded corners] (2.5,1.5)--(4,1.5)--(4.5,2.5)--(3,2.5)--cycle;
\draw[line width=2pt, color=red] (3.5,2) -- (3.5,3.25);
\path (3.25,3.25) node [style=sergio]{\small $p$};
\path (3.25,-1.25) node [style=sergio]{\small $p$};
\path (3.85,2) node [style=sergio]{\large $\mathsf{A}$};
\path (3.85,0) node [style=sergio]{\large $\mathsf{A^*}$};
\path (3.25,1) node [style=sergio]{\small $a$};
\path (4.4,3) node [style=sergio]{\small $\tau_2^u$};
\path (2.6,1) node [style=sergio]{\small $\tau_4^u$};
\path (5,2.35) node [style=sergio]{\small $\tau_3^u$};
\path (2,2.35) node [style=sergio]{\small $\tau_1^u$};
\path (4.4,1) node [style=sergio]{\small $\tau_2^l$};
\path (2.6,-1) node [style=sergio]{\small $\tau_4^l$};
\path (5,0.35) node [style=sergio]{\small $\tau_3^l$};
\path (2,0.35) node [style=sergio]{\small $\tau_1^l$};
\path (0.6,1.25) node [style=sergio]{\small $\rho=$};
\path (1.5,1.25) node [style=sergio]{\small $\cdots$};
\path (5.5,1.25) node [style=sergio]{\small $\cdots$};
\path (4.25,3.5) node [style=sergio]{\small $\cdots$};
\path (2.75,-1.5) node [style=sergio]{\small $\cdots$};
\end{tikzpicture},
\label{2d LPDO}
\end{align}
where the contracted index $a$ is referred to as the Kraus index. 

\subsection{MPO-injectivity and symmetry action of PEPS}
An injective PEPS is also characterized by its local conditions.
Consider an arbitrary region $R$ with boundary $\partial R$, then a PEPS is injective if the map
\begin{align}
    A_R: (\mathbb{C}^D)^{\otimes |\partial R|}\rightarrow (\mathbb{C}^{d_p})^{\otimes |R|}
\end{align}
is injective, where $A_R$ is obtained by contracting the tensors within the region $R$.
For pure states, injective PEPS neither exhibit any intrinsic long-range entanglement or long-range order nor characterize SPT phases.
Any state with nontrivial order is described by a more complex class of PEPS that satisfies the MPO-injectivity condition~\cite{Williamson2016, Sahinoglu2021, Molnar2018}.

A PEPS is MPO-injective if $A$ is injective on its support space, and the corresponding projector $P=A^{+} A$ can be written as a matrix product operator (MPO), where $A^+$ is the pseudo-inverse of $A$.
\begin{align}
\begin{tikzpicture}[use Hobby shortcut,scale=0.7]
\coordinate[] (A1) at (7,1) {};
\coordinate[] (A2) at (7.2,1.4) {};
\coordinate[] (A3) at (9.3003,2.2942) {};
\draw[line width=2pt, dashed] (A1) .. (A2) .. (A3);
\coordinate[] (A5) at (10.3,1) {};
\coordinate[] (A4) at (9.75,2.19) {};
\draw[line width=2pt, dashed] (A3) .. (A4) .. (A5);
\coordinate[] (A6) at (9.15,-0.1) {};
\coordinate[] (A7) at (8.0179,-0.2063) {};
\draw[line width=2pt, dashed] (A5) .. (A6) .. (A7);
\coordinate[] (A8) at (7.15,0) {};
\draw[line width=2pt, dashed] (A7) .. (A8) .. (A1);
\tikzstyle{sergio}=[rectangle,draw=none]
\draw[line width=1pt] (1.5,2) -- (4.5,2);
\draw[line width=1pt] (1.5,0) -- (4.5,0);
\draw[line width=1pt] (2.5,-1) -- (3.5,1);
\draw[line width=1pt] (2.5,1) -- (3.5,3);
\filldraw[fill=white, draw=black, rounded corners] (2,-0.5)--(3.5,-0.5)--(4,0.5)--(2.5,0.5)--cycle;
\draw[line width=2pt, color=red] (3,2) -- (3,0);
\filldraw[fill=white, draw=black, rounded corners] (2,1.5)--(3.5,1.5)--(4,2.5)--(2.5,2.5)--cycle;
\path (3,2) node [style=sergio]{\large $\mathsf{A}^+$};
\path (3.35,0) node [style=sergio]{\large $\mathsf{A}$};
\draw[line width=1pt] (6,2) -- (7,2) -- (7,0) -- (6,0);
\draw[line width=1pt] (11.3,2) -- (10.3,2) -- (10.3,0) -- (11.3,0);
\draw[line width=1pt] (9.8,4.2942) -- (9.3,3.2942) -- (9.3,1.2942) -- (9.8,2.2942);
\draw[line width=1pt] (7.5179,-0.2063) -- (8.0179,0.8) -- (8.0179,-1.2063) -- (7.5179,-2.2063);
\filldraw[fill=white, draw=black] (9.301,2.3038)circle (10pt);
\filldraw[fill=white, draw=black] (10.3,1)circle (10pt);
\filldraw[fill=white, draw=black] (7,1)circle (10pt);
\filldraw[fill=white, draw=black] (8.0276,-0.2063)circle (10pt);
\path (9.3003,2.2942) node [style=sergio]{\small $B_P$};
\path (7,1) node [style=sergio]{\small $B_P$};
\path (10.3,1) node [style=sergio]{\small $B_P$};
\path (8.0179,-0.2063) node [style=sergio]{\small $B_P$};
\path (5.25,1) node [style=sergio]{$=$};
\end{tikzpicture}.
\label{MPO parity}
\end{align}
Here the circle of $B$ operators is combined to be an MPO $P$ acting on all virtual indices of the PEPS.
The explicit form is given by
\begin{align}
P=\sum\limits_{\{\tau_j,\tau_j'\}}\mathrm{Tr}\left(\prod\limits_{j=1}^4B_{\tau_j\tau_j'}\right)|\tau_1,\tau_2,\tau_3,\tau_4)(\tau_1',\tau_2',\tau_3',\tau_4'|,\label{VgBg}
\end{align}
where $B_{\tau_j\tau_j'}$ is a $D\times D$ matrix defined on the virtual index with the following form
\begin{align}
\begin{tikzpicture}[scale=0.6]
\tikzstyle{sergio}=[rectangle,draw=none]
\draw[line width=2pt, dashed] (-1,0) -- (1,0);
\draw[line width=1pt] (0,-1) -- (0,1);
\filldraw[fill=white, draw=black] (0,0)circle (15pt);
\path (0,0) node [style=sergio]{$B_P$};
\path (4.5,-0.25) node [style=sergio]{ $=\sum\limits_{\tau\tau^{\prime}\alpha\beta}(B_P)_{\tau\tau^{\prime}}^{\alpha\beta}|\tau\rangle|\alpha)(\beta|\langle \tau^{\prime}|$};
\end{tikzpicture},
\end{align}
where $|\tau\rangle$ and $|\tau^{\prime}\rangle$ are vectors applied on the virtual index of PEPS, while $|\alpha)$ and $|\beta)$ are internal indices of the MPO $P$.
For the MPO to ensure stability under RG transformations, it must satisfy some axioms, including pulling-through, trivial-loop, and X-inverse conditions~\cite{Williamson2016, Sahinoglu2021}.

In particular, SPT phases in $(2+1)D$ systems are characterized by PEPS with a single-blocked MPO projector $P$~\cite{Perez2007}, or simply denoted as single-blocked MPO-injective PEPS.
Similar to the $(1+1)D$ case, this class of PEPS can exclude the existence of any topological order or any long-range order defined by~\cite{Rispler2015, Williamson2016}
\begin{align}
    C(i, j)=\braket{\psi|O_iO_j|\psi} - \braket{\psi|O_i|\psi}\braket{\psi|O_i|\psi}\sim \mathrm{e}^{-|i-j|/\xi}.
\end{align}

\begin{figure*}
\centering
\includegraphics[width=0.8\linewidth]{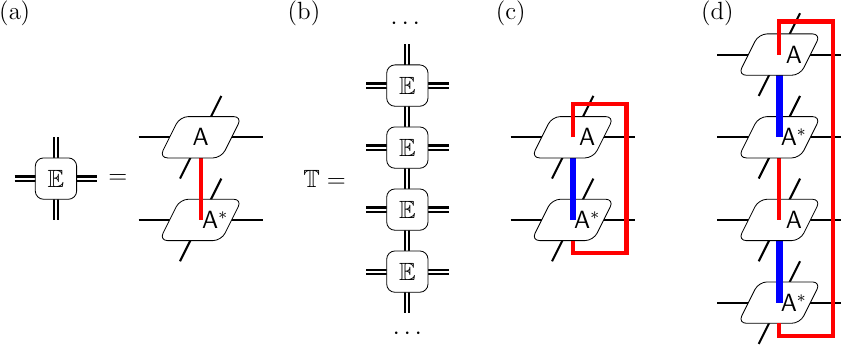}
\caption{Transfer Matrix of PEPS.
(a) Double tensor $\mathbb{E}$ for a single site.
(b) Transfer matrix $\mathbb{T}$ after contracting an entire column of $\mathbb{E}$.
(c) Double tensor for weak injectivity in Eq.~\eqref{weak injective 2D}.
(d) Double tensor for strong injectivity in Eq.~\eqref{strong injective 2D}.}
\label{Fig: PEPS_TM}
\end{figure*}

Here, we point out that the inverse direction remains an open problem.
In principle, the correlation length $\xi$ in the $(2+1)D$ systems is characterized by a $(1+1)D$ transfer matrix $\mathbb{T}$ (shown in Fig.~\ref{Fig: PEPS_TM}(b)) obtained by contracting the double tensor $\mathbb{E}$ (shown in Fig.~\ref{Fig: PEPS_TM}(a)) for an entire row or column, whose dimension scales exponentially with the system length scale $N$.
The existence of SSB requires the spectrum of $\mathbb{T}$ to be gapless, i.e., $\Delta \sim 1/N$~\cite{Rispler2015}, to trigger a long-range correlation length $\xi\sim 1/\Delta\sim N$, a property that cannot be directly written on the local tensors.
Therefore, it is difficult to claim the exact existence of a long-range order for a general non-injective PEPS.

In $(2+1)D$, the graphical representation of the symmetry operator $U(k)$ on a square lattice PEPS is given by~\cite{Williamson2016}
\begin{align}
\begin{tikzpicture}[use Hobby shortcut,scale=0.7]
\coordinate[] (A1) at (7.85,2);
\coordinate[] (A2) at (8.05,2.4);
\coordinate[] (A3) at (10.1503,3.2942);
\draw[line width=2pt, dashed] (A1) .. (A2) .. (A3);
\coordinate[] (A5) at (11.15,2);
\coordinate[] (A4) at (10.6,3.19);
\draw[line width=2pt, dashed] (A3) .. (A4) .. (A5);
\coordinate[] (A6) at (10,0.9);
\coordinate[] (A7) at (8.8679,0.7937);
\draw[line width=2pt, dashed] (A5) .. (A6) .. (A7);
\coordinate[] (A8) at (8,1);
\draw[line width=2pt, dashed] (A7) .. (A8) .. (A1);
\tikzstyle{sergio}=[rectangle,draw=none]
\draw[line width=1pt] (2.5,2) -- (5.5,2);
\draw[line width=1pt] (3.5,1) -- (4.5,3);
\filldraw[fill=white, draw=black, rounded corners] (3,1.5)--(4.5,1.5)--(5,2.5)--(3.5,2.5)--cycle;
\draw[line width=2pt, color=red] (4,2) -- (4,4);
\filldraw[fill=white, draw=black] (4,3)circle (10pt);
\path (4,3) node [style=sergio]{\small $U_k$};
\draw[line width=1pt] (7,2) -- (12,2);
\draw[line width=1pt] (8.5,0) -- (10.5,4);
\filldraw[fill=white, draw=black, rounded corners] (8.5,1.5)--(10,1.5)--(10.5,2.5)--(9,2.5)--cycle;
\draw[line width=2pt, color=red] (9.5,2) -- (9.5,4);
\filldraw[fill=white, draw=black] (10.151,3.3038)circle (10pt);
\filldraw[fill=white, draw=black] (11.15,2)circle (10pt);
\filldraw[fill=white, draw=black] (7.85,2)circle (10pt);
\filldraw[fill=white, draw=black] (8.8776,0.7937)circle (10pt);
\path (10.1503,3.2942) node [style=sergio]{\small $B_k$};
\path (7.85,2) node [style=sergio]{\small $B_k$};
\path (11.15,2) node [style=sergio]{\small $B_k$};
\path (8.8679,0.7937) node [style=sergio]{\small $B_k$};
\path (6.25,2) node [style=sergio]{$=$};
\end{tikzpicture},
\label{peps symmetry}
\end{align}
where similar to the projector $P$, $V(k)$ is also written as an MPO constructed from tensor $B(k)$ for each $k$.

In particular, the injectivity of PEPS requires that $V(1) = \mathbb{I}^{\otimes 4}$, i.e., a trivial MPO for $g=1$, from which one can also demonstrate that $V(k)$ can be factorized into a tensor product of local symmetry actions on each virtual index~\cite{Perez2010}, thereby excluding any nontrivial SPT order~\cite{Williamson2016}.
On the other hand, for a single-blocked MPO injective PEPS, although there are different linearly independent solutions of $V(k)$ satisfying the above equation, projecting them onto the virtual support space ensures the uniqueness of $V(k)$ (up to a phase).
In other words, by imposing an additional constraint $PV(k) = V(k)$, the symmetry transformation $V(k)$ in Eq.~\eqref{peps symmetry} is uniquely defined for each $k$ and characterizes the SPT order in the PEPS.

\subsection{Weak and strong MPO-injectivity}
Now we generalize the injectivity condition to LPDO in $(2+1)D$ systems.
\paragraph*{Definition} An LPDO in $(2+1)D$ systems is weakly injective if the corresponding purified PEPS in Eq.~\eqref{2d PEPS} with the local tensor
\begin{align}
A_{\{\tau_j\}}^{p,a}: (\mathbb{C}^D)^{\otimes |\partial R|}\rightarrow (\mathbb{C}^{d_p}\otimes \mathbb{C}^{d_{\kappa}})^{\otimes |R|}\label{weak injective 2D}
\end{align} 
is injective, while it is weakly MPO-injective if $A^{p, a}$ is injective on its support space and the corresponding projector $P^{p, a} = A^{p, a\,+}A^{p, a}$ is represented by an MPO satisfying the axioms.
An LPDO is strongly injective if the corresponding double state $\sket{\rho}$ with the local tensor
\begin{align}
    A^{p, p^{\prime}}_{\tau_j^u, \tau_j^l} \equiv \sum_{a} A_{\{\tau_j^u\}}^{p, a}A_{\{\tau_j^l\}}^{*p^{\prime}, a}: (\mathbb{C}^D)^{\otimes 2|\partial R|}\rightarrow (\mathbb{C}^{d_p})^{\otimes 2|R|}\label{strong injective 2D}
\end{align} is injective, while it is strongly MPO-injective if $A^{p, p^{\prime}}$ is injective on its support space and the corresponding projector $P^{p, p^{\prime}} = A^{p, p^{\prime}\,+}A^{p, p^{\prime}}$ is represented by an MPO satisfying the axioms.

Similar to the $(1+1)D$ case, weak single-blocked MPO-injectivity excludes the existence of long-range order in the linear two-point correlation function
\begin{align}
    C^{(1)}(i, j) \equiv \mathrm{Tr}\left(\rho O_i O_j\right) - \mathrm{Tr}\left(\rho O_j\right)\mathrm{Tr}\left(\rho O_j\right)\sim \mathrm{e}^{-|i-j|/\xi}
\end{align} guaranteed by a finite gap in the transfer matrix $\mathbb{T}$ composed by the double tensor $\mathbb{E}$ shown in Fig.~\ref{Fig: PEPS_TM}(c), while strong single-blocked MPO-injectivity excludes the existence of long-range order in the R\'enyi-2 two-point correlation function
\begin{align}
\begin{aligned}
    C^{(2)}(i, j)&\equiv \frac{\mathrm{Tr}\left(\rho O_i O_j\rho O_i O_j\right) - \mathrm{Tr}\left(\rho O_i \rho O_i\right)\mathrm{Tr}\left(\rho O_j\rho O_j\right)}{\mathrm{Tr}(\rho^2)}\\
    &\sim \mathrm{e}^{-|i-j|/\xi^{\prime}}
\end{aligned}
\end{align}
determined by the transfer matrix $\mathbb{T}$ from the double tensor $\mathbb{E}$ shown in Fig.~\ref{Fig: PEPS_TM}(d).
In summary, regardless of spatial dimension, a short-range entangled LPDO should be single-blocked MPO-injective in both senses to ensure that there is no topological order or long-range order in the density matrix.

\subsection{Symmetry in the LPDO representation}
Next, let us consider the strong and weak symmetry actions on the upper side of an LPDO.
A strong symmetry $U(k)$ does not induce transformations in the Kraus index, whose graphical descriptions are as follows
\begin{align}
\begin{tikzpicture}[use Hobby shortcut,scale=0.7]
\draw[line width=3pt, color=blue] (9.5,1.5) -- (9.5,0);
\coordinate[] (A1) at (7.85,2);
\coordinate[] (A2) at (8.05,2.4);
\coordinate[] (A3) at (10.1503,3.2942);
\draw[line width=2pt, dashed] (A1) .. (A2) .. (A3);
\coordinate[] (A5) at (11.15,2);
\coordinate[] (A4) at (10.6,3.19);
\draw[line width=2pt, dashed] (A3) .. (A4) .. (A5);
\coordinate[] (A6) at (10,0.9);
\coordinate[] (A7) at (8.8679,0.7937);
\draw[line width=2pt, dashed] (A5) .. (A6) .. (A7);
\coordinate[] (A8) at (8,1);
\draw[line width=2pt, dashed] (A7) .. (A8) .. (A1);
\tikzstyle{sergio}=[rectangle,draw=none]
\draw[line width=1pt] (2.5,2) -- (5.5,2);
\draw[line width=1pt] (3.5,1) -- (4.5,3);
\draw[line width=3pt, color=blue] (4,2) -- (4,0.5);
\filldraw[fill=white, draw=black, rounded corners] (3,1.5)--(4.5,1.5)--(5,2.5)--(3.5,2.5)--cycle;
\draw[line width=2pt, color=red] (4,2) -- (4,4);
\path (3.75,4) node [style=sergio]{\small $p$};
\path (3.75,0.5) node [style=sergio]{\small $a$};
\filldraw[fill=white, draw=black] (4,3)circle (10pt);
\path (4,3) node [style=sergio]{\small $U_k$};
\draw[line width=1pt] (7,2) -- (12,2);
\draw[line width=1pt] (8.5,0) -- (10.5,4);
\filldraw[fill=white, draw=black, rounded corners] (8.5,1.5)--(10,1.5)--(10.5,2.5)--(9,2.5)--cycle;
\draw[line width=2pt, color=red] (9.5,2) -- (9.5,4);
\filldraw[fill=white, draw=black] (10.151,3.3038)circle (10pt);
\filldraw[fill=white, draw=black] (11.15,2)circle (10pt);
\filldraw[fill=white, draw=black] (7.85,2)circle (10pt);
\filldraw[fill=white, draw=black] (8.8776,0.7937)circle (10pt);
\path (10.1503,3.2942) node [style=sergio]{\small $B_k$};
\path (7.85,2) node [style=sergio]{\small $B_k$};
\path (11.15,2) node [style=sergio]{\small $B_k$};
\path (8.8679,0.7937) node [style=sergio]{\small $B_k$};
\path (6.25,2) node [style=sergio]{$=$};
\path (9.25,4) node [style=sergio]{\small $p$};
\path (9.75,0) node [style=sergio]{\small $a$};
\end{tikzpicture},
\label{2d exact}
\end{align}
promising that the density matrix is invariant if we only apply a strong symmetry on one side of the density matrix.
Here, the transformation in the virtual indices $V(k)$ is still an MPO constructed from $B(k)$.
Similar to the $(1+1)D$ case, the weak single-blocked MPO-injectivity condition of an LPDO guarantees the uniqueness (up to a phase) of $V(k)$ with the constraint $P^{p, a} V(k)=V(k)$, since the actions on the physical and Kraus indices are fixed.

On the other hand, the graphical representation of the weak symmetry $U(g)$ on an LPDO is
\begin{align}
\begin{tikzpicture}[use Hobby shortcut,scale=0.7]
\draw[line width=3pt, color=blue] (9.5,1.5) -- (9.5,-1);
\coordinate[] (A1) at (7.85,2);
\coordinate[] (A2) at (8.05,2.4);
\coordinate[] (A3) at (10.1503,3.2942);
\draw[line width=2pt, dashed] (A1) .. (A2) .. (A3);
\coordinate[] (A5) at (11.15,2);
\coordinate[] (A4) at (10.6,3.19);
\draw[line width=2pt, dashed] (A3) .. (A4) .. (A5);
\coordinate[] (A6) at (10,0.9);
\coordinate[] (A7) at (8.8679,0.7937);
\draw[line width=2pt, dashed] (A5) .. (A6) .. (A7);
\coordinate[] (A8) at (8,1);
\draw[line width=2pt, dashed] (A7) .. (A8) .. (A1);
\tikzstyle{sergio}=[rectangle,draw=none]
\draw[line width=1pt] (2.5,2) -- (5.5,2);
\draw[line width=1pt] (3.5,1) -- (4.5,3);
\draw[line width=3pt, color=blue] (4,2) -- (4,0.5);
\filldraw[fill=white, draw=black, rounded corners] (3,1.5)--(4.5,1.5)--(5,2.5)--(3.5,2.5)--cycle;
\draw[line width=2pt, color=red] (4,2) -- (4,4);
\path (3.75,4) node [style=sergio]{\small $p$};
\path (3.75,0.5) node [style=sergio]{\small $a$};
\filldraw[fill=white, draw=black] (4,3)circle (10pt);
\path (4,3) node [style=sergio]{\small $U_g$};
\draw[line width=1pt] (7,2) -- (12,2);
\draw[line width=1pt] (8.5,0) -- (10.5,4);
\filldraw[fill=white, draw=black, rounded corners] (8.5,1.5)--(10,1.5)--(10.5,2.5)--(9,2.5)--cycle;
\draw[line width=2pt, color=red] (9.5,2) -- (9.5,4);
\filldraw[fill=white, draw=black] (10.151,3.3038)circle (10pt);
\filldraw[fill=white, draw=black] (11.15,2)circle (10pt);
\filldraw[fill=white, draw=black] (7.85,2)circle (10pt);
\filldraw[fill=white, draw=black] (8.8776,0.7937)circle (10pt);
\path (10.1503,3.2942) node [style=sergio]{\small $B_g$};
\path (7.85,2) node [style=sergio]{\small $B_g$};
\path (11.15,2) node [style=sergio]{\small $B_g$};
\path (8.8679,0.7937) node [style=sergio]{\small $B_g$};
\path (6.25,2) node [style=sergio]{$=$};
\path (9.25,4) node [style=sergio]{\small $p$};
\path (9.75,-1) node [style=sergio]{\small $a$};
\filldraw[fill=white, draw=black] (9.5,0)circle (10pt);
\path (9.5,0) node [style=sergio]{\small $M_g$};
\end{tikzpicture},
\label{2d average}
\end{align}
where $M(g)$ is an additional unitary operator acting on the Kraus index.
Similar to the $(1+1)D$ case, due to the nontrivial $M(g)$ action on the Kraus index, the density matrix $\rho$ is not invariant if we apply a weak symmetry on one side of the density matrix, while $\rho$ is invariant only if we apply weak symmetry operators on both sides of the density matrix, meaning that $M(g)^{\dag}M(g) = I$ forms a unitary representation.
Furthermore, strong and weak single-blocked MPO-injectivity conditions together guarantee that $V(g)$ is unique (up to a phase) for a given $U(g)$ with the constraints $P^{p, a} V(g)=V(g)$ and $P^{p, p^{\prime}} V(g)\otimes V(g)^*=V(g)\otimes V(g)^*$ although maybe $M(g)$ is not, through an argument analogous to the $(1+1)D$ case.

\subsection{$(2+1)D$ SPT with fermion parity symmtry}
In this section, we review the classification of $(2+1)D$ pure SPT states in terms of PEPS representation~\cite{Bultinck2017B} for both bosonic and fermionic systems. 

For bosonic SPT states, the on-site symmetry action $U(k)$ is shown in Eq.~\eqref{peps symmetry}.
We sequentially apply two weak symmetry operators $U(k_2)$ and $U(k_1)$ that will leave two surrounding MPOs $B(k_2)$ and $B(k_1)$, which are essentially the same as applying the symmetry operator $U(k_1k_2)$ leaving a surrounding MPO $B(k_1k_2)$.
This equivalence leads to the following composition rule of $B(k)$ as
\begin{align}
\begin{tikzpicture}[scale=0.8]
\tikzstyle{sergio}=[rectangle,draw=none]
\draw[line width=2pt, dashed] (2.5,3.75) -- (4.5,3.75);
\draw[line width=2pt, dashed] (2.5,2.25) -- (4.5,2.25);
\draw[line width=1pt] (3.5,1.5) -- (3.5,4.5);
\filldraw[fill=white, draw=black] (3.5,3.75)circle (11pt);
\path (3.5,3.75) node [style=sergio]{\small $B_{k_1}$};
\filldraw[fill=white, draw=black] (3.5,2.25)circle (11pt);
\path (3.5,2.25) node [style=sergio]{\small $B_{k_2}$};
\draw[line width=2pt, dashed] (6,3) -- (5,3);
\draw[line width=2pt, dashed] (1,3) -- (2,3);
\filldraw[fill=white, draw=black, rounded corners] (5.25,3)--(4.5,1.75)--(4.5,4.25)--cycle;
\path (4.8,3) node [style=sergio]{\small $\lambda_\beta$};
\filldraw[fill=white, draw=black, rounded corners] (1.75,3)--(2.5,1.75)--(2.5,4.25)--cycle;
\path (2.2,3) node [style=sergio]{\small $\lambda_\alpha$};
\path (6.5,3) node [style=sergio]{\small $=$};
\draw[line width=2pt, dashed] (9,3) -- (11,3);
\draw[line width=1pt] (10,4) -- (10,2);
\filldraw[fill=white, draw=black] (10,3)circle (16pt);
\path (7.75,3) node [style=sergio]{\small $\lambda(k_1,k_2)$};
\path (10,3) node [style=sergio]{\small $B_{k_1k_2}$};
\end{tikzpicture},
\label{MPO fusion}
\end{align}
where $\lambda_\alpha$ and $\lambda_\beta$ are the fusion tensor on virtual indices to project out the redundant virtual degree of freedom for the MPO $V(k_1)V(k_2)$, and $\lambda(k_1,k_2)\in U(1)$. 

Subsequently, we consider the sequential application of three symmetry operators to the physical index: $U(k_3)$, $U(k_2)$, and $U(k_1)$, which is essentially the same as applying $U(k_1k_2k_3)$.
There are different orders concerning the fusion rule of the MPO tensor $B(k)$ that result in a phase ambiguity of the fusion tensor $\lambda_\beta$, namely
\begin{align}
\begin{tikzpicture}[scale=0.8]
\tikzstyle{sergio}=[rectangle,draw=none]
\draw[line width=2pt, dashed] (-2.5,3.5) -- (-2,3.5);
\draw[line width=2pt, dashed] (-2.5,2.5) -- (-2,2.5);
\draw[line width=2pt, dashed] (-2.5,1.5) -- (-1,1.5);
\draw[line width=2pt, dashed] (-1,3) -- (-1.5,3);
\filldraw[fill=white, draw=black, rounded corners] (-1.25,3)--(-2,2)--(-2,4)--cycle;
\path (-1.7,3) node [style=sergio]{\small $\lambda_\beta$};
\path (2.25,2.5) node [style=sergio]{\small $={\nu}_3(k_1,k_2,k_3)$};
\draw[line width=2pt, dashed] (0.5,2.5) -- (-0.2,2.5);
\filldraw[fill=white, draw=black, rounded corners] (0,2.5)--(-1,1)--(-1,4)--cycle;
\path (-0.5,2.5) node [style=sergio]{\small $\lambda_\beta$};
\draw[line width=2pt, dashed] (4,1.5) -- (4.5,1.5);
\draw[line width=2pt, dashed] (4,2.5) -- (4.5,2.5);
\draw[line width=2pt, dashed] (5.5,2) -- (5,2);
\filldraw[fill=white, draw=black, rounded corners] (5.25,2)--(4.5,1)--(4.5,3)--cycle;
\path (4.8,2) node [style=sergio]{\small $\lambda_\beta$};
\draw[line width=2pt, dashed] (5.5,3.5) -- (4,3.5);
\draw[line width=2pt, dashed] (7,2.5) -- (6,2.5);
\filldraw[fill=white, draw=black, rounded corners] (6.5,2.5)--(5.5,1)--(5.5,4)--cycle;
\path (6,2.5) node [style=sergio]{\small $\lambda_\beta$};
\end{tikzpicture},
\label{phase ambiguity}
\end{align}
where ${\nu}_3(k_1,k_2,k_3)\in U(1)$.
For simplicity, we abbreviate the above graph by a $F$-symbol as
\begin{align}
\begin{tikzpicture}[scale=0.8]
\tikzstyle{sergio}=[rectangle,draw=none]
\draw[line width=1pt] (-1.5,0.5) -- (-0.5,1.5);
\draw[line width=1pt] (-1.5,0.5) -- (-1.5,0);
\draw[line width=1pt] (-2.5,1.5) -- (-1.5,0.5);
\draw[line width=1pt] (-1.5,1.5) -- (-2,1);
\draw[line width=1pt] (4,0.5) -- (5,1.5);
\draw[line width=1pt] (4,0.5) -- (4,0);
\draw[line width=1pt] (3,1.5) -- (4,0.5);
\draw[line width=1pt] (4,1.5) -- (4.5,1);
\path (1.25,1) node [style=sergio]{\small $={\nu}_3(k_1,k_2,k_3)$};
\path (-2.5,1.75) node [style=sergio]{\scriptsize $k_1$};
\path (-1.5,1.75) node [style=sergio]{\scriptsize $k_2$};
\path (-0.5,1.75) node [style=sergio]{\scriptsize $k_3$};
\path (-1.5,-0.25) node [style=sergio]{\scriptsize $k_1k_2k_3$};
\filldraw[fill=red, draw=red] (4,0.5)circle (3pt);
\filldraw[fill=red, draw=red] (-1.5,0.5)circle (3pt);
\filldraw[fill=red, draw=red] (-2,1)circle (3pt);
\filldraw[fill=red, draw=red] (4.5,1)circle (3pt);
\path (4,-0.25) node [style=sergio]{\scriptsize $k_1k_2k_3$};
\path (3,1.75) node [style=sergio]{\scriptsize $k_1$};
\path (4,1.75) node [style=sergio]{\scriptsize $k_2$};
\path (5,1.75) node [style=sergio]{\scriptsize $k_3$};
\end{tikzpicture}.
\label{F-move}
\end{align}

\begin{figure*}
\centering
\includegraphics[width=0.8\linewidth]{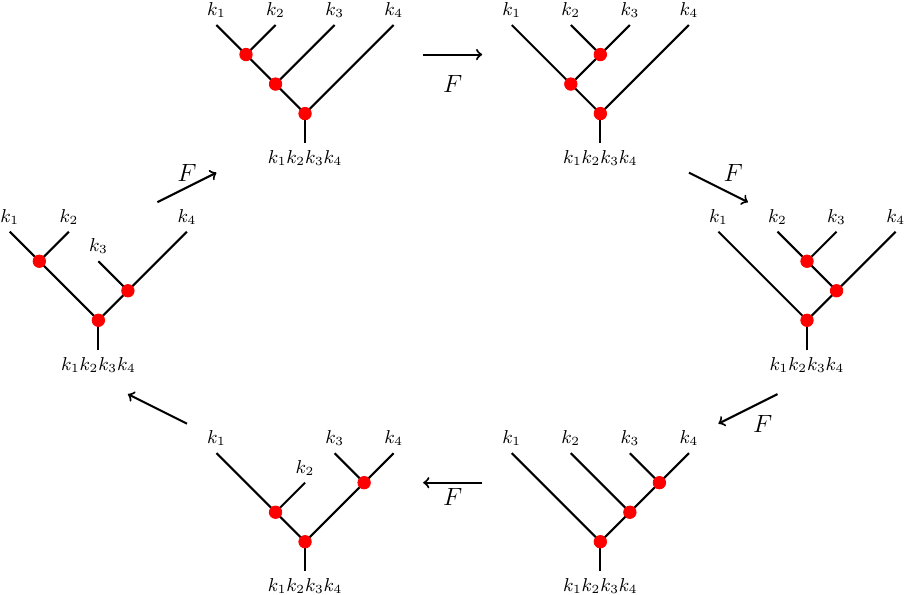}
\caption{Pentagon equation of the $F$-move defined in Eq. \eqref{F-move}.}
\label{pentagon}
\end{figure*}

Toward the consistency condition of the phase factor $\nu_3$, we should further consider the sequential applications of four symmetry operators, $U(k_4)$, $U(k_3)$, $U(k_2)$, and $U(k_1)$.
Different $F$-moves play the role of altering the fusion ordering of the MPO operators $B(k_4)$, $B(k_3)$, $B(k_2)$, and $B(k_1)$ on the virtual indices that finally return to the original fusion other, as depicted by a pentagon equation in Fig.~\ref{pentagon}.
It indicates that the phase ambiguity of the fusion tensor $\lambda_\beta$ satisfies a 3-cocycle condition, namely
\begin{align}
\frac{\nu_3(k_1,k_2,k_3)\nu_3(k_1,k_2k_3,k_4)\nu_3(k_2,k_3,k_4)}{\nu_3(k_1k_2,k_3,k_4)\nu_3(k_1,k_2,k_3k_4)}=1.
\end{align}
Therefore, the different $(2+1)D$ bosonic SPT states represented by PEPS are classified by a 3-cocycle $\nu_3\in\H^3[G, U(1)]$. 

In the following, we generalize the PEPS construction and classification of $(2+1)D$ SPT states to fermionic systems.
Meanwhile, we consider the simplest $G_f = G_b\times\Z_2^f$ symmetry, i.e., the total symmetry group is the direct product of two subgroups, or equivalently, the trivial extension $\omega_2$.
First of all, similar to the $(1+1)D$ case, we should define every physical and virtual index in a supervector space with a graded structure in Eq.~\eqref{super vector space}, and the symmetry action on the physical index is also given by Eq.~\eqref{tilde U} and~\eqref{bosonic symmetry}.
Next, by two sequential applications of $U(h)$ and $U(g)$ ($g, h\in G_b$), the composition rule of $B(g)$ should be 
\begin{align}
\begin{tikzpicture}[scale=0.7]
\tikzstyle{sergio}=[rectangle,draw=none]
\draw[line width=2pt, dashed] (2.5,3.75) -- (4.5,3.75);
\draw[line width=2pt, dashed] (2.5,2.25) -- (4.5,2.25);
\draw[line width=1pt] (3.5,1.5) -- (3.5,4.5);
\filldraw[fill=white, draw=black] (3.5,3.75)circle (11pt);
\path (3.5,3.75) node [style=sergio]{\small $B_{g}$};
\filldraw[fill=white, draw=black] (3.5,2.25)circle (11pt);
\path (3.5,2.25) node [style=sergio]{\small $B_{h}$};
\draw[line width=2pt, dashed] (6,3) -- (5,3);
\draw[line width=2pt, dashed] (1,3) -- (2,3);
\filldraw[fill=white, draw=black, rounded corners] (5.25,3)--(4.5,1.75)--(4.5,4.25)--cycle;
\path (4.8,3) node [style=sergio]{\small $\lambda_\beta$};
\filldraw[fill=white, draw=black, rounded corners] (1.75,3)--(2.5,1.75)--(2.5,4.25)--cycle;
\path (2.2,3) node [style=sergio]{\small $\lambda_\alpha$};
\path (6.5,3) node [style=sergio]{\small $=$};
\draw[line width=2pt, dashed] (10,3) -- (12,3);
\draw[line width=1pt] (11,4) -- (11,2);
\filldraw[fill=white, draw=black] (11,3)circle (16pt);
\path (8.3,3) node [style=sergio]{\small $P_f^{\omega_2(g,h)}\lambda(g,h)$};
\path (11,3) node [style=sergio]{$B_{gh}$};
\end{tikzpicture}.
\label{fermionic MPO fusion}
\end{align}

Similar to the bosonic case, we then consider three sequentially applied symmetry operators, $U(k)$, $U(h)$, and $U(g)$, which is essentially the same as applying $U(ghk)$.
Therefore, we can define a phase ambiguity $\nu_3$ for the bosonic symmetry $G_b$ from two equivalent fusion orders as in the following
\begin{align}
\begin{tikzpicture}[scale=0.8]
\tikzstyle{sergio}=[rectangle,draw=none]
\draw[line width=2pt, dashed] (-2.5,3.5) -- (-2,3.5);
\draw[line width=2pt, dashed] (-2.5,2.5) -- (-2,2.5);
\draw[line width=2pt, dashed] (-2.5,1.5) -- (-1,1.5);
\draw[line width=2pt, dashed] (-1,3) -- (-1.5,3);
\filldraw[fill=white, draw=black, rounded corners] (-1.25,3)--(-2,2)--(-2,4)--cycle;
\path (-1.7,3) node [style=sergio]{\small $\lambda_\beta$};
\path (2.25,2.5) node [style=sergio]{\small $={\nu}_3(g,h,k)$};
\draw[line width=2pt, dashed] (0.5,2.5) -- (-0.2,2.5);
\filldraw[fill=white, draw=black, rounded corners] (0,2.5)--(-1,1)--(-1,4)--cycle;
\path (-0.5,2.5) node [style=sergio]{\small $\lambda_\beta$};
\draw[line width=2pt, dashed] (4,1.5) -- (4.5,1.5);
\draw[line width=2pt, dashed] (4,2.5) -- (4.5,2.5);
\draw[line width=2pt, dashed] (5.5,2) -- (5,2);
\filldraw[fill=white, draw=black, rounded corners] (5.25,2)--(4.5,1)--(4.5,3)--cycle;
\path (4.8,2) node [style=sergio]{\small $\lambda_\beta$};
\draw[line width=2pt, dashed] (5.5,3.5) -- (4,3.5);
\draw[line width=2pt, dashed] (7,2.5) -- (6,2.5);
\filldraw[fill=white, draw=black, rounded corners] (6.5,2.5)--(5.5,1)--(5.5,4)--cycle;
\path (6,2.5) node [style=sergio]{\small $\lambda_\beta$};
\end{tikzpicture},
\label{phase ambiguity_Gb}
\end{align}
where ${\nu}_3(g,h,k)\in U(1)$, or depicted with the $F$-symbol as
\begin{align}
\begin{tikzpicture}[scale=0.8]
\tikzstyle{sergio}=[rectangle,draw=none]
\draw[line width=1pt] (-1.5,0.5) -- (-0.5,1.5);
\draw[line width=1pt] (-1.5,0.5) -- (-1.5,0);
\draw[line width=1pt] (-2.5,1.5) -- (-1.5,0.5);
\draw[line width=1pt] (-1.5,1.5) -- (-2,1);
\draw[line width=1pt] (4,0.5) -- (5,1.5);
\draw[line width=1pt] (4,0.5) -- (4,0);
\draw[line width=1pt] (3,1.5) -- (4,0.5);
\draw[line width=1pt] (4,1.5) -- (4.5,1);
\path (1.25,1) node [style=sergio]{\small $={\nu}_3(g,h,k)$};
\path (-2.5,1.75) node [style=sergio]{\scriptsize $g$};
\path (-1.5,1.75) node [style=sergio]{\scriptsize $h$};
\path (-0.5,1.75) node [style=sergio]{\scriptsize $k$};
\path (-1.5,-0.25) node [style=sergio]{\scriptsize $ghk$};
\filldraw[fill=red, draw=red] (4,0.5)circle (3pt);
\filldraw[fill=red, draw=red] (-1.5,0.5)circle (3pt);
\filldraw[fill=red, draw=red] (-2,1)circle (3pt);
\filldraw[fill=red, draw=red] (4.5,1)circle (3pt);
\path (4,-0.25) node [style=sergio]{\scriptsize $ghk$};
\path (3,1.75) node [style=sergio]{\scriptsize $g$};
\path (4,1.75) node [style=sergio]{\scriptsize $h$};
\path (5,1.75) node [style=sergio]{\scriptsize $k$};
\end{tikzpicture}.
\label{F-move_Gb}
\end{align}
The consistency condition for $\nu_3$ will be discussed later.

For fermionic systems, the boundary MPO can carry a nonzero fermion parity $n_1(g)$, defined as
\begin{align}
    P_f^{\otimes 4}V(g) = (-1)^{n_1(g)} V(g)P_f^{\otimes 4}.\label{Vg parity}
\end{align}
Following the composition of the fermionic boundary MPO in Eq.~\eqref{fermionic MPO fusion}, we obtain a fermion parity conservation as
\begin{align}
n_1(g)+n_1(h)-n_1(gh)=0~(\mathrm{mod}~2). \label{Parity Bg}
\end{align}

Furthermore, the conservation of the fermion parity of the $B_g$ operators leaves an ambiguity: the fermion parity of $\lambda_\alpha(g,h)$ and $\lambda_\beta(g,h)$ should be even in total, but they can be both even or odd individually.
We label one of them, e.g., that of $\lambda_\beta(g,h)$, by $n_2(g,h)$ and the red dots in Eq.~\eqref{F-move_Gb}, defined as
\begin{align}
\begin{tikzpicture}[scale=0.9]
\tikzstyle{sergio}=[rectangle,draw=none]
\draw[line width=2pt, dashed] (3.5,4.5) -- (5,4.5);
\draw[line width=2pt, dashed] (3.5,3) -- (5,3);
\filldraw[fill=white, draw=black] (4.25,4.5)circle (10pt);
\filldraw[fill=white, draw=black] (4.25,3)circle (10pt);
\draw[line width=2pt, dashed] (7.25,3.75) -- (5.5,3.75);
\filldraw[fill=white, draw=black, rounded corners] (5.75,3.75)--(5,2.5)--(5,5)--cycle;
\path (5.335,3.75) node [style=sergio]{\large $\lambda_\beta$};
\filldraw[fill=white, draw=black] (6.5,3.75)circle (10pt);
\path (4.25,4.5) node [style=sergio]{$P_f$};
\path (4.25,3) node [style=sergio]{$P_f$};
\path (6.5,3.75) node [style=sergio]{$P_f$};
\draw[line width=2pt, dashed] (10.5,4.5) -- (10,4.5);
\draw[line width=2pt, dashed] (10.5,3) -- (10,3);
\draw[line width=2pt, dashed] (11,3.75) -- (11.5,3.75);
\filldraw[fill=white, draw=black, rounded corners] (11.25,3.75)--(10.5,2.5)--(10.5,5)--cycle;
\path (8.75,3.75) node [style=sergio]{ $=(-1)^{n_2(g, h)}$};
\path (10.835,3.75) node [style=sergio]{\large $\lambda_\beta$};
\end{tikzpicture}.
\label{TODO}
\end{align}
By left-applying $P_f^{\otimes3}$ to both sides of Eq.~\eqref{phase ambiguity_Gb}, we obtain another fermion parity conservation as 
\begin{align}
n_2(g,h)+n_2(gh,k)=n_2(g,hk)+n_2(h,k)~(\mathrm{mod}~2).\label{Parity lambda}
\end{align}

\begin{figure*}
\centering
\includegraphics[width=0.8\linewidth]{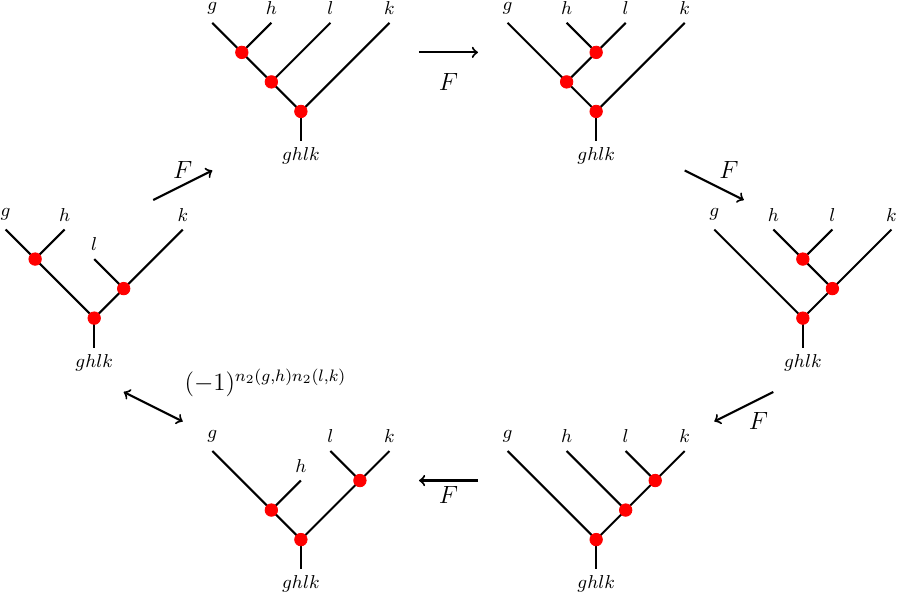}
\caption{Super-pentagon equation of the $F$-move defined in Eq.~\eqref{F-move_Gb}.
The move in the left-bottom corner depicts an exchange of two fusion tensors, which may lead to a fermion sign.}
\label{pentagon_F}
\end{figure*}

Finally, we discuss the consistency condition of phase ambiguity $\nu_3$.
Here, as an illustrative example, we consider the simplest case with $n_1(g)\equiv 0$ for all bosonic symmetry $g$, covering the Gu-Wen SPT classes but excluding the existence of any Majorana-type MPO~\cite{Bultinck2017B}.
Similar to the bosonic case, we consider the sequential applications of four symmetry operators, namely $U(l)$, $U(k)$, $U(h)$, and $U(g)$.
We further take several $F$-moves to alter the fusion ordering of the MPO operators $B(l)$, $B(k)$, $B(h)$, and $B(g)$ on the virtual indices and then back to the original ordering, there will be a super-pentagon equation as illustrated in Fig.~\ref{pentagon_F}.
In particular, in the left-bottom corner of Fig.~\ref{pentagon_F}, the exchange of two fusion tensors leads to an additional fermion sign: $(-1)^{n_2(g,h)n_2(k,l)}$. 
Gathering everything together, the phase ambiguity $\nu_3$ of the $F$-move should satisfy a twisted 3-cocycle condition, namely
\begin{align}
&\frac{\nu_3(g,h,l)\nu_3(g,hl,k)\nu_3(h,l,k)}{\nu_3(gh,l,k)\nu_3(g,h,lk)}=(-1)^{n_2(g,h)n_2(l,k)},
\label{2d nu3}
\end{align}
which requires $(-1)^{n_2(g,h)n_2(l,k)}$ to be a 4-coboundary in $\H^4[G_b, U(1)]$.
This twisted 3-cocycle condition will be much more complicated for a nontrivial $n_1$ or $\omega_2$~\cite{Wang2018, Wang2020}.

We summarize the PEPS classification and topological invariants of $(2+1)D$ fermionic SPT phases:
\begin{enumerate}[1.]
\item $n_1\in\H^1[G_b, h^2(\Z_2^f)]$: fermion parity of the boundary MPO $B(g)$ for bosonic symmetry with the 1-cocycle condition in Eq.~\eqref{Parity Bg}.
\item $n_2\in\H^2[G_b, h^1(\Z_2^f)]$: fermion parity of the fusion tensor $\lambda_\beta$ of the boundary MPOs $B(g)$ with the 2-cocycle condition in Eq.~\eqref{Parity lambda}, together with the requirement that $(-1)^{n_2(g,h)n_2(l,k)}$ be a 4-coboundary in $\H^4[G_b, U(1)]$.
\item $\nu_3\in\H^3[G_b, U(1)]$: phase ambiguity of the $F$-moves of the fusion tensors with the twisted 3-cocycle condition in Eq.~\eqref{2d nu3} for the case $n_1(g)\equiv 0$.
\end{enumerate}

We emphasize that each set of classification data ($n_1$ and $n_2$) obtained from the PEPS construction implies a distinct decorated domain wall configuration.
When a truncated symmetry operator $U_\Gamma(g)=\prod_{j\in\Gamma}U_j(g)$ is applied within a finite region $\Gamma$, it induces an MPO $B_g$ on the boundary of $\Gamma$, where the total fermion parity is characterized by $n_1(g)$.
Alternatively, if we define fermionic injective PEPS within the finite region $\Gamma$ under an open boundary condition, a global symmetry transformation $U(g)$ corresponds to the MPO $B_g$ is acted on the boundary MPS defined on $\partial\Gamma$.
In particular, when $n_1(g)=1$, the MPO $B_g$ twists the total fermion parity of the boundary MPS, indicating the presence of a Kitaev Majorana chain on $\partial \Gamma$.

Then we apply a truncated symmetry operator $U_\Gamma(g)=\prod_{j\in\Gamma}U_j(g)$ within a finite regime $\Gamma$, and similarly, another truncated symmetry operator $U_{\Gamma'}(h)$ within another finite regime $\Gamma'$.
At the junction $\partial\Gamma\cap\partial\Gamma'$, where the boundary MPOs $B_g$ and $B_h$ intersect, there exists a fusion rule specific to these boundary MPOs, i.e.,
\begin{align}
\begin{tikzpicture}[scale=0.7]
\tikzstyle{sergio}=[rectangle,draw=none]
\draw[line width=2pt, dashed] (1.75,4.25) -- (4.25,4.25);
\draw[line width=1pt] (2.75,5) -- (2.75,3.5);
\draw[line width=2pt, dashed] (2.25,2.75) -- (4.25,2.75);
\draw[line width=1pt] (3.25,2) -- (3.25,3.5);
\filldraw[fill=white, draw=black] (2.75,4.25)circle (11pt);
\path (2.75,4.25) node [style=sergio]{\small $B_{g}$};
\filldraw[fill=white, draw=black] (3.25,2.75)circle (11pt);
\path (3.25,2.75) node [style=sergio]{\small $B_{h}$};
\draw[line width=2pt, dashed] (7,3.5) -- (4.75,3.5);
\filldraw[fill=white, draw=black, rounded corners] (5,3.5)--(4.25,2.25)--(4.25,4.75)--cycle;
\path (4.575,3.5) node [style=sergio]{\small $\lambda_\beta$};
\draw[line width=1pt] (6,4.5) -- (6,2.5);
\filldraw[fill=white, draw=black] (6,3.5)circle (16pt);
\path (6,3.5) node [style=sergio]{\small $B_{gh}$};
\end{tikzpicture},
\end{align}
and the fermion parity of the fusion tensor $\lambda_\beta$ is labeled by $n_2(g, h)$.
This means that there might be a complex fermion decorated at the junction of the symmetry domain walls $\partial\Gamma\cap\partial\Gamma'$.

\subsection{$(2+1)D$ ASPT with strong fermion parity symmetry}
Now we turn to studying the ASPT phases protected by strong symmetry $K$ and weak symmetry $G$.
We discuss a simplest but physically relevant example with a strong $K=\Z_2^f$ fermion parity symmetry and a weak $G=G_b$ symmetry, with trivial extension $G_b\times\Z_2^f$.

First, we consider the fermion parity of the $B_g$ operator (labeled by $n_1(g)$) with the definition in Eq.~\eqref{Vg parity}.
Because the fermion parity $\Z_2^f$ is a strong symmetry, the fermion parity of $B_g$ on the upper and lower sides of the $(2+1)D$ LPDO are well defined individually and the complex fermion cannot tunnel between two sides.
We sequentially apply $U(h)$ and $U(g)$ on the physical indices ($g,h\in G_b$) that is equivalent to the application of $U(gh)$, according to Eq.~\eqref{2d average}, there must be an MPO fusion defined in Eq.~\eqref{fermionic MPO fusion} on each nearby virtual index, which implies the following 1-cocycle condition for the fermion parity conservation of the MPO $B(g)$, namely
\begin{align}
n_1(g)+n_1(h)+n_1(gh)=0~(\mathrm{mod}~2).\label{Parity_Bg}
\end{align}
Furthermore, for fermionic systems, each fusion tensor $\lambda_\beta(g, h)$ can carry a nonzero fermion parity labeled by $n_2(g,h)$ as defined in Eq. \eqref{TODO}, where the fermion parity conservation of the $F$-move gives the following 2-cocycle condition
\begin{align}
n_2(g,h)+n_2(gh,k)=n_2(g,hk)+n_2(h,k)~(\mathrm{mod}~2).\label{Parity_lambda}
\end{align}

On the contrary, the phase ambiguity of $F$-move demonstrated in Eq.~\eqref{phase ambiguity} at the upper and lower sides of an LPDO have to be canceled, i.e., there is no phase ambiguity of $F$-move.
This is consistent with the $(1+1)D$ case that there is no nontrivial ASPT state solely protected by weak symmetry.
Therefore, we do not need to care about the condition of $\nu_3$ in Eq.~\eqref{2d nu3} discussed for pure state SPT, as the weak symmetry $G_b$ must be applied simultaneously to the upper and lower sides of an LPDO and all the phase ambiguity will be canceled out.
Consequently, the current classification applies not only to the case of $n_1(g)\equiv 0$, but also accommodates the existence of Majorana-type MPOs $V(g)$ with $n_1(g)=1$, as the super-pentagon equation and the cocycle condition of $\nu_3$ become irrelevant in this context.

We summarize the LPDO classification of $(2+1)D$ ASPT phases with a strong $\Z_2^f$ fermion parity and a weak $G_b$ symmetry.
The topological invariants of these ASPT states are 
\begin{enumerate}[1.]
    \item $n_1\in\H^1(G_b,\Z_2)$: Fermion parity of the $B_g$ operator satisfying the 1-cocycle condition in Eq.~\eqref{Parity_Bg};
    \item $n_2\in\H^2(G_b,\Z_2)$: Fermion parity of the fusion tensor $\lambda_\beta$ satisfying the 2-cocycle condition in Eq.~\eqref{Parity_lambda}, but without the coboundary condition as in the pure state case.
\end{enumerate}
We note that there is no obstruction of the above data.
In particular, each $n_2$ data such that $(-1)^{n_2(g,h)n_2(l,k)}$ is a nontrivial 4-cocycle in $\H^4[G_b, U(1)]$ corresponds to an intrinsic ASPT state. 

Similar to the fermion SPT pure states, we can also extract the decorated domain wall patterns directly from the classification data: a nontrivial $n_1$ labels a fermionic ASPT state with Majorana chain decoration on the $G_b$ domain wall, and a nontrivial $n_2$ labels a fermionic ASPT state with complex fermion decoration on the junction of the $G_b$ domain wall.

\subsection{Example: $(2+1)D$ intrinsic ASPT}
In this section, we provide an example illustrating the tensor construction of an intrinsic ASPT phase in $(2+1)D$ systems.
Our approach is guided by the decorated domain wall picture~\cite{Chen2014} and integrates the intrinsic ASPT density matrix proposed for the $(1+1)D$ systems in the previous section.

\begin{figure*}
\centering
\includegraphics[width=\linewidth]{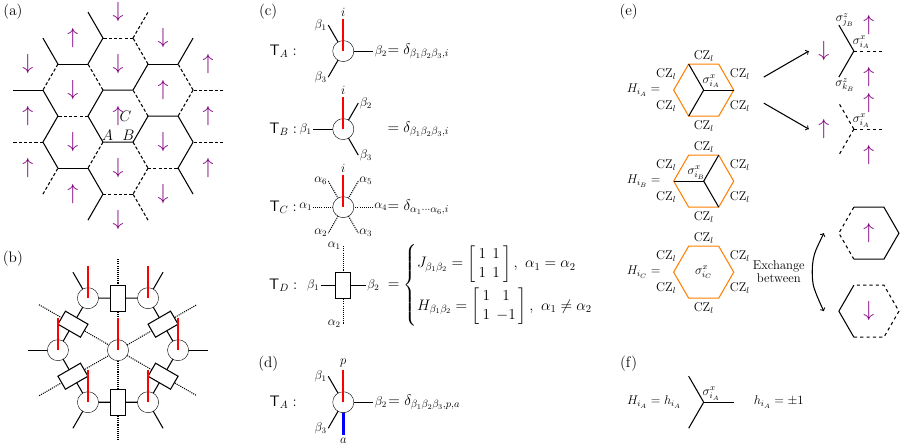}
\caption{Tensor construction of intrinsic ASPT phase in $(2+1)D$ systems.
(a) Honeycomb Lattice and domain wall of $C$ spins.
(b) Tensor structure for $(2+1)D$ SPT phase protected by $\Z_2^A\times\Z_2^B \times \Z_2^C$.
(c) Tensor elements for $(2+1)D$ SPT phase protected by $\Z_2^A\times\Z_2^B \times \Z_2^C$.
(d) Tensor elements for $(2+1)D$ intrinsic ASPT phase protected by $\Z_2^C\times\Z_4^{A, B}$.
Here, $\mathsf{T}_{B/C/D}$ remain unchanged, while $\mathsf{T}_{A}$ possesses an extra Kraus index.
(e) Hamiltonian for $(2+1)D$ SPT phase protected by $\Z_2^A\times\Z_2^B \times \Z_2^C$.
$H_{i_A}$ (or $H_{i_B}$) degrades to $\sigma_{j_B}^z\sigma_{i_A}^x\sigma_{k_B}^z$ or $\sigma_{i_A}^x$ according to the configuration of $C$ domain walls.
$H_{i_C}$ flips the spin at site $i_C$ and transforms the quantum state on surrounding links between trivial and cluster chains.
(f) Disordered Hamiltonian for $(2+1)D$ intrinsic ASPT phase protected by $\Z_2^C\times\Z_4^{A, B}$.
Here, $H_{i_B}$ and $H_{i_C}$ remain unchanged, while $H_{i_A}$ is replaced by a random local potential term that determines the configuration of $A$ spins.
}
\label{Fig: 2D ASPT}
\end{figure*}

\subsubsection{$(2+1)D$ SPT phase protected by $\Z_2^A\times\Z_2^B \times \Z_2^C$}
We start our discussion from the construction of a nontrivial spin-$\frac{1}{2}$ SPT phase protected by $\Z_2^A\times\Z_2^B\times\Z_2^C$ symmetry defined on a honeycomb lattice, where spins are positioned on both plaquettes and vertices~\cite{Yoshida2016, Tantivasadakarn2023}.
To be more specific, we assign spins $A$ and $B$ to two inequivalent sets of sites on the honeycomb lattice, while $C$ spins are placed on the plaquettes, as shown in Fig.~\ref{Fig: 2D ASPT}.
The three $\Z_2$ symmetries are defined as $\hat{O}_A = \prod_{i}\sigma^x_{i_A}$, $\hat{O}_B = \prod_{i}\sigma^x_{i_B}$, and $\hat{O}_C = \prod_{i}\sigma^x_{i_C}$, respectively.
The wavefunction comprises an equal-weight superposition of all possible $C$-spin configurations, each attached by a nontrivial $(1+1)D$ SPT state protected by $\Z_2^A\times \Z_2^B$ along the domain walls of $C$ spins (solid lines in Fig.~\ref{Fig: 2D ASPT} (a)), while the rest vertices form a trivial symmetric product state (dashed lines in Fig.~\ref{Fig: 2D ASPT} (a)).

The tensor network structure for this model is shown in Fig.~\ref{Fig: 2D ASPT} (b), where there are four types of tensors.
Each vertex $A$ or $B$ is associated with a rank-$4$ tensor $\mathsf{T}_{A/B}$ featuring three virtual indices connecting neighboring vertices and a physical index. 
Meanwhile, each plaquette $C$ is represented by a rank-$7$ tensor $\mathsf{T}_C$ with six virtual indices linking surrounding edges and a physical index.
Additionally, we place a rank-$4$ tensor $\mathsf{T}_D$ on each edge with four virtual indices connecting neighboring vertices $A$, $B$, and two adjacent plaquettes $C$.
The tensor elements for these four types of tensors are shown in Fig.~\ref{Fig: 2D ASPT} (c).
In particular, the Hadamard gate $H$ is placed on the domain walls to generate the $(1+1)D$ cluster state (together with the delta function on each vertex), while the $J$ matrix facilitates the construction of the product symmetric state $\ket{\uparrow}+\ket{\downarrow}\sim \ket{\rightarrow}$.
This construction effectively illustrates its nontrivial topological properties by examining the symmetry actions on the domain wall structure.
When a $\Z_2^C$ operator is applied to a single plaquette, flipping the $C$ spin, it induces a Pauli $X$ action on each virtual index of $\mathsf{T}_C$.
Consequently, the neighboring $\mathsf{T}_D$ tensors experience exchange between $J$ and $H$, reflecting a deformation in the domain wall structure.

\subsubsection{$(2+1)D$ intrinsic ASPT phase protected by $\Z_2^C\times\Z_4^{A, B}$}
Next, we proceed to the construction of intrinsic ASPT in $(2+1)D$ systems.
In this case, we retain the quantum superpositions of different $C$-spin configurations, while the cluster chain on the domain walls of $C$ spins is replaced by the intrinsic $(1+1)D$ SPT density matrix proposed in the previous section.
The symmetries associated with $B$ and $C$ spins remain strong, while the symmetry action on $A$ spins is defined as $U_A = \prod_i \sigma_{i_A}^xe^{i\frac{\pi}{4}(1-\sigma_{i_B}^x)}$, becoming a weak symmetry.
Similarly, the symmetry actions $\hat{O}_B$ and $U_A$ form the group extension
\begin{align}
    1\rightarrow \Z_2^B\rightarrow \Z_4^{A, B}\rightarrow \Z_2^A\rightarrow 1
\end{align}
with $U_A^2 = \hat{O}_B$.
Consequently, the total symmetry group is defined as $\Z_2^C\times \Z_4^{A, B}$.

As for the tensor construction, we note that all quantum superpositions of $A$-spin configurations are replaced by classical mixture, which means that the tensors $\mathsf{T}_{B/C/D}$ remain unchanged and $\mathsf{T}_A$ is replaced by the one depicted in Fig.~\ref{Fig: 2D ASPT} (d), i.e., with one extra Kraus index. It can be easily verified that on the domain walls of $C$ spins, there is a corresponding $(1+1)D$ intrinsic ASPT state protected by $\Z_4^{A, B}$ symmetry, while on other vertices, we are left with a maximally mixed state $\ket{\uparrow}\hspace{-1mm}\bra{\uparrow}_A+\ket{\downarrow}\hspace{-1mm}\bra{\downarrow}_A$ on $A$ spins that is weakly symmetric and a product state $\ket{\uparrow}_B+\ket{\downarrow}_B\sim \ket{\rightarrow}_B$ that is strongly symmetric.
Therefore, our construction yields an intrinsic ASPT phase protected by $\Z_2^C\times \Z_4^{A, B}$ symmetry in $(2+1)D$ systems because there is no $(1+1)D$ nontrivial SPT phase protected by $\Z_4$ symmetry.
As a result, we cannot obtain a corresponding $(2+1)D$ SPT phase protected by $\Z_2\times \Z_4$ with the same decorated domain wall structure.

\subsubsection{Physical realization in a disordered system}
In general, once the LPDO representation for a $(2+1)D$ mixed state is constructed, one may design the corresponding decohered or disordered systems that realize this mixed state, similar to the approach discussed in the $(1+1)D$ case.
However, the circuit preparation of PEPS~\cite{Wei2022} and the parent Hamiltonian method of PEPS~\cite{Perez2008} for various lattices is less general than their $(1+1)D$ version, requiring additional case-by-case efforts to determine whether these methods apply.
Alternatively, for the example considered here, since the tensor construction for the intrinsic $\Z_2^C\times \Z_4^{A, B}$ ASPT phase is inspired by the $\Z_2^A\times\Z_2^B\times\Z_2^C$ SPT state, we directly construct the disordered system that realizes this LPDO by modifying the Hamiltonian of the SPT state.
The Hamiltonian that stabilizes the previous $(2+1)D$ SPT state with domain-wall decoration is given by~\cite{Yoshida2016, Tantivasadakarn2023}
\begin{align}
    H = -\sum_{i} H_{i_A}-\sum_{i} H_{i_B}-\sum_{i} H_{i_C},\label{2+1D SPT}
\end{align}
where $H_{i_{\alpha}} = \sigma^x_{i_{\alpha}}\prod_{l} \textrm{CZ}_l$ and $l$ denotes all links $l$ surrounding the site $i_{\alpha}$, as depicted in Fig.~\ref{Fig: 2D ASPT}(e) by orange lines for $\alpha=A$, $B$, and $C$, respectively.
In this Hamiltonian, $\textrm{CZ}_l$ terms in $H_A$ and $H_B$ characterize the decoration of $(1+1)D$ cluster chains on the domain walls of $C$ spins.
Specifically, $H_{i_A}$ degrades to $\sigma^z_{j_B} \sigma^x_{i_A} \sigma^z_{k_B}$ if $j_B$, $i_A$, and $k_B$ sites constitute the domain wall, while $H_{i_A}$ becomes a single $\sigma^x_{i_A}$ for other sites.
A similar argument holds for $H_{i_B}$ in a symmetric fashion.
Meanwhile, $H_{i_C}$ flips the spin on $i_C$ and each surrounding link $l$ is transformed between the trivial and cluster states by a $\textrm{CZ}_l$ gate.

For the ASPT density matrix, the quantum superposition of $A$ spins is replaced by an ensemble of paramagnetic states, leading to the following disordered Hamiltonians
\begin{align}
    &H_0 = -\sum_i H_{i_B}-\sum_i H_{i_C}, \\ 
    &H_{\rm disorder} = -\sum_{i}h_{i_A}\sigma_{i_A}^z,
\end{align}
where $h_{i_A} = \pm 1$ determines the configuration of $A$ spins, as shown in Fig.~\ref{Fig: 2D ASPT}(f).
It can be verified that $\hat{O}_B = \prod_{i}\sigma^x_{i_B}$ and $\hat{O}_C = \prod_{i}\sigma^x_{i_C}$ commute with both $H_0$ and $H_{\rm disorder}$, while $U_A = \prod_i \sigma_{i_A}^xe^{i\frac{\pi}{4}(1-\sigma_{i_B}^x)}$ commutes only with $H_0$ but anticommutes with $H_{\rm disorder}$, indicating that $U_A$ is a weak symmetry of the disordered system.
In particular, $U_A$ does not commute with $H$ in Eq.~\eqref{2+1D SPT}, suggesting that the original $(2+1)D$ SPT state does not preserve this symmetry.
This indicates that our disordered ensemble belongs to an intrinsic ASPT phase that cannot be realized in closed systems with the same symmetry structure.

\section{Summary and outlook}
In this paper, we demonstrate that LPDOs provide systematic tensor network representations of ASPT phases in mixed quantum states, protected by the collaboration of a strong symmetry and a weak symmetry in the $(1+1)D$ and $(2+1)D$ systems.

We introduce short-range correlated LPDOs characterized by two distinct injectivity conditions: strong injectivity and weak injectivity.
Weak injectivity ensures that the LPDO can be locally transformed into a trivial product state using local quantum channels and prohibits the existence of long-range linear correlation functions of any local operators.
Strong injectivity, on the other hand, ensures the absence of long-range R\'enyi-2 correlation functions of any local operators.
Taking advantage of these injectivity conditions, we reproduce the full classification of the $(1+1)D$ ASPT phases for generic symmetry groups.
Additionally, we present explicit LPDO forms for $(1+1)D$ ASPT density matrices, including an intrinsic $\Z_4$ ASPT state lacking a pure state realization.

We expand our LPDO construction of ASPT phases to $(2+1)D$ open quantum systems, where we define the strong and weak injectivity conditions of $(2+1)D$ LPDOs.
Similar to the $(1+1)D$ cases, weak injectivity signifies short-ranged linear correlation functions of any local operator, while strong injectivity implies short-ranged R\'enyi-2 correlation functions of any local operator.
As a representative case, we investigate the classification of ASPT states protected by a strong $\Z_2^f$ symmetry and a weak bosonic $G_b$ symmetry.
Our exploration is ended by the explicit construction of a $(2+1)D$ intrinsic ASPT phase on the honeycomb lattice protected by $\Z_2\times \Z_4$ symmetry without any pure-state counterpart.

We emphasize that our approach not only reproduces previous classification results of ASPT phases, but also offers a clearer and more concise formalism with the symmetry action directly embedded in the local tensor.
This formulation naturally reveals the picture of decorated domain walls within the LPDO construction. Moreover, our method provides a systematic approach to constructing tensor network representations of ASPT phases for those with pure-state counterparts, with a novel example of intrinsic ASPT phases.
Moreover, the physical realization of LPDO in decohered and disordered systems provides a pathway for the experimental demonstration of those novel ASPT phases.
This approach bridges the gap between the advanced theoretical study of complex topological phases and their practical experimental realization, making these topological phenomena accessible for investigation in real-world systems.

We end this work with some open questions:
\begin{enumerate}[1.]
\item \textit{Fixed-point tensor of intrinsic ASPT phases}:
In this work, we have laid out some general methods to construct fixed-point tensors for ASPT phases with pure-state SPT correspondence. However, despite several examples that have already been proposed in this work, the way to construct the fixed-point tensor for general intrinsic ASPT is still missing.
One possible approach could involve constructing the fixed-point tensor of an SPT state within the double state formalism.
However, the challenge lies in utilizing the conditions of Hermiticity and positivity to convert this tensor back into an LPDO.
This line of investigation holds the potential to deepen our understanding of the structure of quantum phases in doubled space, shedding light on the underlying mechanisms of intrinsic ASPT phases.
\item \textit{Boundary anomaly of ASPT states}:
The extension of the 't Hooft anomaly from global symmetry to mixed states remains an unresolved issue~\cite{Hsin2024, Lessa2024A, Wang2024}.
Here, our primary focus lies in tensor network states with periodic boundary conditions.
Leveraging the LPDO constructions of ASPT states developed in this work, we can delve into the boundary physics of these states by placing the LPDO on systems with open boundary conditions.
\item \textit{Topologically ordered mixed states}: 
In this work, our focus has been restricted to injective LPDOs when we explore the construction and classification of fixed-point tensors for ASPT states.
However, an intriguing avenue for future research lies in investigating LPDO tensor network states with topological order by gauging our injective LPDO in $(2+1)D$.
Specifically, an interesting question to explore is the distinction between gauging weak symmetry and gauging strong symmetry, which has the potential to deepen our understanding of topological orders in mixed states~\cite{Sohal2025, Ellison2025}.
\end{enumerate}

\acknowledgments{We thank Meng Cheng, Ke Ding, Sung-Sik Lee, Zhu-Xi Luo, Ruben Verresen, Chong Wang, Xiao-Gang Wen, Yichen Xu, and Carolyn Zhang for stimulating discussions. Y. Guo, H.-R. Zhang, and S. Yang are supported by the National Natural Science Foundation of China (NSFC) (Grant No. 12174214 and No. 12475022) and the Innovation Program for Quantum Science and Technology (Project 2021ZD0302100). J.-H. Zhang and Z. Bi are supported by the startup fund from the Pennsylvania State University.}

\textit{Note} -- During the completion of this work, we were aware of an independent work~\cite{Yimu2024} which also addresses the tensor network formalism of mixed-state symmetry-protected topological phases. 

\bibliography{references}
\appendix
\renewcommand{\theequation}{S\arabic{equation}} \setcounter{equation}{0}
\renewcommand{\thefigure}{S\arabic{figure}} \setcounter{figure}{0}

\section{LPDO construction of mixed state with SW-SSB}\label{Sec.SWSSB}
In Ref.~\cite{Lessa2024B}, the authors proposed a class of short-range correlated density matrices in the sense of linear correlation function in Eq.~\eqref{Linear_corr}, while exhibiting a long-range R\'enyi-2 correlation in Eq.~\eqref{Renyi2} of operators charged under strong symmetry but neutral under weak symmetry, which defines as a strong-to-weak spontaneous symmetry breaking (SW-SSB) state.
A prototypical example of a density matrix that exhibits SW-SSB can be written as
\begin{align}
    \rho = \frac{1}{2^N}\left[\prod_i \sigma_i^0 + \prod_i \sigma_i^x\right] \equiv \rho_1+\rho_2.
\end{align}
Roughly speaking, there are two macroscopic parts in the mixture whose double state mimics the well-known GHZ state, implying a possible symmetry-breaking pattern in the density matrix.
To be more specific, this mixed state exhibits a strong symmetry $K=\prod_i \sigma_i^x$ that satisfies $K\rho = \rho$.
However, such a symmetry action induces a mapping between two components in the density matrix, i.e., $K\rho_1 = \rho_2$ and $K\rho_2=\rho_1$, with each component only possessing the weak symmetry $K\rho_i K^{\dagger} = \rho_i$ for $i=1, 2$.
Therefore, the strong symmetry is spontaneously broken into weak symmetry in this density matrix.
To detect this pattern, we consider the following R\'enyi-2 correlator
\begin{align}
    \mathcal{C}^{(2)}(i, j) = \frac{\mathrm{Tr}\left(\rho \sigma_i^z \sigma_j^z\rho \sigma_i^z \sigma_j^z\right) - \mathrm{Tr}\left(\rho \sigma_i^z \rho \sigma_i^z\right)\mathrm{Tr}\left(\rho \sigma_j^z\rho \sigma_j^z\right)}{\mathrm{Tr}(\rho^2)}
\end{align}
where we have $\sigma_i^z\rho_1 \sigma_i^z = \rho_1$, while $\sigma_i^z\rho_2\sigma_i^z=-\rho_2$, from which we derive
\begin{align}
    \mathcal{C}^{(2)}(i, j) = \frac{\mathrm{Tr}\left(\rho^2\right) - \left[\mathrm{Tr}\left(\rho_1^2-\rho_2^2\right)\right]^2}{\mathrm{Tr}(\rho^2)} = 1.
\end{align}

To construct the LPDO representation of the above density matrix, we note the fact that it can be purified to a cluster state $\ket{\psi}$ in Eq.~\eqref{cluster}
\begin{align}
    \rho = \mathrm{Tr}_{\tau}[\ket{\psi}\hspace{-0.5mm}\bra{\psi}],
\end{align}
since if the ket and bra have the same domain wall configuration, they must be completely aligned or anti-aligned for every spin.
Therefore, the LPDO representation can be directly inherited from the cluster state MPS in Eq.~\eqref{MPS_cluster} by converting the physical index of $\tau$ spin into the Kraus index.
Moreover, since the purified cluster state is represented by an injective MPS, the LPDO satisfies the weak injectivity condition by definition.
To demonstrate the violation of the strong injectivity condition, we recall the symmetry action of the local tensor as
\begin{align}
\begin{tikzpicture}[scale=0.8]
\tikzstyle{sergio}=[rectangle,draw=none]
\filldraw[fill=white, draw=black, rounded corners] (0.25,-0.5)--(1.75,-0.5)--(1.75,0.5)--(0.25,0.5)--cycle;
\draw[line width=1pt] (1.75,0) -- (2.25,0);
\draw[line width=1pt] (0.25,0) -- (-0.25,0);
\draw[line width=2pt, color=red] (1,0.5) -- (1,2);
\path (1,0) node [style=sergio]{\large $\mathsf{A}$};
\path (2.75,0) node [style=sergio]{$=$};
\filldraw[fill=white, draw=black] (1,1.25)circle (10pt);
\path (1,1.25) node [style=sergio]{$\sigma_x$};
\filldraw[fill=white, draw=black, rounded corners] (4.75,-0.5)--(6.25,-0.5)--(6.25,0.5)--(4.75,0.5)--cycle;
\draw[line width=1pt] (4.75,0) -- (3.25,0);
\filldraw[fill=white, draw=black] (4,0)circle (10pt);
\path (4,0) node [style=sergio]{$X$};
\path (5.5,0) node [style=sergio]{\large $\mathsf{A}$};
\draw[line width=1pt] (6.25,0) -- (7.75,0);
\filldraw[fill=white, draw=black] (7,0)circle (10pt);
\path (7,0) node [style=sergio]{$X$};
\draw[line width=2pt, color=red] (5.5,0.5) -- (5.5,1);
\draw[line width=3pt, color=blue] (5.5,-1) -- (5.5,-0.5);
\draw[line width=3pt, color=blue] (1,-1) -- (1,-0.5);
\end{tikzpicture},
\end{align}
and
\begin{align}
\begin{tikzpicture}[scale=0.8]
\tikzstyle{sergio}=[rectangle,draw=none]
\filldraw[fill=white, draw=black, rounded corners] (0.25,-0.5)--(1.75,-0.5)--(1.75,0.5)--(0.25,0.5)--cycle;
\draw[line width=1pt] (1.75,0) -- (2.25,0);
\draw[line width=1pt] (0.25,0) -- (-0.25,0);
\draw[line width=2pt, color=red] (1,0.5) -- (1,1);
\draw[line width=3pt, color=blue] (1,-2) -- (1,-0.5);
\path (1,0) node [style=sergio]{\large $\mathsf{A}$};
\path (2.75,0) node [style=sergio]{$=$};
\filldraw[fill=white, draw=black] (1,-1.25)circle (10pt);
\path (1,-1.25) node [style=sergio]{$\tau_x$};
\filldraw[fill=white, draw=black, rounded corners] (4.75,-0.5)--(6.25,-0.5)--(6.25,0.5)--(4.75,0.5)--cycle;
\draw[line width=1pt] (4.75,0) -- (3.25,0);
\filldraw[fill=white, draw=black] (4,0)circle (10pt);
\path (4,0) node [style=sergio]{$Z$};
\path (5.5,0) node [style=sergio]{\large $\mathsf{A}$};
\draw[line width=1pt] (6.25,0) -- (7.75,0);
\filldraw[fill=white, draw=black] (7,0)circle (10pt);
\path (7,0) node [style=sergio]{$Z$};
\draw[line width=2pt, color=red] (5.5,0.5) -- (5.5,1);
\draw[line width=3pt, color=blue] (5.5,-1) -- (5.5,-0.5);
\end{tikzpicture}.
\end{align}
The first one is the standard form for a strong symmetry of the density matrix, while the second one leads to the following property
\begin{align}
    \begin{tikzpicture}[scale=0.8]
\tikzstyle{sergio}=[rectangle,draw=none]
\filldraw[fill=white, draw=black, rounded corners] (-0.25,-1.5)--(1.25,-1.5)--(1.25,-0.5)--(-0.25,-0.5)--cycle;
\filldraw[fill=white, draw=black, rounded corners] (-0.25,-3)--(1.25,-3)--(1.25,-2)--(-0.25,-2)--cycle;
\draw[line width=1pt] (1.25,-1) -- (1.75,-1);
\draw[line width=1pt] (-0.25,-1) -- (-0.75,-1);
\draw[line width=1pt] (1.25,-2.5) -- (1.75,-2.5);
\draw[line width=1pt] (-0.25,-2.5) -- (-0.75,-2.5);
\draw[line width=2pt, color=red] (0.5,-0.5) -- (0.5,0);
\draw[line width=3pt, color=blue] (0.5,-1.5) -- (0.5,-2);
\draw[line width=2pt, color=red] (0.5,-3) -- (0.5,-3.5);
\path (0.5,-1) node [style=sergio]{\large $\mathsf{A}$};
\path (0.5,-2.5) node [style=sergio]{\large $\mathsf{A^*}$};
\path (2.5,-1.75) node [style=sergio]{$=$};
\filldraw[fill=white, draw=black, rounded corners] (4.75,-1.5)--(6.25,-1.5)--(6.25,-0.5)--(4.75,-0.5)--cycle;
\filldraw[fill=white, draw=black, rounded corners] (4.75,-3)--(6.25,-3)--(6.25,-2)--(4.75,-2)--cycle;
\draw[line width=1pt] (4.75,-1) -- (3.25,-1);
\filldraw[fill=white, draw=black] (4,-1)circle (10pt);
\path (4,-1) node [style=sergio]{$Z$};
\path (5.5,-1) node [style=sergio]{\large $\mathsf{A}$};
\draw[line width=1pt] (6.25,-1) -- (7.75,-1);
\filldraw[fill=white, draw=black] (7,-1)circle (10pt);
\path (7,-1) node [style=sergio]{$Z$};
\draw[line width=1pt] (4.75,-2.5) -- (3.25,-2.5);
\filldraw[fill=white, draw=black] (4,-2.5)circle (10pt);
\path (4,-2.5) node [style=sergio]{$Z$};
\path (5.5,-2.5) node [style=sergio]{\large $\mathsf{A^*}$};
\draw[line width=1pt] (6.25,-2.5) -- (7.75,-2.5);
\filldraw[fill=white, draw=black] (7,-2.5)circle (10pt);
\path (7,-2.5) node [style=sergio]{$Z$};
\draw[line width=2pt, color=red] (5.5,-3) -- (5.5,-3.5);
\draw[line width=3pt, color=blue] (5.5,-1.5) -- (5.5,-2);
\draw[line width=2pt, color=red] (5.5,-0.5) -- (5.5,0);
\end{tikzpicture}.
\end{align}
Here, the local tensor for the double state remains unchanged under the symmetry transformation on only virtual indices.
As a result, the double transfer matrix $\mathbb{E}^2$ in Eq.~\eqref{strong transfer matrix} has degenerate dominant eigenvalues, leading to the long-range order in R\'enyi-2 correlation function evaluated in the given density matrix.
This example illustrates the necessity to impose the strong injectivity condition on the LPDO for our classification of ASPT phases to prevent such a hidden symmetry-breaking pattern that cannot be discovered by only linear correlators.
On the other hand, it also provides a general method to construct a mixed state with SW-SSB: we can start from a symmetric state with two degrees of freedom on one site, and trace out one of them that has a nontrivial symmetry action on virtual indices.
The resulting LPDO will not be strongly injective and exhibit an SW-SSB pattern.
\end{document}